\documentclass[twocolumn,showpacs,preprintnumbers,nofootinbib,prd,
superscriptaddress,10pt]{revtex4-1}

\usepackage{amsmath,amssymb}
\usepackage[normalem]{ulem}
\usepackage{textcomp}
\usepackage{hyperref}
\usepackage{bm}
\usepackage{graphicx}
\usepackage{psfrag}
\usepackage[usenames,dvipsnames]{xcolor}
\usepackage[utf8]{inputenc}

\graphicspath{{./figures/}}
\allowdisplaybreaks[4]

\newcommand{\sub}[1]{_{\text{#1}}}
\newcommand{\super}[1]{^{\text{#1}}}
\newcommand{\uvec}[1]{\bm{\hat{#1}}}
\newcommand{\dvec}[1]{\dot{\bm{#1}}}

\newcommand{\duvec}[1]{\dot{\bm{\hat{#1}}}}

\newcommand{\ord}[1]{\mathcal{O} \left( #1 \right)}

\begin{document}

\title{Fourier domain gravitational waveforms for precessing eccentric binaries}

\date{\today}

\author{Antoine Klein}
\affiliation{Department of Physics and Astronomy, The University of Mississippi, University, Mississippi 
38677, USA}
\affiliation{CENTRA, Departamento de F\'isica, Instituto Superior T\'ecnico, Universidade de 
Lisboa, Avenida Rovisco Pais 1, 1049 Lisboa, Portugal}
\affiliation{CNRS, UMR 7095, Institut d'Astrophysique de Paris, 98 bis Bd Arago, 75014 Paris, 
France}

\author{Yannick Boetzel}
\affiliation{Physik-Institut, Universit\"at Z\"urich, Winterthurerstrasse 190, 8057 Z\"urich, 
Switzerland}

\author{Achamveedu Gopakumar}
\affiliation{Department of Astronomy and Astrophysics, Tata Institute of Fundamental Research, 
Mumbai 400005, India}

\author{Philippe Jetzer}
\affiliation{Physik-Institut, Universit\"at Z\"urich, Winterthurerstrasse 190, 8057 Z\"urich, 
Switzerland}

\author{Lorenzo de Vittori}
\affiliation{Physik-Institut, Universit\"at Z\"urich, Winterthurerstrasse 190, 8057 Z\"urich, 
Switzerland}

\begin{abstract}
We build two families of inspiral waveforms for precessing binaries on eccentric orbits in the 
Fourier domain. To achieve this, we use a small eccentricity expansion of the waveform amplitudes 
in order to separate the periastron precession timescale from the orbital timescale, and use a 
shifted uniform asymptotics transformation to compute the Fourier transform in the presence of 
spin-induced precession. We show that the resulting waveforms can yield a median faithfulness above 
0.993 when compared to an equivalent time domain waveform with an initial eccentricity of $e_0 
\approx 0.3$. We also show that when the spins are large, using a circular waveform can potentially 
lead to significant biases in the recovery of the parameters, even when the system has fully 
circularized, particularly when the accumulated number of cycles is large. This is an effect of the 
residual eccentricity present when the objects forming the binary have nonvanishing spin 
components in the orbital plane.
\end{abstract}


\maketitle

\section{Introduction}

The recent discoveries of gravitational wave (GW) signals by the LIGO and Virgo Collaborations have
opened a new observation window on the Universe~\cite{LIGO,Virgo, GEO600,GW150914, GW151226, 
GW170104, GW170814, GW170817, GW170608}, through which the potential for new discoveries in 
astrophysics is truly tremendous. So far, those events have been analyzed with the assumption that 
the systems that produced them were evolving on circular orbits. Indeed, it has been a well-known 
fact that the emission of gravitational waves by binary systems has the tendency to circularize 
their orbits~\cite{peters-1964}. Nevertheless, it has been argued that certain astrophysical 
scenarios could lead to stellar-origin black holes binaries having high initial 
eccentricities~\cite{postnov-lrr, shappee-2012, antonini-2016, antonini-2017}, so they would 
still be measurable when the signal reaches the frequency window of the space-based GW detector 
LISA~\cite{nishizawa-2016-1, nishizawa-2016-2, breivik-2016}. Furthermore, recent results have 
shown that eccentricity measurements by LIGO could be used to constrain stellar-mass black hole 
formation mechanisms~\cite{antonini-2017, petrovich-2017, samsing-2017, rodriguez-2017, hoang-2018, 
samsing-2018}. It has been estimated that large biases in the recovery of the parameters of the 
first direct detection GW150914 could have occurred if the initial eccentricity in the detector was 
$e_0 \gtrsim 0.05$~\cite{GW150914-systematics} Supermassive black hole binaries could also have 
important eccentricities in the late inspiral, if triple systems are a significant ingredient of 
supermassive black hole evolution~\cite{blaes-2002, hoffman-2007, amaro-seoane-2010, 
bonetti-2017-1, bonetti-2017-2, bonetti-2017-3}. Furthermore, in some spin configurations, it has 
been shown that the eccentricity of the system never truly vanishes, but reaches a stationary value 
where it ceases to decrease through the emission of GWs~\cite{klein-2010}.

This has motivated the development of waveforms that include the effects of 
a nonzero eccentricity in GW binary signals. The first steps towards this goal rely on the 
derivation of quasi-Keplerian equations describing the orbits~\cite{memmesheimer-2004}, the 
derivation of the evolution equations for the orbital elements~\cite{damour-2004, 
koenigsdoerffer-2006, arun-2008, arun-2008-2, arun-2009}, and the derivation of GW polarization 
amplitudes~\cite{mishra-2015}. The effects of individual spins were later added to this 
approach~\cite{gergely-1998, gergely-1999, gergely-2002, mikoczi-2005, keresztes-2005, klein-2010}. 
Using these solutions, several waveforms have been developed. Yunes \emph{et al.}~\cite{yunes-2009} 
proposed an analytic eccentric waveform in the post-Newtonian (PN) postcircular approximation, by 
solving for the Fourier phase of a binary signal analytically at Newtonian order using a small 
eccentricity expansion. Cornish and Key~\cite{cornish-2011, cornish-err, key-2011} and 
Gopakumar and Sch\"afer~\cite{gopakumar-2011} independently developed a numerical waveform in the 
time domain by solving the 1.5PN equations of motion numerically together with the spin-orbit 
precession equations, and using 1.5PN accurate amplitudes. Huerta \emph{et al.}~\cite{huerta-2014} 
expanded the analytical work of Yunes et al. by including the most important eccentricity-dependent 
terms up to 3.5PN order and at eighth order in the initial eccentricity for non-spinning systems. 
Tanay \emph{et al.}~\cite{tanay-2016} later computed the full 2PN Fourier phase for nonspinning systems 
at second order in the eccentricity. Moore \emph{et al.}~\cite{moore-2016} then expanded this result to 
3PN order. Huerta \emph{et al.}~\cite{huerta-2017-1,huerta-2017-2} and Hinder \emph{et al.}~\cite{hinder-2017} 
combined those results with numerical relativity to produce an eccentric inspiral-merger-ringdown 
waveform for nonspinning binaries. Recently, Hinderer and Babak~\cite{hinderer-2017} and Cao and 
Han~\cite{cao-2017} developed an eccentric waveform using a new approach in the effective one-body 
(EOB) formalism.

In this work, we further develop the formalism of post-Newtonian eccentric waveforms to include the 
effects of spin-induced precession in the Fourier domain. The advantage of Fourier domain waveforms 
over time domain ones is that they provide a much more computationally efficient way of computing a 
GW signal. Indeed, in order to produce a time domain waveform, one has to compute an equally spaced 
time series of the signal before computing its Fourier transform to use in detection or parameter 
estimation algorithms. The relevant timescale for this time series is the inverse of the maximum 
orbital frequency, which, being very short, makes this process computationally very expensive. 
Having a waveform available directly in the Fourier domain circumvents this problem and greatly 
reduces the computational cost of GW data analysis. In order to construct such a waveform, we solve 
the evolution equations for the orbital elements together with the orbit-averaged spin precession 
equations numerically at 3PN order, including spin effects at 2PN order. Using a quasi-Keplerian 
description of the orbit, we employ instantaneous nonspinning amplitudes to construct the resulting 
GW polarizations. We then use a shifted uniform asymptotics (SUA) technique~\cite{klein-2014} to 
compute the waveforms in the Fourier domain. The resulting waveform has the advantage that the 
phasing is computed without any expansion for small eccentricities and thus can be be very 
faithful compared to corresponding time domain waveforms for moderate to large eccentricities ($e 
\lesssim 0.4$). However, the amplitudes require a small-eccentricity expansion, and thus we do not 
expect the present waveforms to be faithful for 
arbitrarily large eccentricities.

In Sec.~\ref{sec:WF}, we derive two different families of eccentric waveforms. Due to the 
similarity between the orbital timescale and the periastron-to-periastron timescale, we derive the 
first family by expanding the Fourier domain waveform into combined harmonics of the mean orbital 
phase and of the mean anomaly. We then derive the second family by further expanding the resulting 
Fourier phase and time-frequency relations for small differences between the two similar phases.
In Sec.~\ref{sec:sim}, we describe simulations that we performed to compute the faithfulness 
between our Fourier domain waveforms and a corresponding time domain waveform, including a detailed 
summary of how these different waveforms are constructed. We also compare a circular waveform to 
probe which domain of the parameter space allows for such circular waveforms to be effectively used 
for parameter estimation of binary signals. 
We give concluding remarks in Sec.~\ref{sec:conc}. Throughout this paper, we use geometric 
units where $G=c=1$.

\section{Waveform}\label{sec:WF}

In the presence of spins, the orbit of a binary system is, in general, not restricted to an orbital 
plane~\cite{barker-1975}. Indeed, interactions between the spins and the orbit cause them to 
precess. However, in the post-Newtonian regime, the timescale on which this precession occurs is 
well separated from the other timescales present in the problem. We can therefore approximate the 
spin-orbital precession to be occurring much more slowly than the orbit, which allows us to 
describe it using a so-called quasi-Keplerian parametrization inside an orbital plane that stays 
perpendicular to the orbital angular momentum as the latter precesses. A quasi-Keplerian 
parametrization of the orbit of a spinning binary system is known at 3PN order for the 
nonspinning part~\cite{memmesheimer-2004,koenigsdoerffer-2006}, and at 2PN order for the 
spin-dependent part~\cite{klein-2010}. In this work, we restrict the quasi-Keplerian orbital 
description at 2PN for the computation of the polarization amplitudes. We can express the orbit at 
2PN order as
\begin{subequations}
\begin{align}
	r &= a ( 1 - e_r \cos u) +f_r(v) \,, \\
	\phi &= (1 + k) v + f_\phi(v) \,, \label{eq:phiofv} \\
	\tan \frac{v}{2} &= \sqrt{\frac{1 + e_\phi}{1 - e_\phi}}\tan \frac{u}{2} \,, \label{eq:vofu}\\
	l &= u - e_t \sin u + f_t(u,v) \,, \label{eq:lofu}\\
	\dot{l} &= n \,,
\end{align}
\end{subequations}
where $(r, \phi)$ is a polar representation of the separation vector in the orbital plane, $a$ is 
the semimajor axis; $u$ is the eccentric anomaly; $v$ is the true anomaly; $l$ is the mean 
anomaly; $n$ is the mean motion; $e_r$, $e_\phi$ and $e_t$ are eccentricity parameters; and the 
functions $f_i$ are general relativistic corrections given 
by~\cite{memmesheimer-2004,damour-2004,klein-2010}
\begin{subequations}
\begin{align}
	f_r(v) =&\; \sum_{i=0}^2 b_{r,i} \cos (2v - 2\psi_i) \,, \\
	f_\phi(v) =&\; \sum_{k=2}^3 a_{\phi,k} \sin(k v) \nonumber\\
		&+ \sum_{k=1}^2 \sum_{i=0}^2 b_{\phi,k,i} \sin(k v - 2\psi_i) \,,\\
	f_t(u,v) =&\; g_t (v - u) + a_{t} \sin(v) \,,
\end{align}
\label{eq:qK-relcorr}
\end{subequations}
where $\psi_i$ is the angle between the periastron line and the projection of spin $i$ onto the 
orbital plane (see Fig.~\ref{fig:angles}), $\psi_0 = (\psi_1 + \psi_2)/2$ and the constants $a_A$, 
$b_A$ and $g_t$ are listed in Appendix~\ref{sec:qK-param}. We complemented the spinning solution of 
~\cite{klein-2010} by including quadrupole-monopole terms as described in Appendix 
~\ref{sec:quad-mono}.

\begin{figure}[!ht]
\begin{center}
	\includegraphics[width=0.4\textwidth]{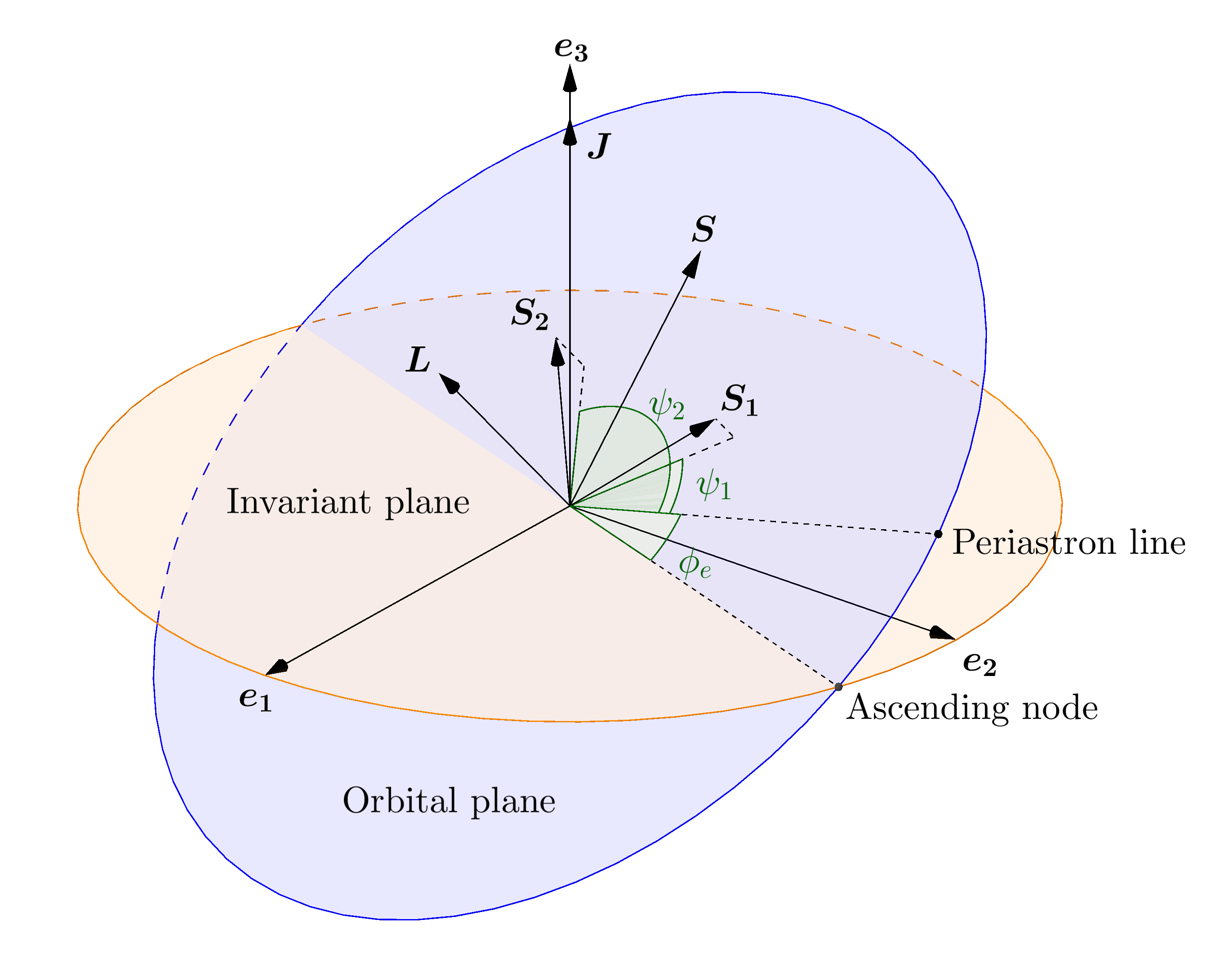}
	\caption{Angles used in the definition of the relativistic corrections defined in Eq.~\eqref{eq:qK-relcorr}. The orbital plane is perpendicular to the Newtonian orbital angular momentum $\bm{L}$, and the invariant plane is perpendicular to the conserved total angular momentum $\bm{J}$. The angle $\phi_e$ locates the periastron line with respect to the fixed invariant plane, and the angles $\psi_i$ are subtended by the periastron line and the projections of the spins onto the orbital plane.}
	\label{fig:angles}
\end{center}
\end{figure}

The orbital phase $\phi$ and the mean anomaly $l$ can be decomposed as the sum of a linearly 
growing part and a periodic part~\cite{damour-2004},
\begin{subequations}
\begin{align}
	\phi &= \lambda + W_\phi \,, \\
	\dot{\lambda} &= (1 + k)n \,, \\
	\dot{l} &= n \,, \\
	W_\phi &= (1 + k) (v - l) + f_\phi(v) \,.
\end{align}
\end{subequations}
We choose to express our equations in terms of the post-Newtonian parameter $y$ and the 
eccentricity parameter $e$ defined by
\begin{subequations}
\begin{align}
	y &= \frac{\left[ M (1 + k) n \right]^{1/3}}{\sqrt{1 - e_t^2}} \,, \\
	e &= e_t \,.
\end{align}
\end{subequations}
The constants in the quasi-Keplerian parametrization are given in terms of these parameters in 
Appendix~\ref{sec:qK-param}.

As the system emits gravitational waves, the orbital frequency and the eccentricity will evolve 
according to the following equations~\cite{damour-2004,klein-2010}
\begin{subequations}
\begin{align}
	M \frac{d y}{dt} &= \left( 1 - e^2	\right)^{3/2} \nu y^9 \left( a_0 + \sum_{n=2} a_n y^n 
		\right) \,, \label{eq:ydot} \\
	M \frac{de^2}{dt} &= - \left( 1 - e^2 \right)^{3/2} \nu y^8 \left( b_0 + \sum_{n=2} b_n y^n 
		\right) \,, \label{eq:e2dot}
\end{align}
\end{subequations}
where the constants $a_i$ and $b_i$ are given at 3PN order for nonspinning systems and at 2PN 
order for spinning systems in Appendix~\ref{sec:evol-eq}. Here we also complemented the spinning 
solution of~\cite{klein-2010} by including quadrupole-monopole terms as described in 
Appendix~\ref{sec:quad-mono}. We found that the minimum value for the eccentricity $e^2\sub{min}$ 
found in~\cite{klein-2010} is unchanged by the addition of quadrupole-monopole effects, with
\begin{align}
	e^2\sub{min} &= \frac{5 y^4}{304} \sigma(-1, 1, 0, 2, -2, 0) \nonumber\\
	&= \frac{5 y^4}{304} \left| \bm{s}_\perp^{(-)} \right|^2 \,, 
		\label{eq:e2min}
\end{align}
where the 2PN spin-spin coupling $\sigma$ can be found in Appendix~\ref{sec:quad-mono}, and
\begin{align}
	\bm{s}_\perp^{(-)} &= \left[ \bm{s}_1 - \uvec{L} \left( \uvec{L} \cdot \bm{s}_1 \right) \right] 
		- \left[ \bm{s}_2 - \uvec{L} \left( \uvec{L} \cdot \bm{s}_2 \right) \right] ,
\end{align}
where $\uvec{L}$ is a normal to the orbital plane.

Note that we found a typo in~\cite{klein-2010}, where the constant factor in $e^2\sub{min}$ should 
read $5/304$ instead of $5/340$. This minimum eccentricity depends on the spin orientations: it is 
maximal when the spins lie inside the orbital plane and are opposite to one another, and it vanishes 
when the projections of $\bm{s}_1$ and $\bm{s}_2$ onto the orbital plane are equal to each other. 
The maximum value it can take is independent of the mass ratio; it is $e^2\sub{min} = 5 y^4/304$, 
which evaluates to $e\sub{min} \approx 0.021$ at the ISCO defined by $y = 6^{-1/2}$, and it is 
multiplied by a factor $(f/f\sub{ISCO})^{2/3}$ earlier in the inspiral. Note that this minimum 
eccentricity, being a spin effect, is unrelated to a similar effect observed in extreme mass-ratio 
inspirals around Schwarzschild black holes in~\cite{cutler-1994}, and also unrelated to another 
effect due to orbital effects derived in~\cite{loutrel-2018}, which is of order $e^2\sub{min} \sim 
y^{10}$ and is independent of the spins.

The 2PN orbit-averaged equations of precession are given by~\cite{barker-1975, racine-2008}
\begin{subequations}
\begin{align}
	M \duvec{L} &= - \left( 1 - e^2 \right)^{3/2} y^6 \left( \bm{\Omega}_1 + \bm{\Omega}_2 \right) 
		\,, \label{eq:Lhatdot} \\
	M\dvec{s}_1 &= \left( 1 - e^2 \right)^{3/2} \mu_2 y^5 \bm{\Omega}_1 \,, \label{eq:s1dot} \\
	M\dvec{s}_2 &= \left( 1 - e^2 \right)^{3/2} \mu_1 y^5 \bm{\Omega}_2 \,, \label{eq:s2dot}
\end{align}
\end{subequations}
where we defined the reduced spins $\bm{s}_i = \bm{S}_i / m_i$, the reduced individual masses 
$\mu_i = m_i/M$, and the precession vectors $\bm{\Omega}_i$ are given by
\begin{align}
	\bm{\Omega}_i =&\; \bigg\{ \left[ 2 \mu_i + \frac{3}{2} \mu_j - \frac{3}{2} y\uvec{L} \cdot 
		\left( \bm{s} + (q_i - 1) \bm{s}_i \right) \right] \uvec{L} \nonumber\\
		&+ \frac{1}{2} y \bm{s}_j \bigg\} \times \bm{s}_i \,,
\end{align}
where $i,j \in \{1, 2\}$, $i \neq j$, and the $q_i$ are quadrupole parameters defined in such a way that $q_i=1$ for black holes.

The gravitational waveform emitted by a binary system on such an orbit has been computed at 3PN 
order for nonspinning binaries, omitting tail effects~\cite{mishra-2015}. The result has the 
following structure:
\begin{subequations}
\begin{align}
	h(t) =&\; F_+ (t) h_+ (t) + F_\times (t) h_\times (t) \,,\label{eq:hoft} \\
	h_{+,\times}(t) =&\; \sum_{n \in \mathbb{Z}} H_{+,\times}^{(n)} (y, e, e \cos u, e \sin u) 	
		e^{in (\phi + \phi_T)} \,, \label{eq:hpcoft}
\end{align}
\end{subequations}
where $F_+$ and $F_\times$ are antenna pattern functions~\cite{cutler-1998}, and the Thomas 
precession phase $\phi_T$ is given by~\cite{apostolatos-1994}
\begin{align}
	 \dot{\phi}_T &= \frac{\uvec{L} \cdot \uvec{N}}{1 - \left(\uvec{L} \cdot \uvec{N} \right)^2} 
		 \left( \uvec{L} \times \uvec{N} \right) \cdot \duvec{L} \,, \label{eq:phiTdot}
\end{align}
with respect to a given sky location vector $\uvec{N}$.

In order to compute the Fourier transform of this signal, we need to separate the orbital timescale 
from the precessional one, and express the orbital timescale dependence in terms of linearly 
growing phases. To do so, we follow~\cite{boetzel-2017} and compute an inversion of the 
PN-accurate Kepler equation \eqref{eq:lofu} as
\begin{align}
	 u &= l + \sum_{s=1}^{\infty} A_s \sin(sl) \,,
\end{align}
with the Fourier coefficients $A_s$ given by
\begin{align}
	 A_s &= \frac{2}{s} J_s(s e) + \sum_{j=1}^{\infty} \alpha_j \left[ J_{s+j} (s e) - 
		 J_{s-j}(se)\right] \,.
\end{align}
The PN-accurate constants $\alpha_j$ can be computed from~\cite{memmesheimer-2004} and are 
given in Eq. (18) of~\cite{boetzel-2017}. Similarly, we can find a Fourier expansion of the true 
anomaly $v$ and the orbital phase $\phi$ in terms of the mean anomaly $l$:
\begin{subequations}
\begin{align}
	v &= l + \sum_{s=1}^{\infty} B_s \sin(sl) \,,\\
	\phi &= \lambda + \sum_{s=1}^{\infty} C_s \sin(sl) \,.
\end{align}
\end{subequations}
The Fourier coefficients $A_s$, $B_s$ and $C_s$ can be found up to~$\ord{y^4, e^5}$ in 
Appendix~\ref{sec:qK-inv}. Using this solution, we can then express
\begin{subequations}
\begin{align}
	e^{i k u} &= \sum_{p \in \mathbb{Z}} \epsilon_{p}^{ku}\; e^{-i p l} \,, \\ 
	e^{i k v} &= \sum_{p \in \mathbb{Z}} \epsilon_{p}^{kv}\; e^{-i p l} \,,\\
	e^{i k \phi} &= e^{i k \lambda} \sum_{s \in \mathbb{Z}} \mathcal{P}_{p}^{k\phi}\; e^{-i p l} \,,
\end{align}
\end{subequations}
where the coefficients $\epsilon_{p}^{ku}$, $\epsilon_{p}^{kv}$ and $\mathcal{P}_{p}^{k\phi}$ are 
given as a Taylor expansion in both $e$ and $y$. We refer to Eqs.~(30), (34), (E11) 
of~\cite{boetzel-2017} for how to calculate these Fourier coefficients.

This small eccentricity expansion allows us to express the waveform as
\begin{align}
	h_{+,\times}(t) &= \sum_{n\in\mathbb{Z}}\sum_{p\in\mathbb{Z}} H_{+,\times}^{(p,n)} 
		e^{-i (n \lambda + p l)} \,,
\end{align}
where we included the Thomas phase $\phi_T$ into the amplitudes $H_{+,\times}^{(p,n)}$, which vary 
on the spin-precession timescale. To separate the periastron precession timescale from the orbital 
timescale, we define $\delta\lambda = \lambda - l$, such that
\begin{subequations}
\begin{align}
	 M \dot{\lambda} &= \left(1 - e^2 \right)^{3/2} y^3 \,, \label{eq:lambdadot}\\
	 \delta\dot{\lambda} &= \frac{k}{1 + k} \dot{\lambda} \,. \label{eq:deltalambdadot}
\end{align}
\end{subequations}
This new angle defines the periastron precession timescale, which is similar to the spin precession 
timescale since $\delta\dot{\lambda}/\dot{\lambda} = \ord{y^2}$.

Using this, we can then further simplify the waveform with
\begin{subequations}\label{eq:waveform+coeff}
\begin{align}
	h_{+,\times}(t) &= \sum_{n \in \mathbb{Z}} H_{+,\times}^{(n)} e^{-i n \lambda} \,,	
		\label{eq:waveform}\\
	H_{+, \times}^{(n)} &= \frac{M \nu y^2}{d_L} \sum_{m \in \mathbb{Z}} G_{+, \times}^{(m,n)} 
		e^{-i m \delta\lambda} e^{-i(n+m)\phi_T} \,. \label{eq:waveformcoeff}
\end{align}
\end{subequations}
The amplitudes $G_{+,\times}^{(m,n)}$ are given in Appendix~\ref{sec:Gmn} at order $\ord{y^2, e}$ 
\footnote{A {\sc{Mathematica}} version of all amplitudes to order $\ord{y^4, e^{10}}$ is available as 
supplemental material or upon request from 
\href{mailto:boetzel@physik.uzh.ch}{boetzel@physik.uzh.ch}.}.

\subsection{Fourier transform approximations}

Before we compute an approximation of the Fourier transform of our signal, let us introduce 
two useful techniques.

Let us first assume that we have a signal of the form
\begin{align}
	h(t) &= A(t) e^{-i \phi(t)} \,,
\end{align}
with $\dot{\phi}(t)$ a positive and monotonically increasing function of time, and that we want to 
compute its Fourier transform
\begin{align}
	\tilde{h}(f) &= \int_{-\infty}^\infty h(t) e^{2\pi i f t} dt \,.
\end{align}

The stationary phase approximation (SPA) of this Fourier transform consists in Taylor expanding the 
amplitude $A(t)$ and phase $\phi(t)$ around the stationary point $t_0$ defined by the relation 
\begin{align}
 	2 \pi f &= \dot{\phi}(t_0) \,,
\end{align}
keeping only the constant term in the expansion of the amplitude and up to the quadratic order in 
the expansion of the phase. We get
\begin{align}
	h(t) \approx&\; A(t_0) \exp \bigg\{-i \bigg[ \phi(t_0) \nonumber\\
		&+ \dot{\phi}(t_0) (t - t_0) + \frac{1}{2} \ddot{\phi}(t_0) (t - t_0)^2 \bigg] \bigg\} \,.
\end{align}
We can compute the Fourier transform of this approximate signal analytically, and we get
\begin{align}
	\tilde{h}(f) &\approx \sqrt{\frac{2 \pi}{| \ddot{\phi}(t_0) |^2}} A(t_0) e^{i (2 \pi f t_0 - 
		\phi(t_0) - \pi/4)} \,.
\end{align}
This approximation will be accurate if $|\dot{A}/A| \ll |\ddot{\phi}|^{1/2}$ around the stationary point, and if the quadratic approximation is accurate around the stationary point. For a formal derivation, see e.g.~\cite{Bender}.

Let us now suppose that our signal is of the form
\begin{align}
	h(t) &= A e^{-i [\phi_C + B \sin \beta]} \,,
\end{align}
with $\dot{A}/A = \ord{y^8}$, $\dot{\phi}_C = \ord{y^3}$, $B = \ord{y}$, $\dot{\beta} = \ord{y^5}$, 
and that each additional time derivative adds a factor $\ord{y^8}$ to the various quantities 
present in the signal, with $y$ a small expansion parameter. This is the simplified form of a GW 
signal that we expect from a binary system undergoing spin-induced orbital precession, with $y$ 
being a PN expansion parameter. The SPA cannot be directly used in this case, because the two terms 
in the second time derivative of the signal phase
\begin{align}
	\ddot{\phi} &= \ddot{\phi}_C - B \dot{\beta}^2 \sin\beta + \ord{y^{13}}
\end{align}
are of the same PN order and can cancel each other. The shifted uniform asymptotics (SUA) 
method~\cite{klein-2014} offers an approximation of the Fourier transform of such a signal by first 
expanding the signal using Bessel functions as
\begin{align}
	h(t) &= A \sum_k J_k(B) e^{-i(\phi_C + k \beta)} \,,
\end{align}
so its Fourier transform can be approximated by a series of SPA, since $\ddot{\beta} \ll 
\ddot{\phi}_C$. The Fourier transform then becomes
\begin{align}
	\tilde{h}(f) \approx&\; \sum_k A \sqrt{\frac{2 \pi}{\ddot{\phi}_C + k \ddot{\beta}}} \nonumber\\
		 &\times \exp \left[ i \left( 2 \pi f t_k - \phi_C - k \beta - \frac{\pi}{4} \right) 
		 \right] \,, \label{eq:SUASPA}
\end{align}
where the various functions are evaluated for each $k \in \mathbb{Z}$ at the stationary time $t_k$ defined by 
\begin{align}
	2 \pi f &= \dot{\phi}_C(t_k) + k \dot{\beta}(t_k) \,.
\end{align}

The different stationary times can be related to each other by Taylor expanding their defining 
equations around $t_0$ and solving for the difference order by order:
\begin{align}
	t_k - t_0 &= - \frac{k \dot{\beta}(t_0)}{\ddot{\phi}_C(t_0)} + \ord{y^{-4}} \,.
\end{align}
By Taylor expanding Eq.~\eqref{eq:SUASPA} around $t_0$, and keeping only the leading PN order 
amplitude and the phase accurate to order $\ord{y^0}$, we obtain
\begin{subequations}
\begin{align}
	\tilde{h}(f) \approx&\; \tilde{h}_0(f) \tilde{h}\sub{corr}(f) \,, \\
	\tilde{h}_0(f) =&\; \sqrt{\frac{2 \pi}{| \ddot{\phi}(t_0) |^2}} A(t_0) e^{i (2 \pi f t_0 - 
		\phi_C(t_0) - \pi/4)} \,, \\
	\tilde{h}\sub{corr}(f) =&\; \sum_k J_k [B(t_0)] \exp\left[ - k \beta(t_0) + \frac{1}{2} T^2 k^2 
		\dot{\beta}^2(t_0) \right], \\
	T =&\; \sqrt{\frac{1}{\ddot{\phi}_C(t_0)}} \,.
\end{align}
\end{subequations}
After some manipulation, we can resum the Bessel functions in $\tilde{h}\sub{corr}(f)$ as
\begin{align}
	\tilde{h}\sub{corr} (f) &= \sum_{p \geq 0} \frac{\left( - i T^2\right)^p}{2^p p!} 
		\partial_t^{2p} e^{-i B \sin \beta} \,,
\end{align}
where the functions are evaluated at $t=t_0$. Truncating this series at some order $p = k\sub{max}$ 
and using a stencil around $t_0$ to approximate the different order time derivatives, we obtain the 
SUA approximation
\begin{align}
	\tilde{h}\sub{corr} (f) &\approx \sum_{k=-k\sub{max}}^{k\sub{max}} a_{k,k\sub{max}} e^{-i B 
	\sin \beta(t_0 + k T)} \,,
\end{align}
where the constants $a_{k,k\sub{max}}$ satisfy the following linear system of equations:
\begin{subequations}
\label{eq:akdef}
\begin{align}
	&\sum_{k=1}^{k\sub{max}} a_{k,k\sub{max}} + \frac{1}{2} a_{0, k\sub{max}} = 1 \,, \\
	&\sum_{k = 1}^{k\sub{max}} a_{k, k\sub{max}} \frac{k^{2p}}{(2p)!} = \frac{(-i)^p}{2^p p!} \,, 
		\quad p \in \{1, \ldots, k\sub{max} \} \,, \\
	&a_{-k,k\sub{max}} = a_{k,k\sub{max}} \,.
\end{align}
\end{subequations}

To summarize, if we are able to separate the spin-precessional timescale effects from
a carrier phase $\phi_C$ that satisfies $\dot{\phi}_C > 0$ and $\ddot{\phi}_C > 0$ as
\begin{align}
	h(t) &= A(t) e^{-i \phi_C(t)} \,,
\end{align}
where all spin-precessional timescale effects are included in $A(t)$, then we can 
write the SUA approximation of its Fourier transform:
\begin{align}
	\tilde{h}(f) =&\; \sqrt{\frac{2\pi}{\ddot{\phi}_C}} e^{i(2\pi f t_0 - \phi_C(t_0) - \pi/4)} 
		\nonumber\\
 		&\times \sum_{k = -k\sub{max}}^{k\sub{max}} a_{k, k\sub{max}} A(t_0 + k T) \,,
\end{align}
with the constants $a_{k,k\sub{max}}$ satisfying the linear system of Eqs.~\eqref{eq:akdef}, and $T 
= [\ddot{\phi}_C(t_0)]^{-1/2}$.

\subsection{Periastron precession effects}

Let us first derive a waveform in the Fourier domain taht is valid for nonprecessing spins, and add the 
effects of spin-orbit precession later. Putting aside spin-orbit precession, the signal in the time 
domain can be expressed as in Eqs.~(\ref{eq:waveform+coeff}):
\begin{align}
	h_{+,\times}(t) &= \frac{M \nu y^2}{d_L} \sum_{n \in \mathbb{Z}} \sum_{m \in \mathbb{Z}} 
		G_{+,\times}^{(m,n)} e^{- i (n \lambda + m \delta\lambda)} \,.
\end{align}
Using the SPA, we can approximate its Fourier transform by
\begin{subequations}
\begin{align}
	\tilde{h}_{+,\times} (f) =&\; \int h_{+,\times}(t) e^{2 \pi i f t} dt \nonumber\\
		=&\;\frac{M \nu y^2}{d_L} \sum_{n \geq 1}\sum_{m \in \mathbb{Z}} \sqrt{\frac{2 \pi}{n 
		\ddot{\lambda} + m \delta \ddot{\lambda}}} G_{+,\times}^{(m,n)} \nonumber\\
		&\times e^{i \left[2 \pi f t_{n,m} - n \lambda(t_{n,m}) - m \delta \lambda (t_{n,m}) - 
		\pi/4\right]} \,, \label{eq:SPAnm} \\
	2\pi f =&\; n \dot{\lambda}(t_{n,m}) + m \delta \dot{\lambda}(t_{n,m}) \,, \label{eq:tnmdef}
\end{align}
\end{subequations}
where each of the harmonics $(n,m)$ has to be evaluated at a different time. It is worth noting 
here that we assumed that $n \dot{\lambda} + m \delta \dot{\lambda} > 0$, which is not necessarily 
true for every $(n, m)$ pair during the whole inspiral. However, for this assumption to break down, 
one needs negative and sufficiently large $m$, since $\delta \dot{\lambda} / \dot{\lambda} = 
\ord{y^2}$, and the corresponding amplitude will be suppressed by a factor $e^m$. We verified that 
ignoring this fact does not lead to high inaccuracies, at least for initial eccentricities $e_0 
\lesssim 0.4$.

In order to simplify the expression of the Fourier domain waveform and to improve its computational 
efficiency, we look for an expression of the following form:
\begin{subequations}
\begin{align}
	\tilde{h}_{+,\times}(f) &= \frac{M \nu y^2}{d_L} \sum_{n \geq 1} \tilde{h}_{n,0}(f) \sum_{m \in 
		\mathbb{Z}} \tilde{h}_{n,m}\super{PP} (f) \,,\\
	\tilde{h}_{n,0}(f) &= \sqrt{\frac{2 \pi}{n \ddot{\lambda}}} e^{i[2 \pi f t_n - n \lambda(t_n) - 
		\pi/4]} \,, \\
	2 \pi f &= n \dot{\lambda}(t_n) \,, \label{eq:tndef}
\end{align}
\end{subequations}
where $\tilde{h}_{n,0}(f)$ is a waveform harmonic without any periastron precession effects and 
$\tilde{h}_{n,m}\super{PP}(f)$ are corrections to it. In order to evaluate 
$\tilde{h}_{n,m}\super{PP} (f)$, we define
\begin{align}
	 \Delta t_{n,m} &= t_{n,m} - t_n \,,
\end{align}
and Taylor expand Eq.~\eqref{eq:tnmdef} around $t_n$:
\begin{align}
	2\pi f &= \sum_{p \geq 0} \frac{\Delta t_{n,m}^p}{p!} \left. \frac{d^p}{dt^p} \left(n 
		\dot{\lambda} + m \delta \dot{\lambda} \right) \right|_{t=t_n} \,.
\end{align}
We can use this together with Eq.~\eqref{eq:tndef} to solve for the PN expansion of $\Delta 
t_{n,m}$ order by order, and we obtain
\begin{align}
	\Delta t_{n,m} &= \sum_{p = 1}^P \frac{1}{p!} \left(-\frac{m}{n} \right)^p D^{p-1} \left( 
		\frac{\delta \dot{\lambda}^p}{\ddot{\lambda}} \right) \,, \label{eq:Deltat}
\end{align}
where the differential operator $D$ is given by
\begin{align}
	 D &= \frac{1}{\ddot{\lambda}} \frac{d}{dt} \,,
\end{align}
and every function of time is evaluated at $t=t_n$. We have checked that this expression remains 
valid at least up to $P=6$.

Using this, we can then Taylor expand the phase in Eq.~\eqref{eq:SPAnm} around $t_n$ to compute
\begin{subequations}
\begin{align}
	\tilde{h}_{n,m}\super{PP} (f) &= G_{+, \times}^{(m,n)} e^{i \Delta \Psi_{n,m}} \,,\\
	\Delta \Psi_{n,m} &= - m \delta \lambda + n \sum_{p = 2}^{P+1} \frac{1}{p!} \left( -\frac{m}{n} 
		\right)^p D^{p-2} \left( \frac{\delta \dot{\lambda}^p}{\ddot{\lambda}} \right) \,, 
		\label{eq:DeltaPsi}
\end{align}
\end{subequations}
where all functions are once again evaluated at the stationary time $t_n$ defined by Eq.~\eqref{eq:tndef}, and we checked that the latter equation is valid at least up to $P=6$. 
Equations~\eqref{eq:Deltat} and~\eqref{eq:DeltaPsi} are PN expansions in the sense that each increasing 
 order in $m$ is multiplied by a factor of PN order $(\delta \dot{\lambda}/\ddot{\lambda}) 
(d/d_t) = \ord{y^2}$, as both $\delta \dot{\lambda}$ and $\ddot{\lambda}$ evolve on the radiation 
reaction timescale. This implies that the formal expansion in $m$ in these two equations coincides with a PN expansion at order $2P$ beyond leading order.

\subsection{Complete waveform}

We can now add spin precession by using a SUA transformation~\cite{klein-2014} instead of a SPA. We 
start by noting that we can express the waveform in the time domain by
\begin{align}
	h(t) &= \sum_{n,m} \mathcal{A}_{n,m}(t) e^{-i(n\lambda + m\delta\lambda)} \,,
\end{align}
where all spin-precession timescale effects are included in the amplitudes 
\begin{align}
	\mathcal{A}_{n,m}(t) =&\; \frac{M \nu y^2}{d_L} \Big[ F_+(t) G_+^{(n,m)}(t) \nonumber\\
		&+ F_\times(t) G_\times^{(n,m)} (t) \Big] e^{-i (n + m) \phi_T} \,.
\end{align}
This allows us to directly use a SUA transformation. If we restrict the amplitudes to $\ord{y^N, 
e^M}$, we then obtain
\begin{subequations}
\begin{align}
	\tilde{h}(f) =&\; \sum_{n = \max(1, 2-N)}^{2+N} \sum_{m = -M}^{M} \tilde{h}_{n,m} (f) \,, 		
		\label{eq:WF1} \\
	\tilde{h}_{n,m}(f) =&\; \tilde{h}_{n,m}^{(0)}(f) \tilde{h}_{n,m}\super{SP}(f) \,, \\
	\tilde{h}_{n,m}^{(0)}(f) =&\; \sqrt{2\pi}\;T_{n,m} \exp\big[i \big( 2 \pi f t_{n,m} \nonumber\\
		& -n \lambda (t_{n,m}) - m \delta \lambda (t_{n,m}) - \pi/4 \big) \big] \,, \\
	2 \pi f =&\; n \dot{\lambda} (t_{n,m}) + m \delta \dot{\lambda}(t_{n,m}) \,, \\
	T_{n, m} =&\; \left[ n \ddot{\lambda} (t_{n,m}) + m \delta \ddot{\lambda} (t_{n,m}) 
		\right]^{-1/2} \,, \label{eq:SUAtime}\\
	h_{n,m}\super{SP} (f) =&\; \sum_{k = - k\sub{max}}^{k\sub{max}} a_{k, k\sub{max}} 
		\mathcal{A}_{n,m}(t_{n,m} + k T_{n,m}) \,, \label{eq:htSUA}
\end{align}
\end{subequations}
where the constants $a_{k, k\sub{max}}$ satisfy the linear system of equations defined in 
Eq.~\eqref{eq:akdef}. This waveform is in the Fourier domain and consistently includes the effects 
of spin-induced precession and periastron precession. As we will see in the next section, it allows 
for large matches with time domain waveforms with eccentricities $e \lesssim 0.3$ that we can 
consider as moderate in the modeling sense, because only the amplitudes, not the phasing, have been expanded for small 
eccentricities.

The waveform defined by Eq.~\eqref{eq:WF1} suffers from the fact that it includes a double sum, and 
therefore its computational cost rises quickly as the precision of the amplitudes increases. 
However, in order to increase its computational efficiency, we can use a similar strategy as 
described in the previous subsection and expand the waveform in powers of 
$\delta\dot{\lambda}/\dot{\lambda}$.

First, we can approximate the SUA timescale in Eq.~\eqref{eq:SUAtime} by $T_{n,m} \approx T_n = 
T_{n,0}$. Next, we can use Eqs.~\eqref{eq:Deltat} and~\eqref{eq:DeltaPsi} to define $\Delta 
t_{n,m}$ and $\Delta \Psi_{n,m}$ at order $P$:
\begin{subequations}
\label{eq:Porderdef}
\begin{align}
	\Delta t_{n,m} =\;& \sum_{p = 1}^{P} \frac{1}{p!} \left( -\frac{m}{n} \right)^p D^{p-1} \left( 
		\frac{\delta \dot{\lambda}^p}{\ddot{\lambda}} \right) \,, \label{eq:Deltatnm} \\
	\Delta \Psi_{n,m} =\;& - m \delta \lambda + n \sum_{p = 2}^{P+1} \frac{1}{p!} \left(- 
		\frac{m}{n}\right)^p D^{p-2} \left( \frac{\delta \dot{\lambda}^p}{\ddot{\lambda}} 
		\right)\,. \label{eq:DeltaPsinm}
\end{align}
\end{subequations}
We can use Eqs.~\eqref{eq:ydot} and~\eqref{eq:e2dot} together with the chain rule
\begin{align}
	\frac{d}{dt} f(y, e^2) &= \frac{\partial f}{\partial y} \frac{dy}{dt} + \frac{\partial f} 
		{\partial e^2} \frac{de^2}{dt} \,,
\end{align}
to get the necessary derivatives of $\lambda$ and $\delta\lambda$ as PN expanded functions. Thus, 
we can simplify the waveform as
\begin{subequations}
\begin{align}
	\tilde{h}(f) =&\; \sum_{n = \max(1, 2-N)}^{2+N} \tilde{h}_n^{(0)} (f) \tilde{h}_n\super{PP}(f) 
		\,, \label{eq:WF2} \\
	\tilde{h}_{n}^{(0)}(f) =&\; \sqrt{2\pi}\; T_n \exp\Big[i \Big( 2\pi f t_n - n \lambda (t_n) - 
		\frac{\pi}{4} \Big) \Big] \,, \\
	2 \pi f =&\; n \dot{\lambda} (t_n) \,, \\
	T_n =&\; \left[ n \ddot{\lambda} (t_n) \right]^{-1/2} \,, \\
	\tilde{h}_n\super{PP} (f) =&\; \sum_{m = - M}^{M} e^{i \Delta \Psi_{n,m}} \nonumber\\
		&\times \sum_{k = - k\sub{max}}^{k\sub{max}} a_{k, k\sub{max}} \mathcal{A}_{n,m} (t_n + 
		\Delta t_{n,m} + k T_n) \,.
\end{align}
\end{subequations}
Equation~\eqref{eq:WF2} presents a further expanded waveform, and can possibly be made more efficient 
than the one defined by Eq.~\eqref{eq:WF1}, especially for amplitudes of high $(N,M)$ order. Thus 
we get a family of Fourier domain gravitational waveforms for spin-precessing binaries on eccentric 
orbits characterized by the expansion orders $(P, N, M)$.

\section{Comparisons}\label{sec:sim}

We have run different sets of simulations in order to probe under what circumstances our waveforms 
defined in Eqs.~\eqref{eq:WF1} and~\eqref{eq:WF2} are sufficiently faithful to equivalent waveforms 
obtained in the time domain. For all the waveforms used in our comparisons, we use nonspinning 
amplitudes at 2PN order omitting tail terms~\cite{mishra-2015}, and we use evolution equations for 
$y$ and $e^2$ at 3PN nonspinning order and 2PN spinning order, including tail terms, as described 
in Appendix~\ref{sec:evol-eq}. For all Fourier domain waveforms, we use a SUA transformation as 
in~\cite{klein-2014} with $k\sub{max} = 3$.

We use as a reference time domain waveform $h_R$ obtained in the following way:
\begin{itemize}
	\item Equations~\eqref{eq:ydot}-\eqref{eq:s2dot} are solved numerically together with 
		Eqs.~\eqref{eq:lambdadot},~\eqref{eq:deltalambdadot}, and~\eqref{eq:phiTdot} in order to 
		yield solutions for $y(t)$, $e^2(t)$, $\uvec{L}(t)$, $\bm{s}_1(t)$, $\bm{s}_2(t)$, 
		$\lambda(t)$, $\phi_T(t)$, and $\delta\lambda(t)$, from an initial time $t_0$ to a maximum 
		time $t\sub{max}$ defined by the ISCO-like condition $M \dot{\lambda}(t\sub{max}) = 
		6^{-3/2}$.
	\item Time is equally sampled between $t_0$ and $t\sub{max}$ using a sampling time $\Delta t = 
		2\pi/[24 \dot{\lambda}(t\sub{max})]$, in order to ensure that the first 12 waveform 
		harmonics fall below the Nyquist frequency. Equations~\eqref{eq:phiofv}-\eqref{eq:lofu} are solved 
		at each step to get the orbital phase $\phi$ and the eccentric anomaly $u$. 
		Equation~\eqref{eq:lofu} is inverted numerically to yield $u( l = \lambda - \delta \lambda, e = 
		e_t)$.
	\item A waveform signal is constructed using Eqs.~\eqref{eq:hoft} and~\eqref{eq:hpcoft}, and 	
		the solutions for $y(t)$, $e(t)$, $\phi(t)$, $\phi_T(t)$, and $u(t)$. The antenna pattern 
		functions are chosen in the low-frequency limit, for a static detector~\cite{cutler-1998}. 
		The waveform amplitudes are included at 2PN order, with the omission of spin effects and 
		tail terms.
	\item A Tukey window is introduced in order to reduce spectral leakage and a discrete Fourier 	
		transform of the signal is taken to yield the waveform in the Fourier domain.
\end{itemize}

We compare different waveforms to the reference one
\begin{itemize}
	\item A nonexpanded eccentric one (NE$_M$) defined by Eq.~\eqref{eq:WF1} and $N=4$, $M \in 
	\{0, \ldots, 6\}$, i.e. with amplitudes at $N/2=2$PN order and amplitudes expanded at $M$th 
	order in $e$.
	\item An expanded eccentric one (EE$_{M,P}$) defined by Eq.~\eqref{eq:WF2} and $N=4$, $M \in 
	\{0, \ldots, 6\}$, $P \in \{0, \ldots, 3\}$, i.e. with amplitudes at $N/2=2$PN order, 
	amplitudes expanded at $M$th order in $e$, and the waveform expanded at $P$th order in 
	$\delta \dot{\lambda} / \dot{\lambda}$ as in Eqs.~\eqref{eq:Porderdef}.
	\item A circular one (C) with amplitudes at 2PN order, taken from~\cite{klein-2014}.
\end{itemize}
Note that Eqs.~\eqref{eq:Porderdef} imply that the waveforms NE$_{0}$ and EE$_{0,P}$ are identical 
for any $P$.

In order to make our comparisons, we compute the faithfulness $F = \max \mathcal{M}$, defined by 
the match $\mathcal{M}$ maximized over some of the waveform parameters, with
\begin{subequations}
\begin{align}
	\mathcal{M} &= \frac{(h, h_R)}{\sqrt{(h,h)(h_R,h_R)}} \,, \label{eq:match} \\
	(a,b) &= \int_{f\sub{min}}^{f\sub{max}} \tilde{a}(f) \tilde{b}^*(f) df \,, \label{eq:scalprod}
\end{align}
\end{subequations}
where we chose a white detector noise in order to make as few assumptions about the detector as 
possible. For the eccentric waveforms, since they use the same phasing as the reference one, we do 
not maximize over any parameters, while for the circular waveform, we maximize the match over time 
and orbital phase shifts to obtain the faithfulness. We compare the faithfulness obtained this 
way to a fiducial value of $F = 0.993$, corresponding to a faithfulness level at which we can 
estimate that the errors in the recovered parameters due to mismodeling are smaller than the 
statistical errors coming from the detector noise, for $D=10$ intrinsic parameters and a 
signal-to-noise ratio (SNR) of $\rho = 25$~\cite{chatziioannou-2017}. The relation between the 
faithfulness and the SNR at which the mismodeling error becomes likely to exceed the statistical 
error in a GW detection is~\cite{chatziioannou-2017}
\begin{align}
	F &\approx 1 - \frac{D-1}{2 \rho^2} \,.
\end{align}

We ran two different sets of simulations: one to study systems in the late inspiral as 
observed by the LIGO/Virgo network and by LISA in the case of massive black hole binaries 
[denoted by (Xa)], and the other to study systems in the early inspiral as observed by LISA for 
stellar-origin black hole binaries [denoted by (Xb)].
We made six different runs in order to probe the faithfulness of our waveforms as a function of 
the starting eccentricity in different situations:
\begin{itemize}
	\item[(I)] We randomize the initial eccentricity with a log-flat distribution $10^{-5} < e_0 < 	
		0.5$ and the spin magnitudes with a flat distribution $0 < \chi_i < 1$.
	\item[(II)] We randomize the initial eccentricity with a log-flat distribution $10^{-5} < e_0 < 
		0.5$ and spin magnitudes with a flat distribution $0 < \chi_i < 0.1$.
	\item[(III)] We randomize the initial eccentricity with a log-flat distribution $10^{-5} < e_0 
		< 0.5$ and spin magnitudes set to the maximum value $\chi_i = 1$.
	\item[(IV)] We start with zero initial eccentricity and spin magnitudes with a flat 	
		distribution $0 < \chi_i < 1$.
	\item[(V)] We start with zero initial eccentricity and spin magnitudes with a flat 	
		distribution $0 < \chi_i < 0.1$.
	\item[(VI)] We start with zero initial eccentricity and spin magnitudes set to the maximum 
		value $\chi_i = 1$.
\end{itemize}
We thus have twelve runs: (Ia)--(VIa) in the late inspiral case and (Ib)--(VIb) in the early inspiral 
case.

To get the binary parameters used in our runs, we randomize all vector directions with a flat 
distribution on the sphere. Since the distance does not affect the match $\mathcal{M}$ in 
Eq.~\eqref{eq:match}, we fix it at some fiducial value. We randomize the initial orbital phase and 
the initial periastron-ascending node angle $\phi_e$ (see Fig.~\ref{fig:angles}) with a flat 
distribution in $[0, 2\pi]$. Whenever the randomized initial eccentricity is lower than the minimal 
value given in Eq.~\eqref{eq:e2min}, we set $e_0 = e\sub{min}$. Note that cases (IV) to (VI) 
correspond to fully circularized binaries, but Eq.~\eqref{eq:e2min} prevents them from having truly 
zero eccentricity unless the reduced spins have exactly equal support in the orbital plane. In each 
late inspiral run, we start our simulations with an initial eccentricity $e_0$ and at a frequency 
$M \dot{\lambda}\sub{start} = 6^{-3/2}/10$, and stop at $M \dot{\lambda}\sub{stop} = 6^{-3/2}$. We 
also randomize the mass ratio $q = m_2/m_1$ with a log-flat distribution between $1$ and $1/30$, 
and use a fixed total mass $M = 100 M_\odot$, taking advantage of the white detector noise.
In each early inspiral run, we randomize the two masses with a log-flat distribution $10 M_\odot < 
m_i < 100 M_\odot$. We then use the Newtonian time-frequency relation and the initial eccentricity 
to determine the starting frequency such that the system will evolve to have an orbital frequency 
of $f\sub{end} = 1$~Hz after $T = 4$~yr:
\begin{subequations}
\begin{align}
	y\sub{start} &= \left[ \frac{5 M y\sub{end}^8}{5M + 32 \nu T \left(1 - e_0^2\right)^{3/2} 
		\left(8 + 7 e_0^2\right) y\sub{end}^8} \right]^{1/8} \,, \\
	y\sub{end} &= \left[ \frac{2 \pi M f\sub{end}}{\left(1- e_0^2 \right)^{3/2}} \right]^{1/3} \,.
\end{align}
\end{subequations}
We then let the system evolve and stop after four years, and set the maximum frequency 
$f\sub{max}=1$~Hz in Eq.~\eqref{eq:scalprod}.

\subsection{Late inspiral systems}

\begin{figure}[!ht]
\begin{center}
	\includegraphics[width=0.45\textwidth]{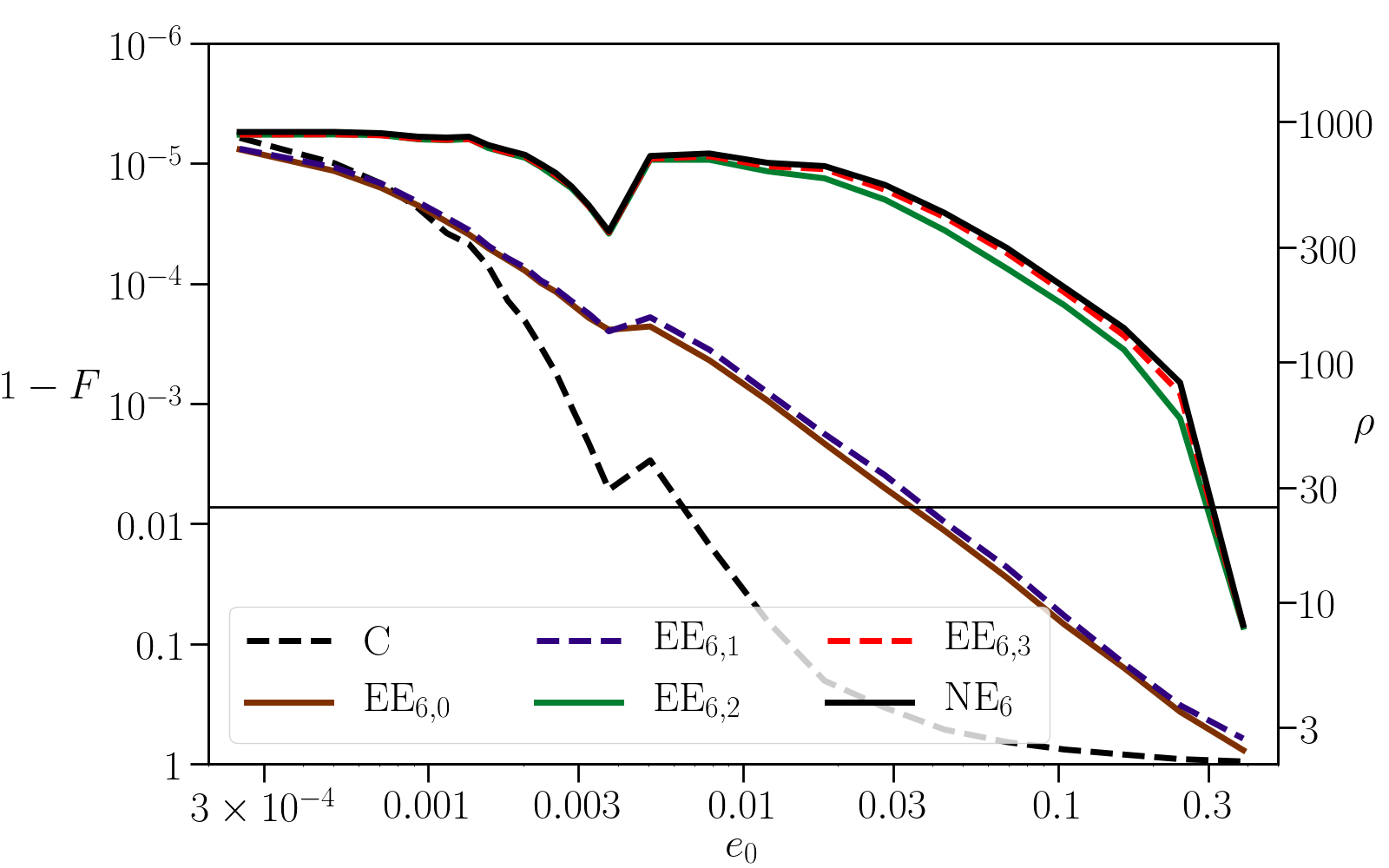}
	\includegraphics[width=0.45\textwidth]{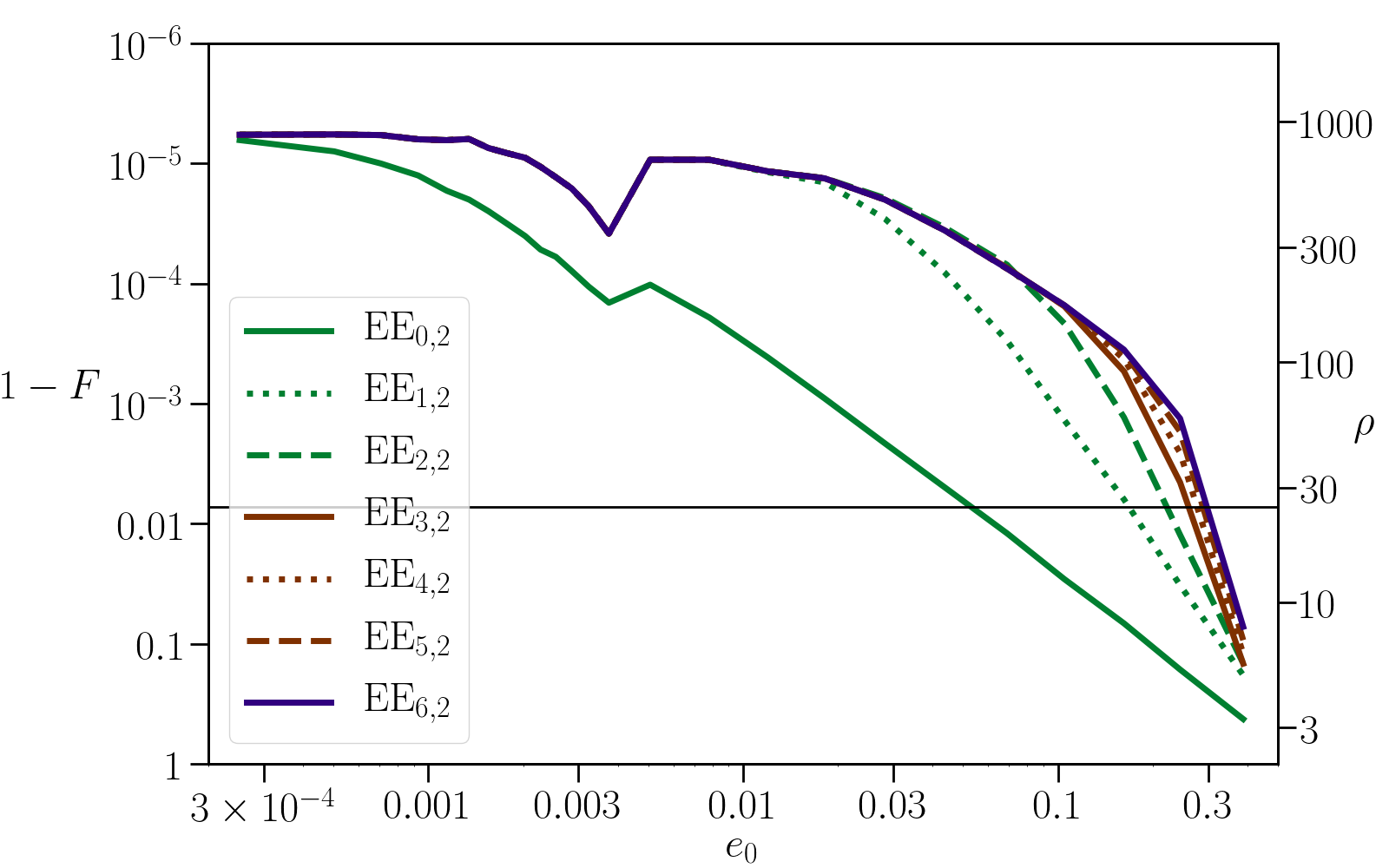}
	\caption{Results from late inspiral run (Ia), with starting eccentricity $10^{-5} < e_0 < 0.5$ 
	and spin magnitudes $0 < \chi_i < 1$. On the top, we show median faithfulness as a function of the 
	starting eccentricity from bottom to top for the circular waveform C, the expanded eccentric 
	waveforms EE$_{6,P}$, with amplitudes at 6th order in the eccentricity and at $P$th order in 
	$\delta\dot{\lambda}/\dot{\lambda}$, $0 \leq P \leq 3$ with increasing $P$ from bottom to top, 
	and the nonexpanded waveform with amplitudes at 6th order in the eccentricity NE$_6$. At the 
	bottom, we show median faithfulness as a function of the starting eccentricity for the expanded 
	eccentric waveforms EE$_{M,2}$, with amplitudes at $M$th order in the eccentricity, $0 \leq M 
	\leq 6$ and increasing $M$ from bottom to top, and at second order in 
	$\delta\dot{\lambda}/\dot{\lambda}$. The left axis shows the unfaithfulness $1-F$, and the 
	right axis shows the corresponding threshold SNR $\rho$ above which we can expect mismodeling 
	errors to exceed the accuracy in a measurement.	In both panels, the thin horizontal black 
	line shows a fiducial faithfulness of 0.993 or a corresponding SNR of 25.}
	\label{fig:faith-ia}
\end{center}
\end{figure}

We present in Fig.~\ref{fig:faith-ia} the results from late inspiral run (Ia), with starting 
eccentricity $10^{-5} < e_0 < 0.5$ and spin magnitudes $0 < \chi_i < 1$. In it, we compare the mean 
faithfulness as a function of the initial eccentricity for different waveforms. The top panel 
shows a comparison between the results for the circular waveform C, the nonexpanded eccentric 
waveform NE$_6$, and the expanded eccentric waveforms $EE_{6,P}$, $0 \leq P \leq 3$, and the bottom 
panel shows a comparison between the expanded eccentric waveforms EE$_{M,2}$, $0 \leq M \leq 
6$. We can see in the top panel that the circular waveform stays above the fiducial faithfulness 
only for initial eccentricities of $e_0 \lesssim 0.008$. Furthermore, the results for the
expanded eccentric waveform become very close to the nonexpanded version starting at second order in 
$\delta \dot{\lambda}/\dot{\lambda}$, and leads to a faithfulness above the fiducial threshold for 
eccentricities below $e_0 \lesssim 0.3$. On the bottom panel, we can see the effects of the 
expansion of the waveform amplitudes for small eccentricities. We can see that the largest starting 
eccentricity for which the median faithfulness stays above the threshold increases with increasing 
order in the expansion. Furthermore, we can see that below a certain starting eccentricity 
depending on the specific order, increasing the expansion order has no effect on the faithfulness, 
as the errors due to this approximation become subdominant.

\begin{figure}[th]
\begin{center}
	\includegraphics[width=0.45\textwidth]{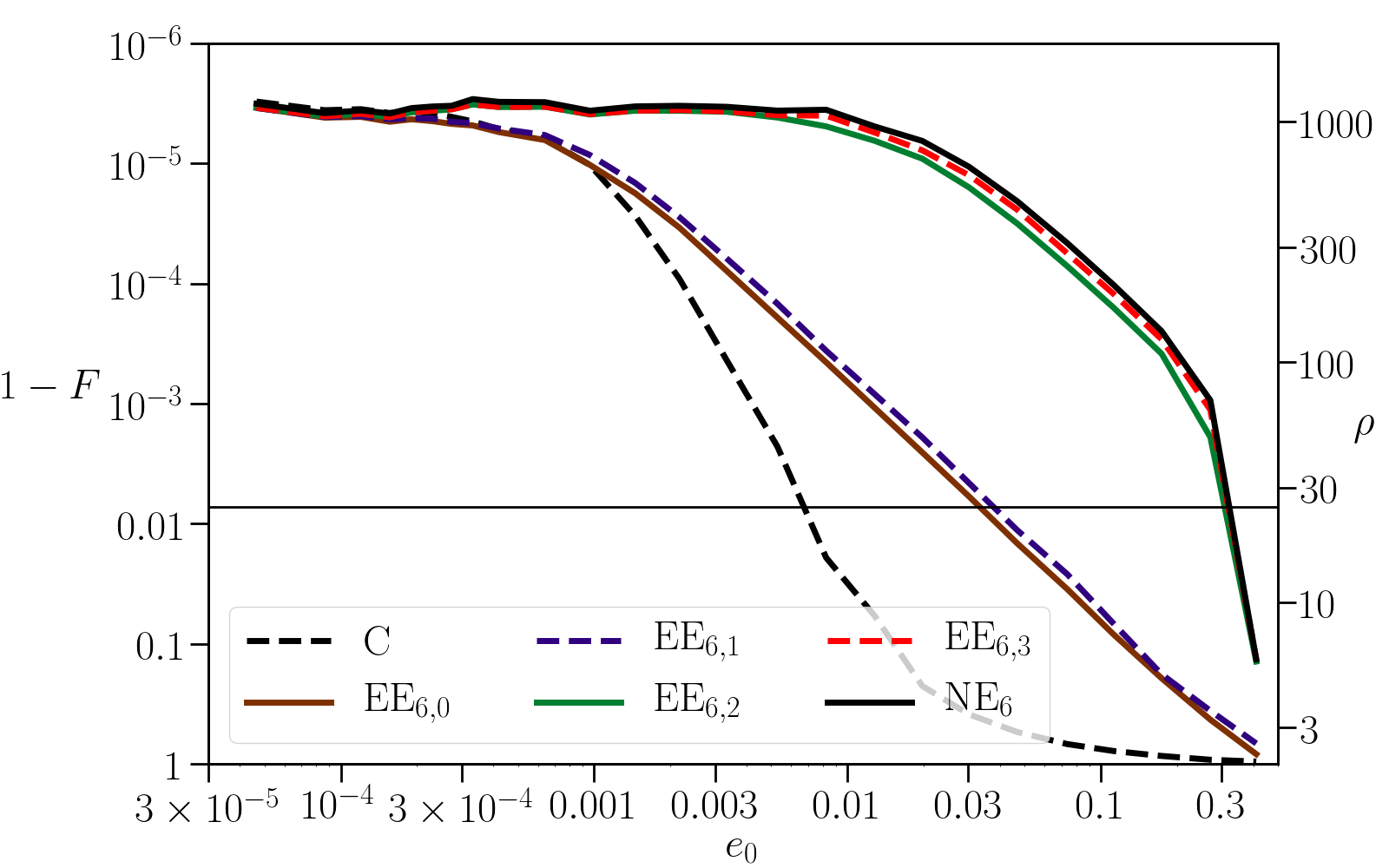}	
	\includegraphics[width=0.45\textwidth]{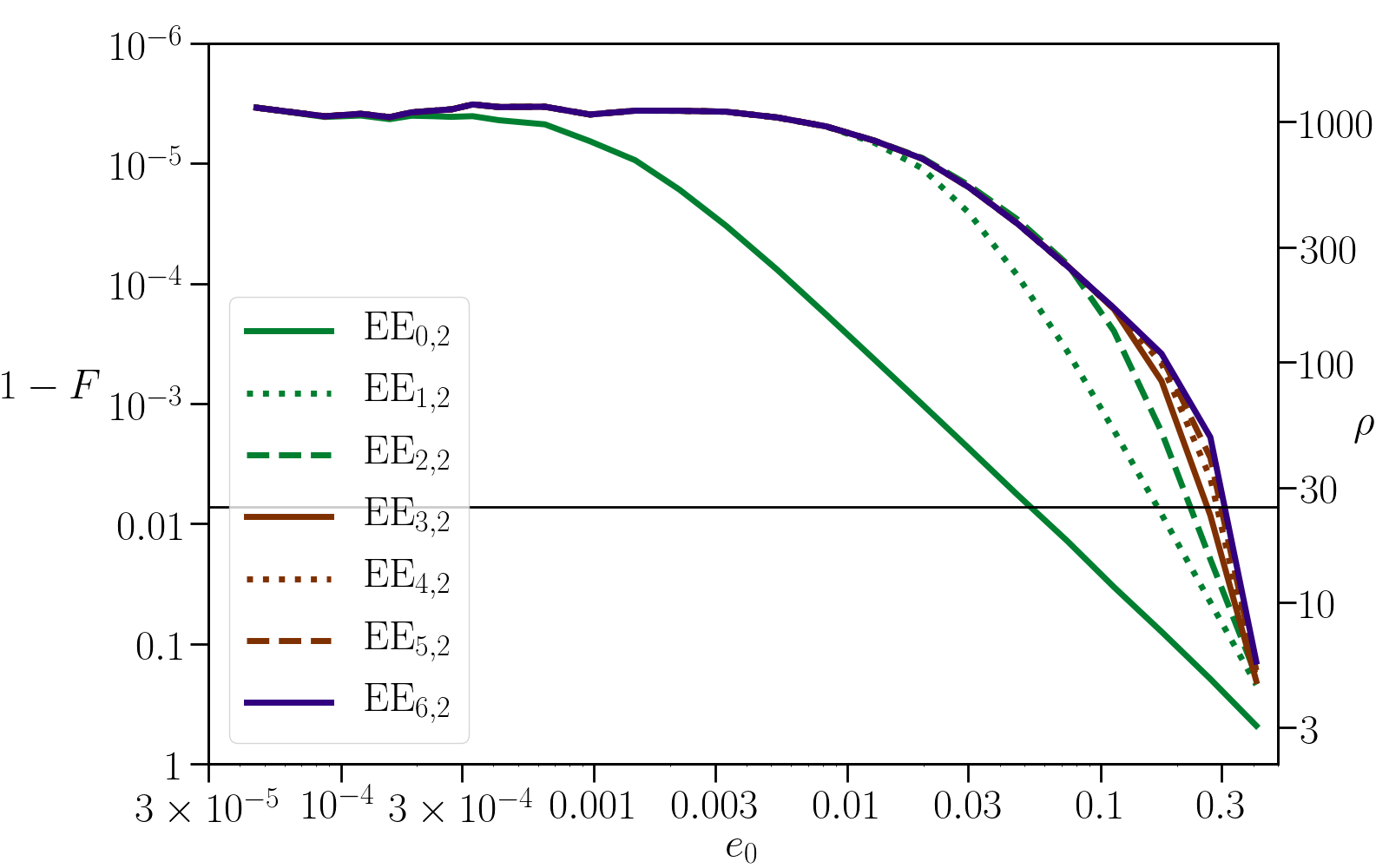}
	\caption{Same as Fig.~\ref{fig:faith-ia}, for late inspiral run (IIa) with starting 
	eccentricity $10^{-5} < e_0 < 0.5$ and spin magnitudes $0 < \chi_i < 0.1$.}
	\label{fig:faith-iia}
\end{center}
\end{figure}

We present in Fig.~\ref{fig:faith-iia} the results from late inspiral run (IIa), with starting 
eccentricity $10^{-5} < e_0 < 0.5$ and spin magnitudes $0 < \chi_i < 0.1$. These results are very similar 
to the results of run (Ia), but due to the reduced spin magnitudes the starting eccentricities 
reach smaller values. On the top panel, we can see that below a starting eccentricity of $e_0 
\lesssim 10^{-3}$, the loss of faithfulness using circular waveforms with respect to our eccentric 
models becomes negligible.

\begin{figure}[th]
\begin{center}
	\includegraphics[width=0.45\textwidth]{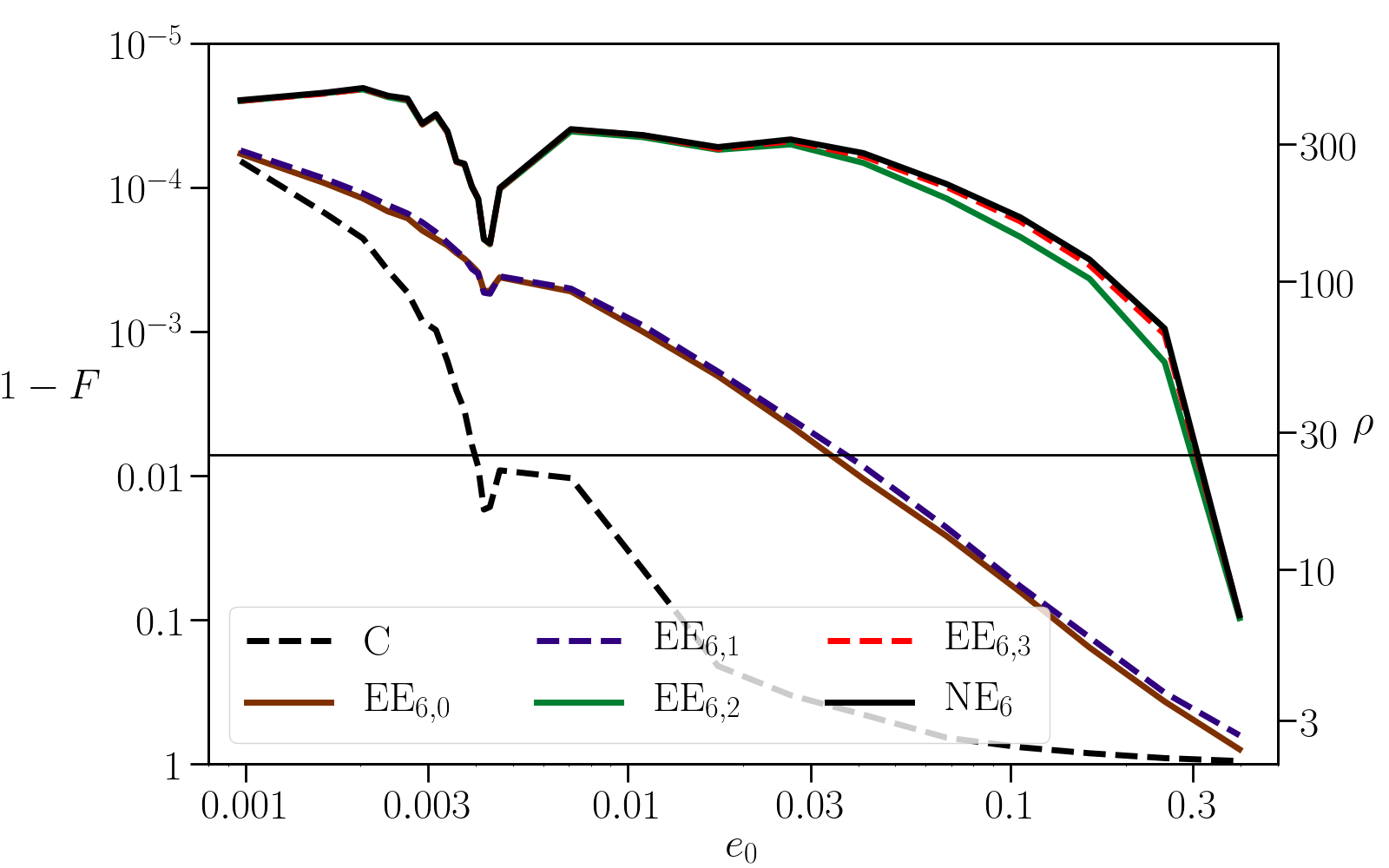}	
	\includegraphics[width=0.45\textwidth]{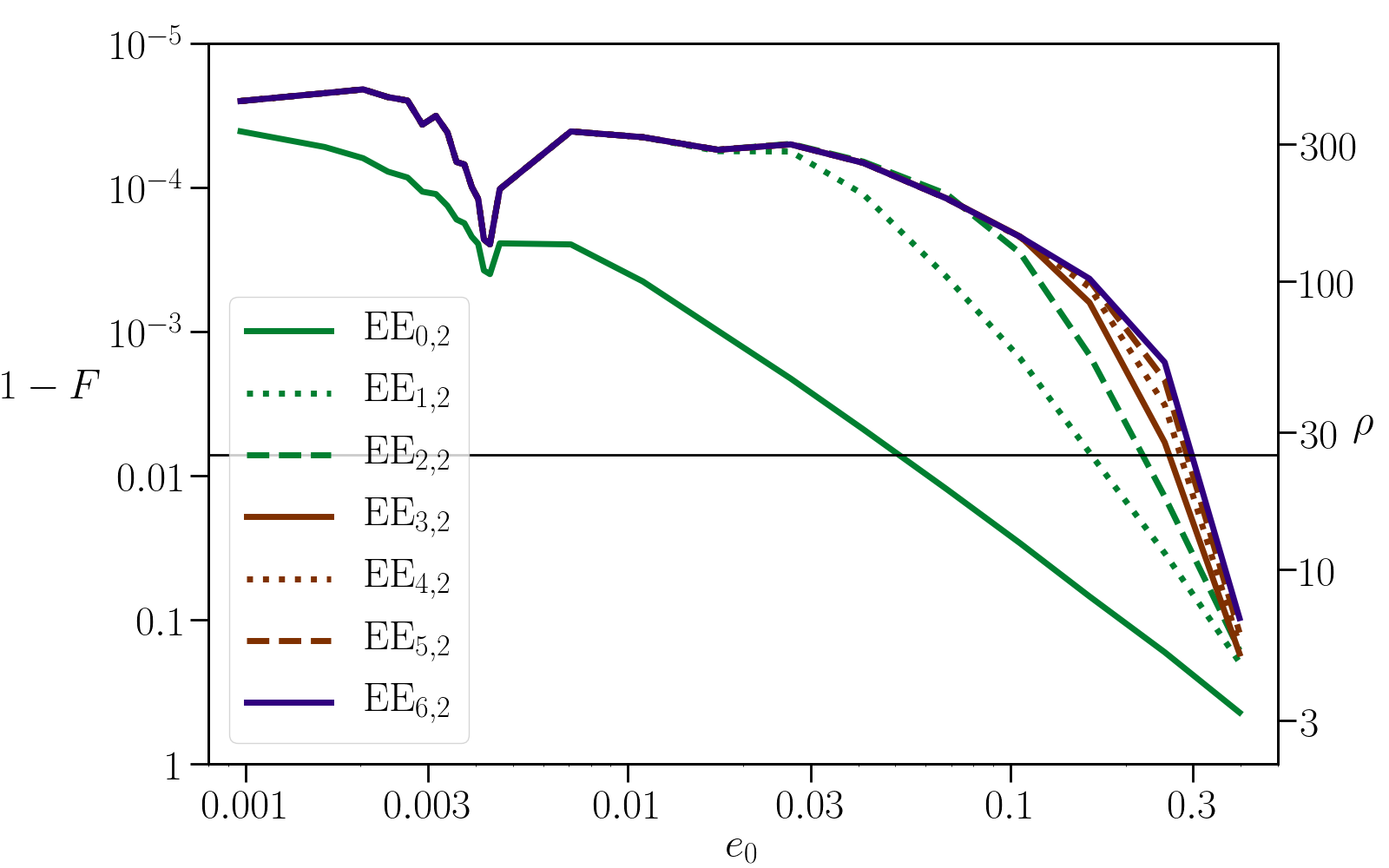}
	\caption{Same as Fig.~\ref{fig:faith-ia}, for late inspiral run (IIIa) with starting 
	eccentricity $10^{-5} < e_0 < 0.5$ and spin magnitudes $\chi_i = 1$.}
	\label{fig:faith-iiia}
\end{center}
\end{figure}

We present in Fig.~\ref{fig:faith-iiia} the results from late inspiral run (IIIa), with starting 
eccentricity $10^{-5} < e_0 < 0.5$ and spin magnitudes $\chi_i = 1$. The results are similar to the ones shown in Figs.~\ref{fig:faith-ia} and~\ref{fig:faith-iia}, but the increased magnitudes of the spins slightly reduce the performance of the circular waveform.
Comparing this figure to Figs.~\ref{fig:faith-ia} and~\ref{fig:faith-iia}, we can conclude that 
the value of the spin magnitudes has little effect on the faithfulness, other than on  the limiting residual eccentricity. 

\begin{figure}[th]
\begin{center}
	\includegraphics[width=0.45\textwidth]{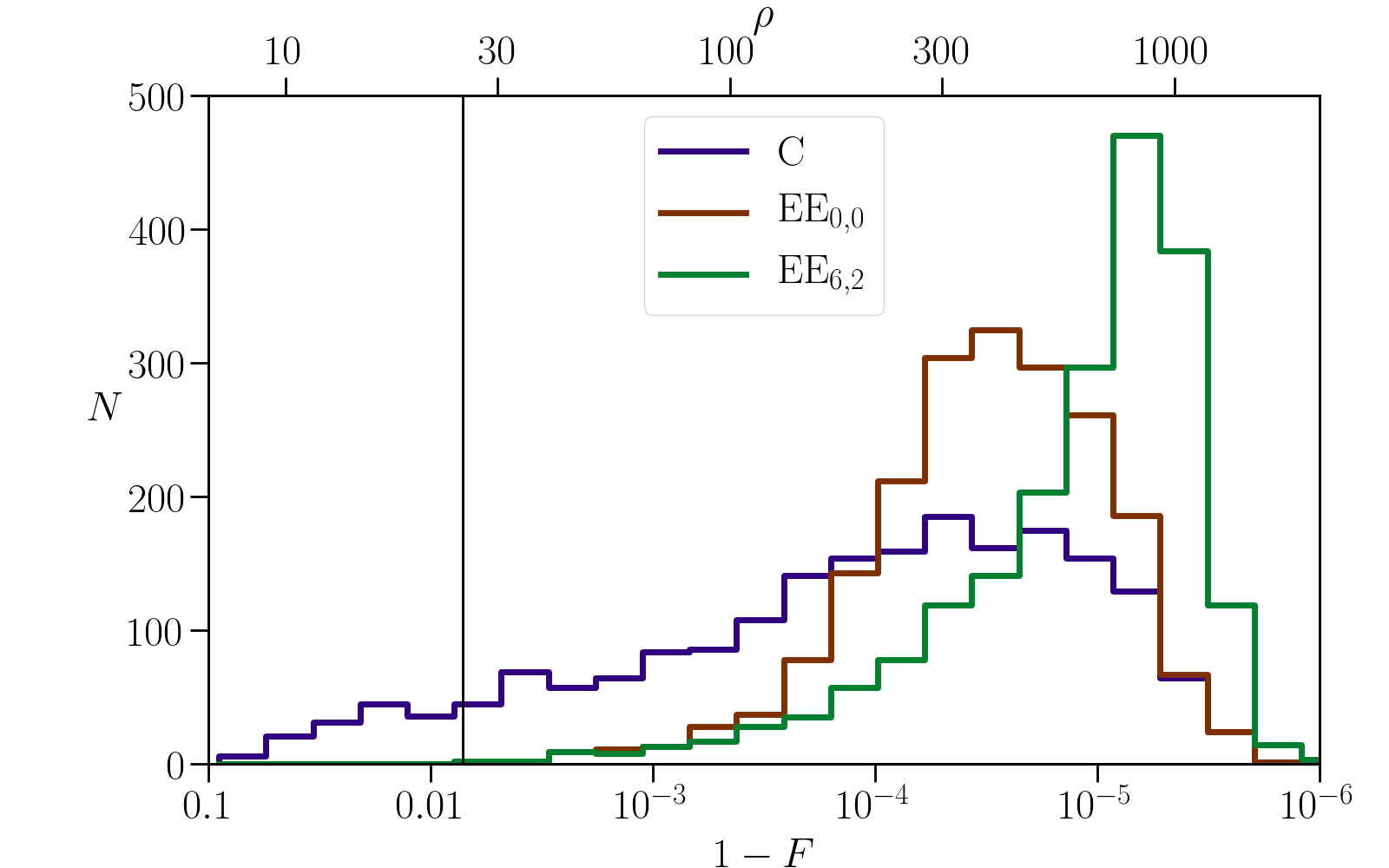}
	\caption{Results from late inspiral run (IVa) with starting eccentricity $e_0 = e\sub{min}$ 
	and spin magnitudes $0 < \chi_i < 1$. The systems simulated here correspond to highly spinning 
	fully circularized binaries. The blue line corresponds to the circular 	waveform C, the red 
	line to the lowest-order expanded eccentric waveform EE$_{0,0}$, and the green line to the 
	highly accurate expanded eccentric waveform EE$_{6,2}$. The bottom axis shows the unfaithfulness $1-F$, and the top axis shows the corresponding threshold SNR $\rho$ above which we can expect mismodeling errors to exceed the accuracy in a measurement. The thin vertical line corresponds to a 
	fiducial faithfulness of 0.993 or a corresponding SNR of 25. Note that due to the eccentricity being taken into account in the phasing, even the lowest-order 
	eccentric waveform EE$_{0,0}$ performs better than the circular waveform C.}
	\label{fig:faith-iva}
\end{center}
\end{figure}
 
We present in Fig.~\ref{fig:faith-iva} the results from late inspiral run (IVa), with starting 
eccentricity $e_0 = e\sub{min}$ and spin magnitudes $0 < \chi_i < 1$. We can see here an 
effect due to the residual eccentricity. Indeed, the circular waveform performs poorly in some 
cases, even when the binaries are fully circularized. In our simulations, 7\% of the 
faithfulness for the circular waveform was below the threshold line, while virtually no 
faithfulness were found below it for waveforms that used eccentric phasing, even with the lowest 
order amplitudes. While this does not represent a large proportion of binaries, this number will 
only increase when considering binaries with higher SNRs.

\begin{figure}[th]
\begin{center}
	\includegraphics[width=0.45\textwidth]{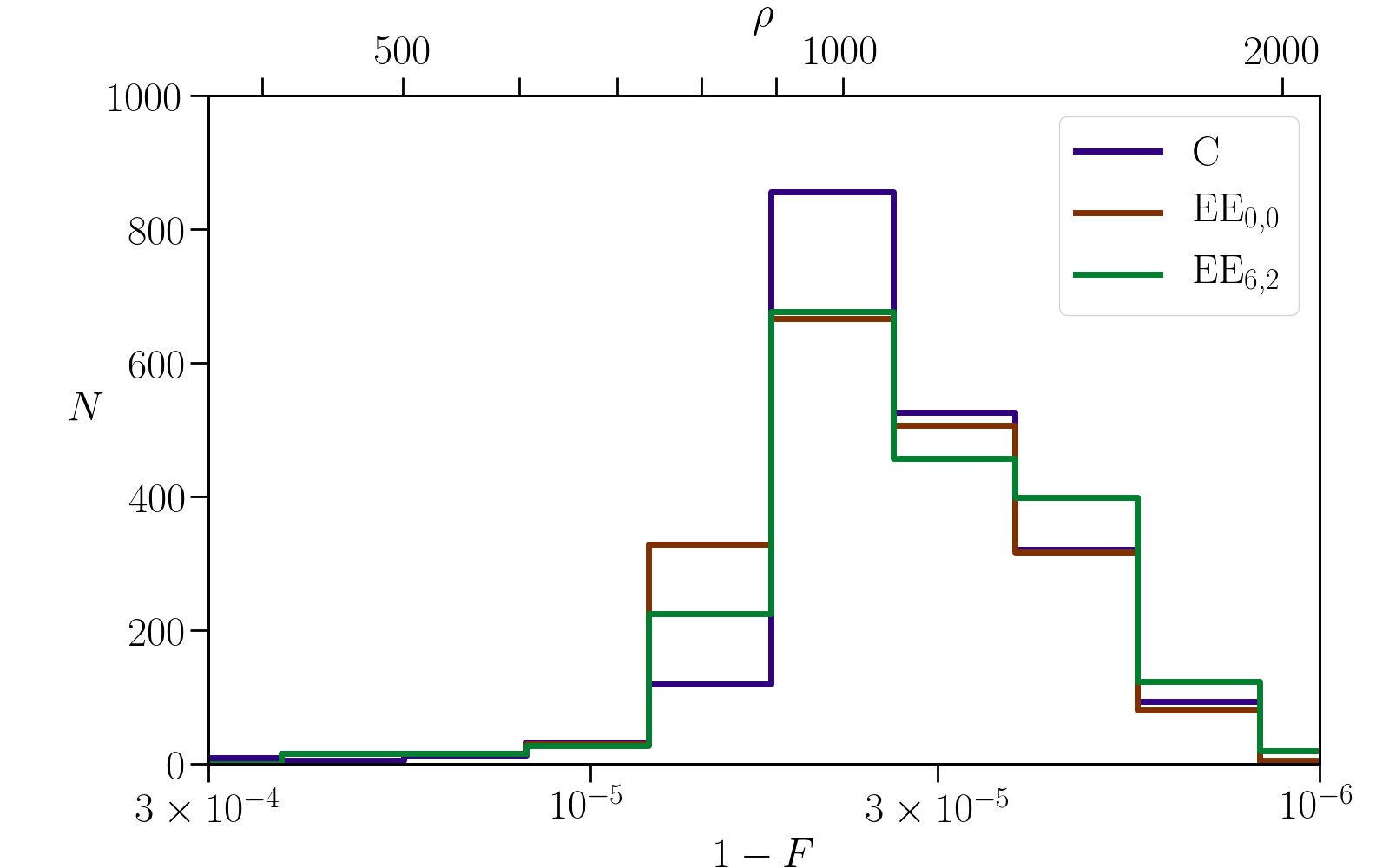}
	\caption{Same as Fig.~\ref{fig:faith-iva}, for late inspiral run (Va) with starting 	
	eccentricity $e_0 = e\sub{min}$ and spin magnitudes $0 < \chi_i < 0.1$. The systems simulated 	
	here correspond to slowly spinning fully circularized binaries.}
	\label{fig:faith-va}
\end{center}
\end{figure}
 
We present in Fig.~\ref{fig:faith-va} the results from late inspiral run (Va), with starting 
eccentricity $e_0 = e\sub{min}$ and spin magnitudes $0 < \chi_i < 0.1$. Comparing with the 
results shown in Fig.~\ref{fig:faith-iva}, we can see that assuming lower spins prevents the 
circular waveforms from having faithfulness below the threshold line. Thus, eccentricity effects in the inspiral can be safely ignored when only the last part of it is visible.
This further shows that the starting eccentricity is the most important factor to influencing the 
accuracy of our waveforms in the late inspiral.

\begin{figure}[th]
\begin{center}
	\includegraphics[width=0.45\textwidth]{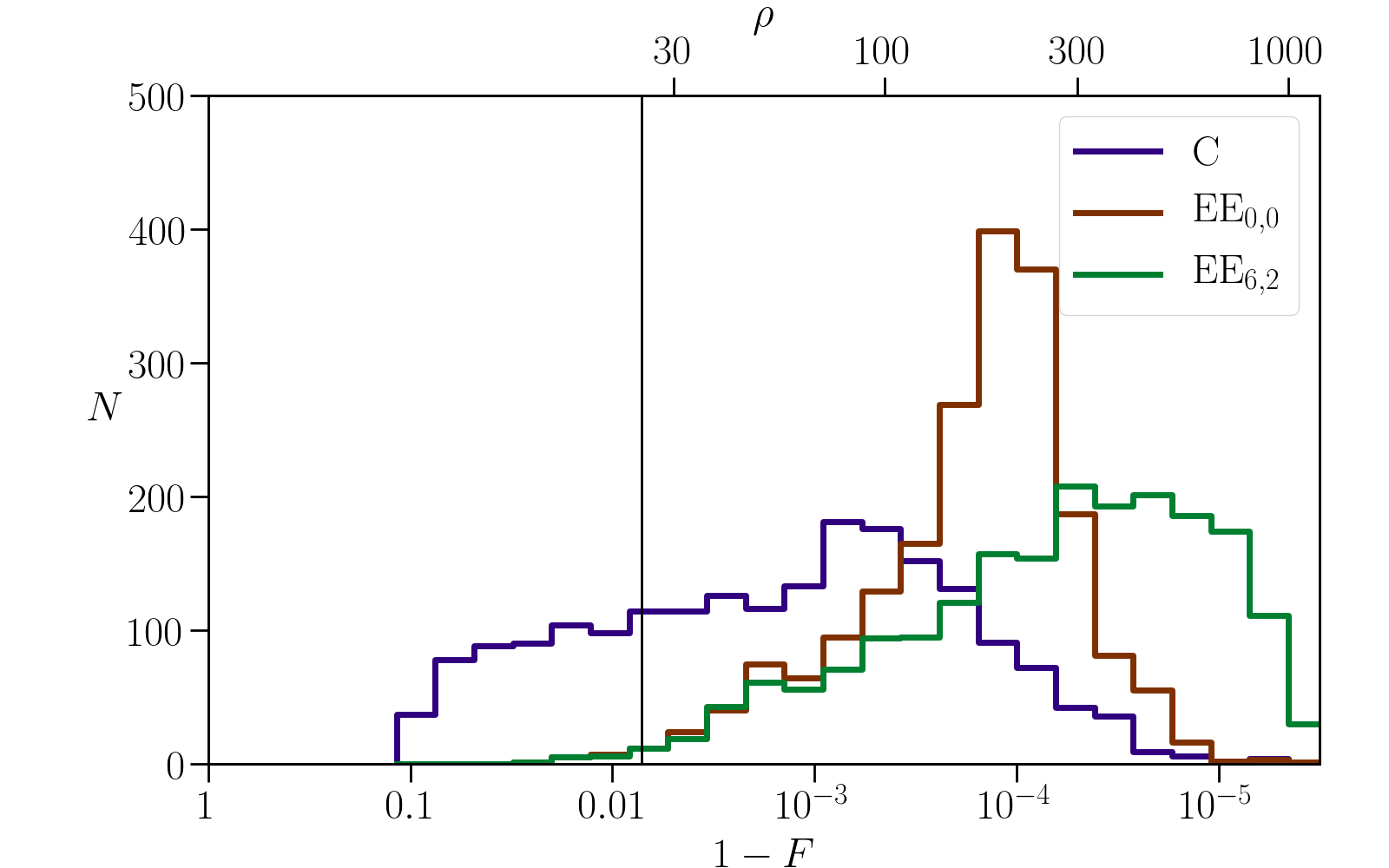}
	\caption{Same as Fig.~\ref{fig:faith-iva}, for late inspiral run (VIa) with starting 	
	eccentricity $e_0 = e\sub{min}$ and spin magnitudes $\chi_i = 1$. The systems simulated 	
	here correspond to maximally spinning fully circularized binaries.}
	\label{fig:faith-via}
\end{center}
\end{figure}
 
We present in Fig.~\ref{fig:faith-via} the results from late inspiral run (VIa), with starting 
eccentricity $e_0 = e\sub{min}$ and spin magnitudes $\chi_i = 1$. The results here are similar to the ones shown in Fig.~\ref{fig:faith-iva}, but more pronounced. The proportion of binaries for which the circular waveform has a faithfulness lying below the threshold line increases to 25~\%, indicating that the inclusion of eccentricity effects might be important even for fully circularized binaries in the last stages of their inspiral when their spins are large.

\subsection{Early inspiral systems}

\begin{figure}[th]
\begin{center}
	\includegraphics[width=0.45\textwidth]{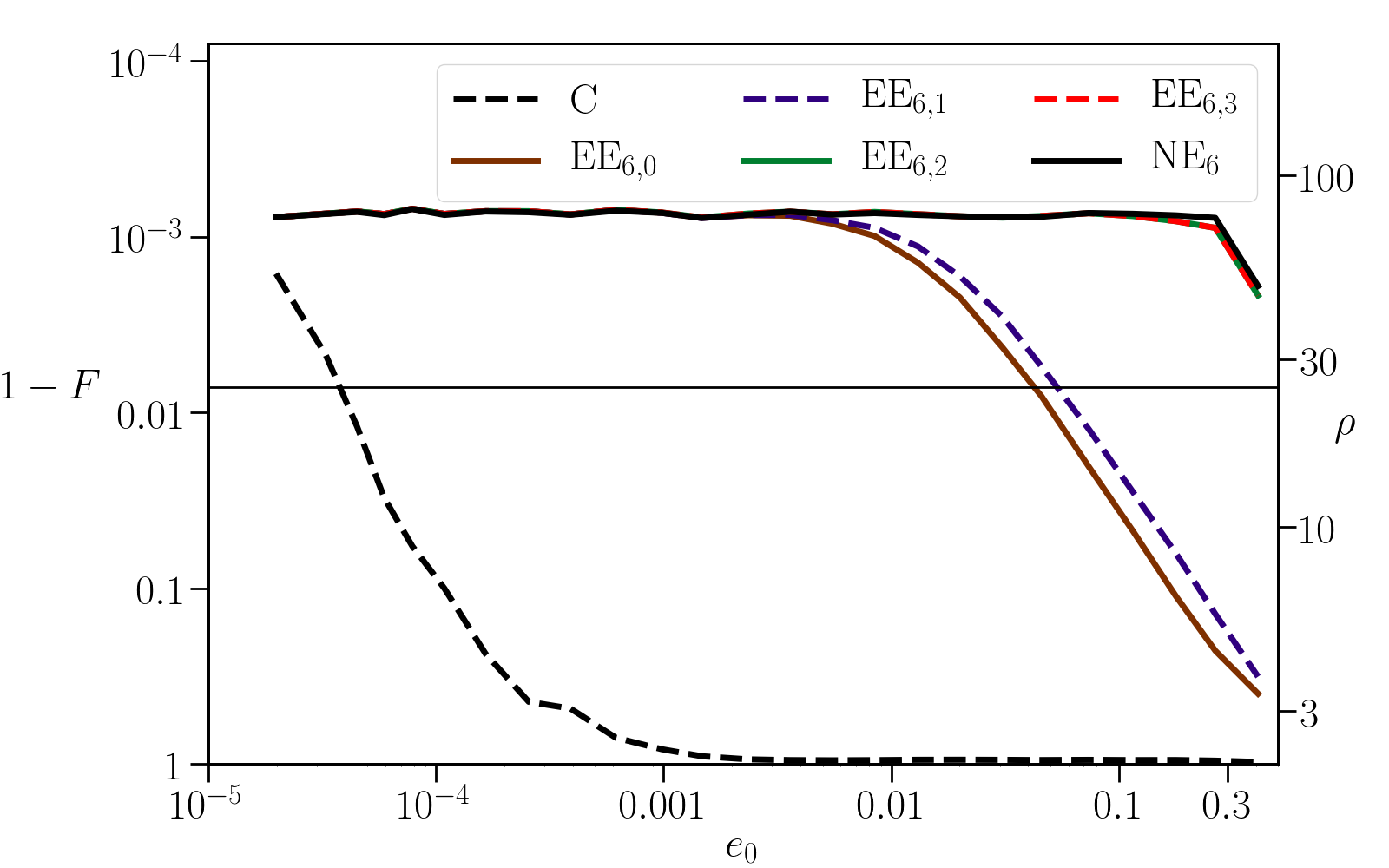}
	\includegraphics[width=0.45\textwidth]{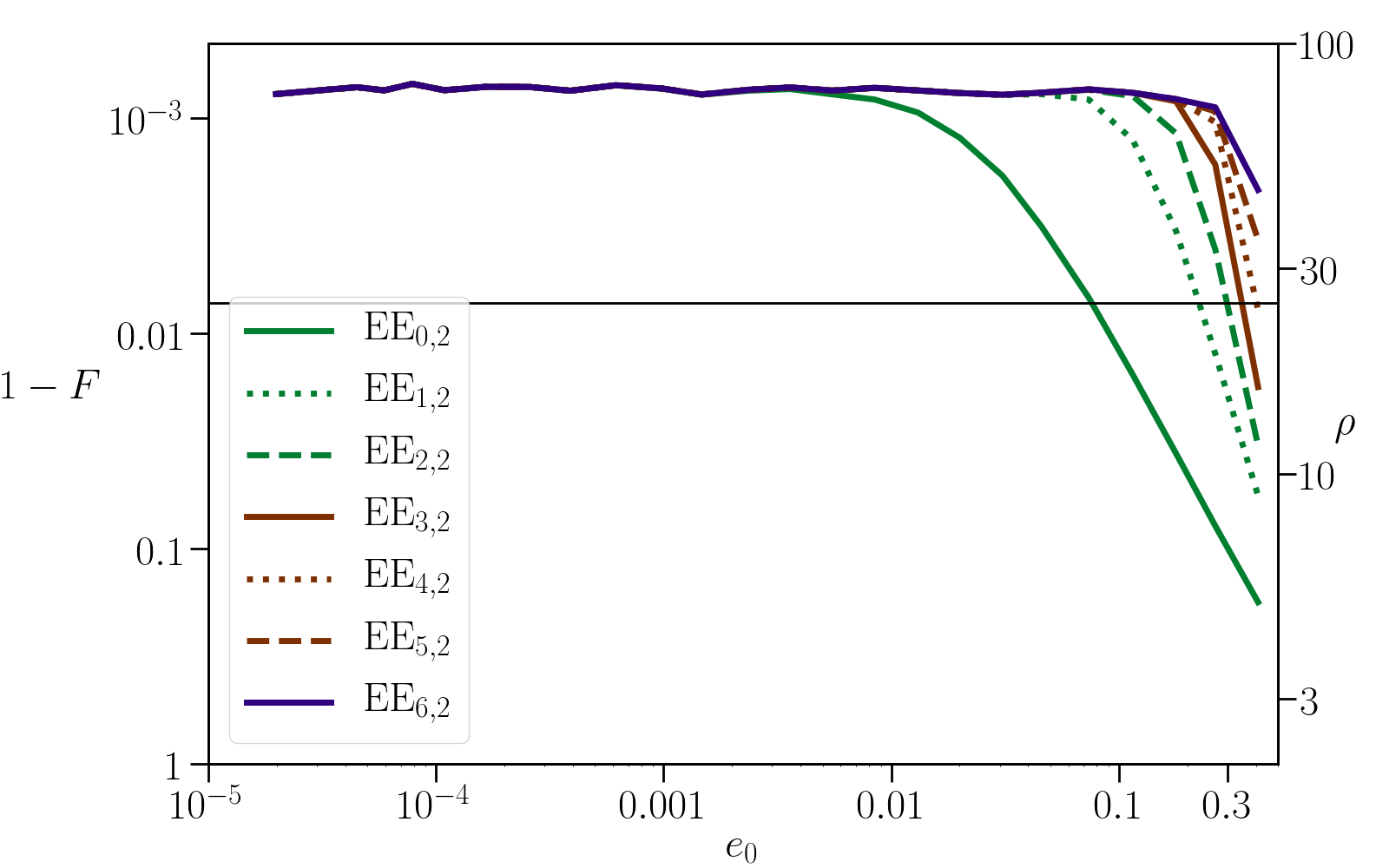}
	\caption{Same as Fig.~\ref{fig:faith-ia}, for early inspiral run (Ib) with starting 	
	eccentricity $10^{-5} < e_0 < 0.5$ and spin magnitudes $0 < \chi_i < 1$.}
	\label{fig:faith-ib}
\end{center}
\end{figure}

We present in Fig.~\ref{fig:faith-ib} the results from early inspiral run (Ib), with starting 
eccentricity $10^{-5} < e_0 < 0.5$ and spin magnitudes $0 < \chi_i < 1$. We can see that, in this 
case, using circular waveforms will likely result in large biases even when the starting 
eccentricity is below $10^{-3}$. The large number of orbital cycles accumulated is such that the 
small difference in the frequency evolution induces very low faithfulness even for very low 
eccentricities. On the other hand, the eccentric waveforms perform better than in the late inspiral 
case. In the top panel, we can see that the low-order EE$_{6,0}$ waveform stays above the 
faithfulness threshold for $e_0 \lesssim 0.05$, while the high-order one EE$_{6,2}$ is above the 
threshold for the whole parameter space that we investigated. In the bottom panel, we can see that 
the waveform with circular amplitudes EE$_{0,2}$ stays above the threshold for $e_0 \lesssim 0.1$, 
while the waveforms EE$_{M,2}$, $M \geq 2$ do so for $e_0 \lesssim 0.3$.

\begin{figure}[th]
\begin{center}
	\includegraphics[width=0.45\textwidth]{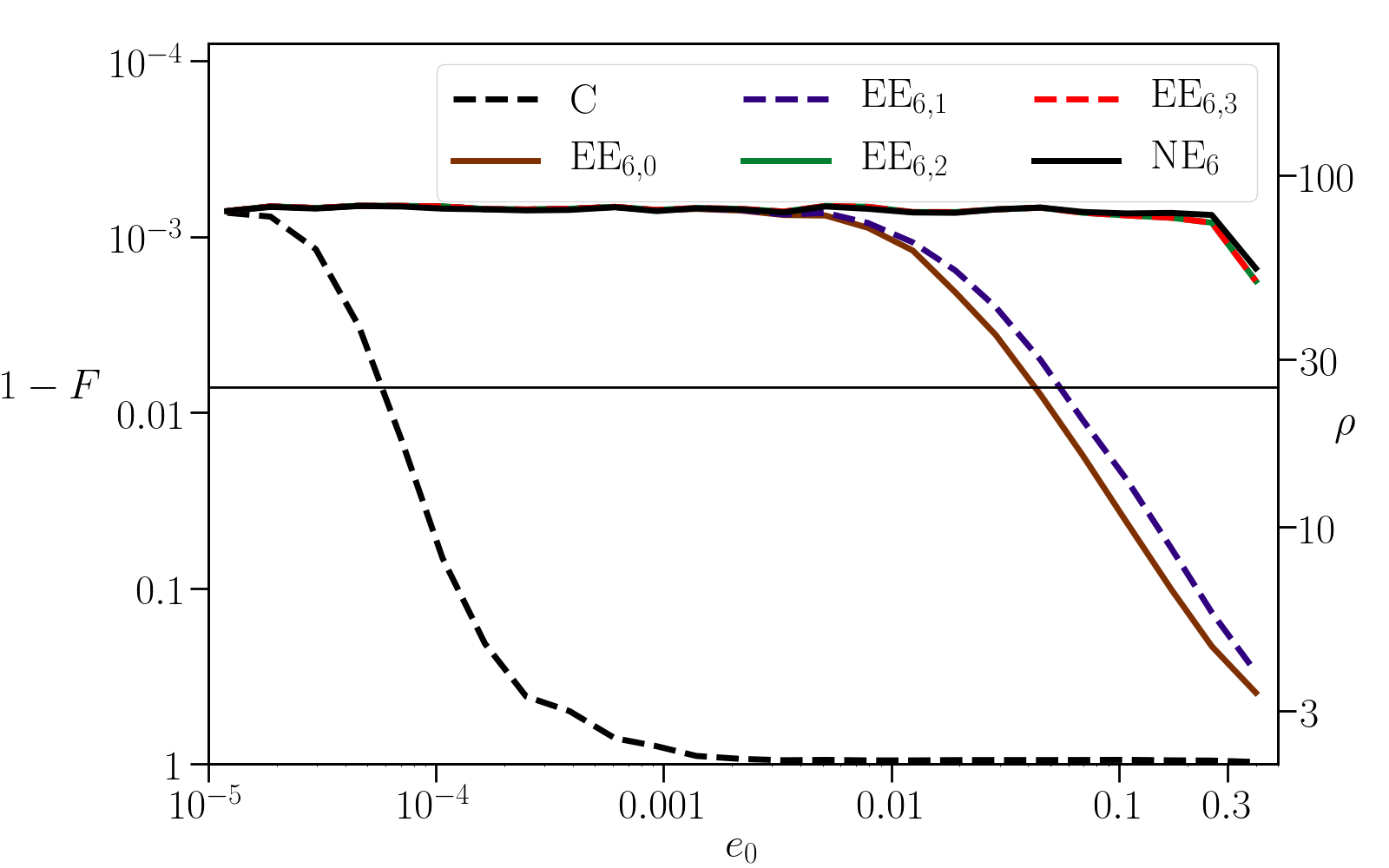}
	\includegraphics[width=0.45\textwidth]{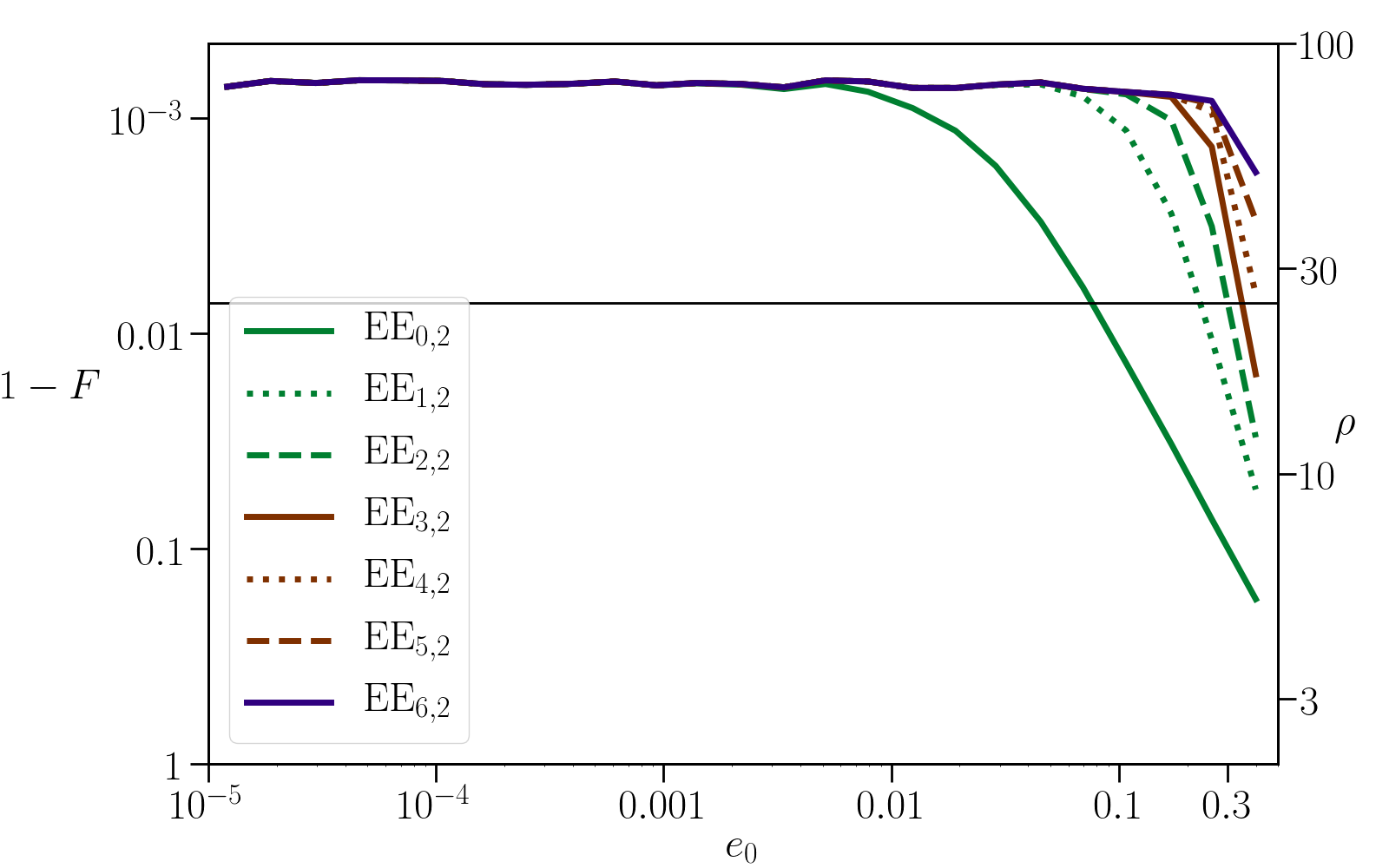}
	\caption{Same as Fig.~\ref{fig:faith-ia}, for early inspiral run (IIb) with starting 	
	eccentricity $10^{-5} < e_0 < 0.5$ and spin magnitudes $0 < \chi_i < 0.1$.}
	\label{fig:faith-iib}
\end{center}
\end{figure}

We present in Fig.~\ref{fig:faith-iib} the results from early inspiral run (IIb), with starting 
eccentricity $10^{-5} < e_0 < 0.5$ and spin magnitudes $0 < \chi_i < 0.1$. We can see that, for 
initial eccentricities $e_0 \gtrsim 10^{-4}$, circular waveforms yield a faithfulness below $F = 
0.9$. Thus, even if they are slowly spinning, the use of circular waveforms for parameter estimation 
for such binaries is likely to yield important biases. Using eccentric waveforms for early inspiral 
systems is therefore crucial in order to ensure accurate parameter recovery, even with initial 
eccentricities as low as $e_0 \sim 10^{-4}$. In the bottom panel, similarly to run (Ib), we can see 
that the waveform with circular amplitudes EE$_{0,2}$ stays above the threshold for $e_0 \lesssim 
0.1$, while the waveforms EE$_{M,2}$, $M \geq 2$ do so for $e_0 \lesssim 0.3$.

\begin{figure}[th]
\begin{center}
	\includegraphics[width=0.45\textwidth]{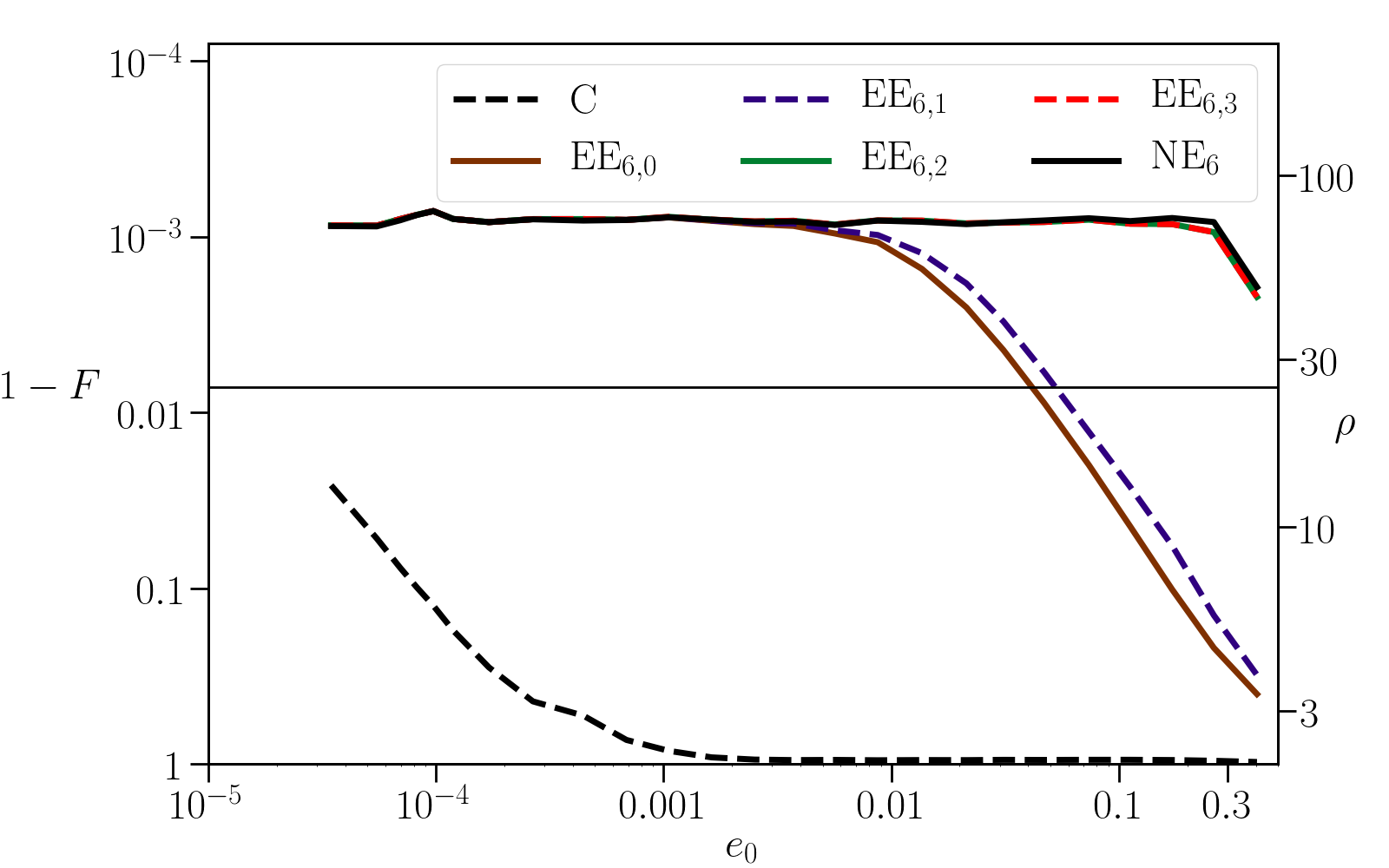}
	\includegraphics[width=0.45\textwidth]{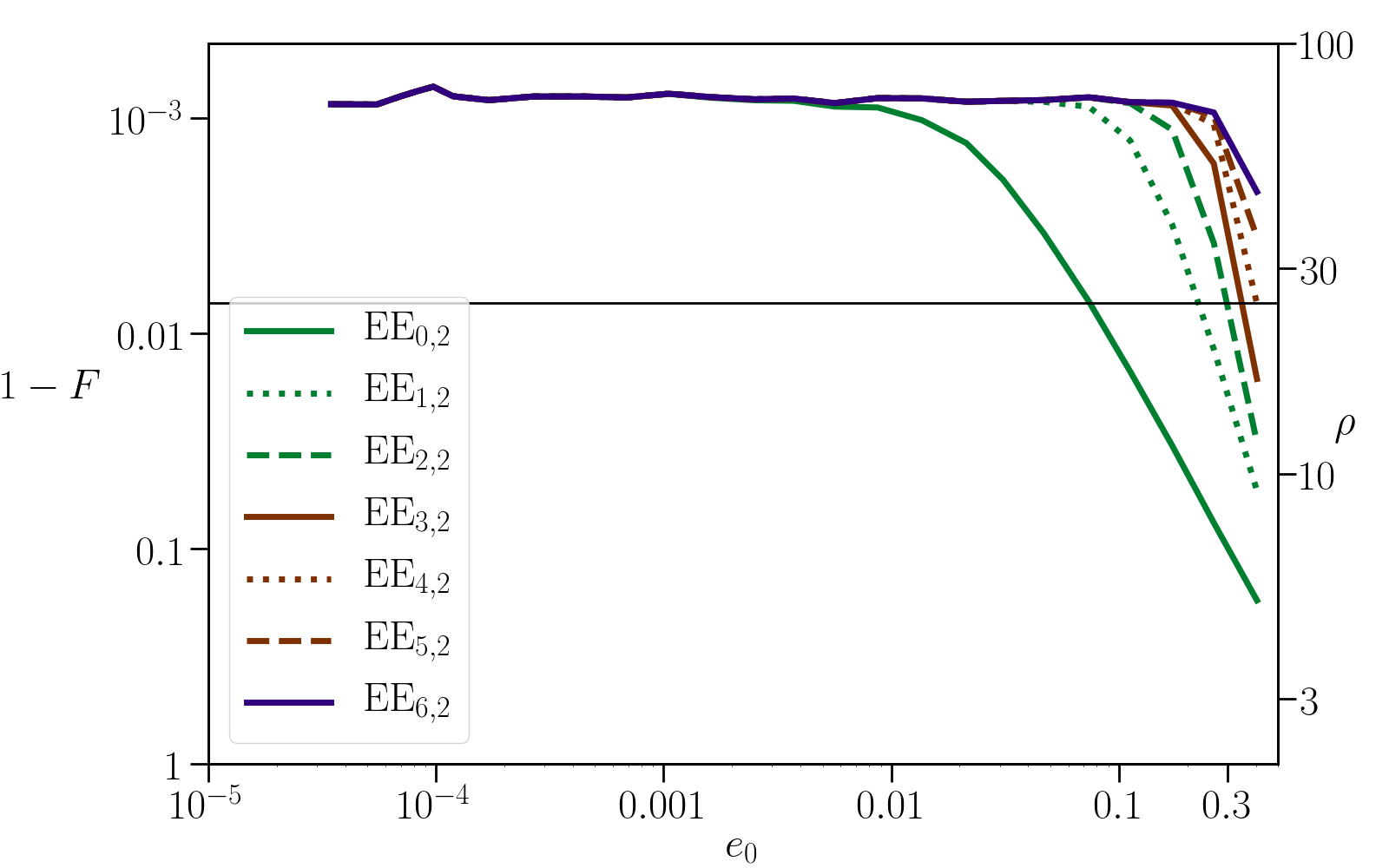}
	\caption{Same as Fig.~\ref{fig:faith-ia}, for early inspiral run (IIIb) with starting 	
	eccentricity $10^{-5} < e_0 < 0.5$ and spin magnitudes $\chi_i = 1$.}
	\label{fig:faith-iiib}
\end{center}
\end{figure}

We present in Fig.~\ref{fig:faith-iiib} the results from early inspiral run (IIIb), with starting 
eccentricity $10^{-5} < e_0 < 0.5$ and spin magnitudes $\chi_i = 1$. While the results for the eccentric waveforms are similar to the ones shown in Figs.~\ref{fig:faith-ib} and~\ref{fig:faith-iib}, the circular waveform never reached a median faithfulness above the threshold line above an initial eccentricity of $e_0 = 3 \times 10^{-5}$. This indicates that highly spinning systems in the early inspiral will require the use of an eccentric model irrespective of their initial eccentricity.

\begin{figure}[th]
\begin{center}
	\includegraphics[width=0.45\textwidth]{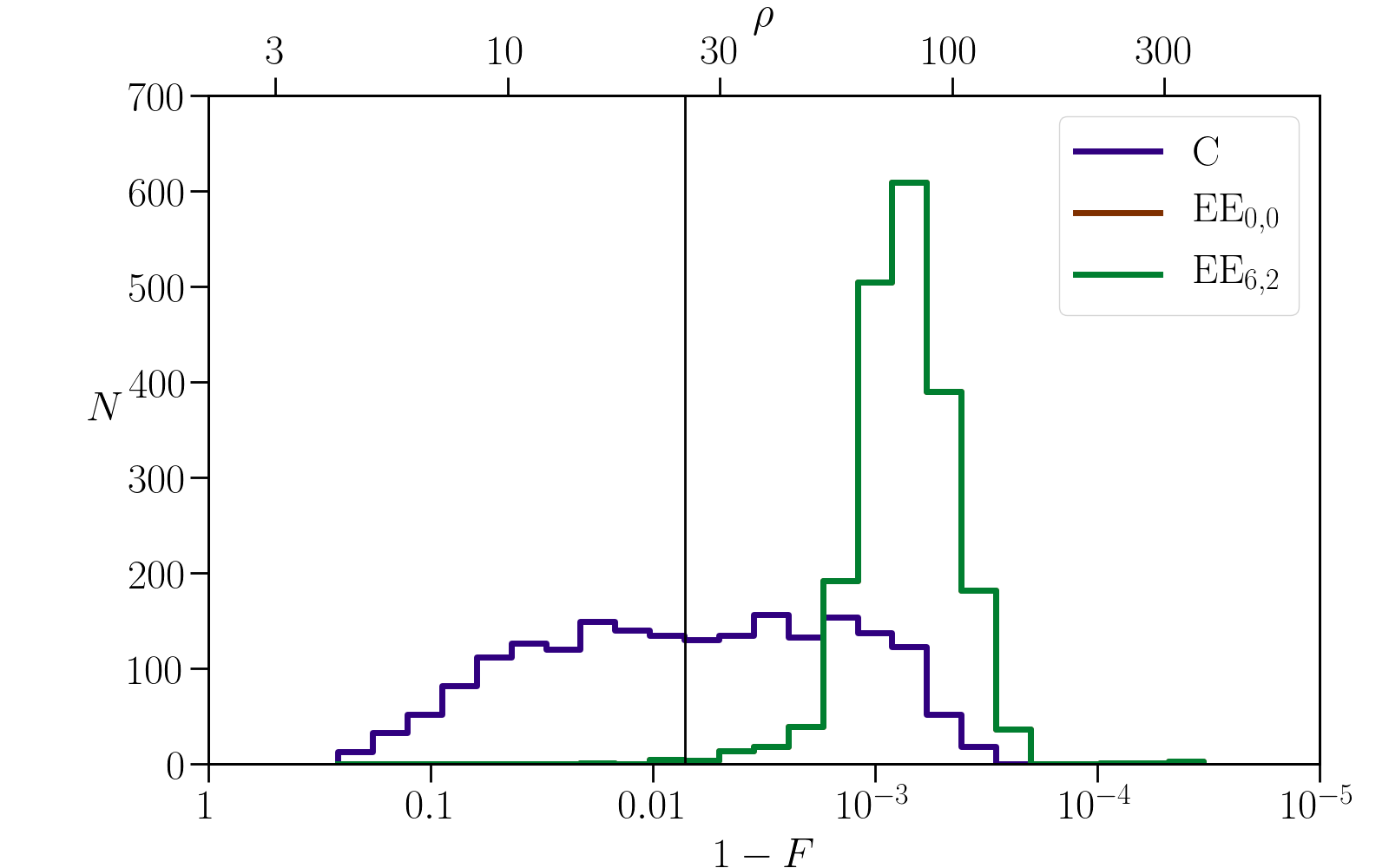}
	\caption{Same as Fig.~\ref{fig:faith-iva}, for early inspiral run (IVb) with starting 
	eccentricity $e_0 = e\sub{min}$ and spin magnitudes $0 < \chi_i < 1$. The systems simulated 	
	here correspond to highly spinning fully circularized binaries.}
	\label{fig:faith-ivb}
\end{center}
\end{figure}

We present in Fig.~\ref{fig:faith-ivb} the results from early inspiral run (IVb), with starting 
eccentricity $e_0 = e\sub{min}$ and spin magnitudes $0 < \chi_i < 1$. We can see that for these 
systems, including the eccentricity in the phasing is important, but the order used in other 
effects matters very little. Indeed, the faithfulness distributions for the two eccentric waveforms 
EE$_{0,0}$ and EE$_{6,2}$ are indistinguishable and have support almost exclusively above 
the faithfulness threshold, whereas the faithfulness distribution for the circular waveforms has 
46~\% support below the threshold.

\begin{figure}[th]
\begin{center}
	\includegraphics[width=0.45\textwidth]{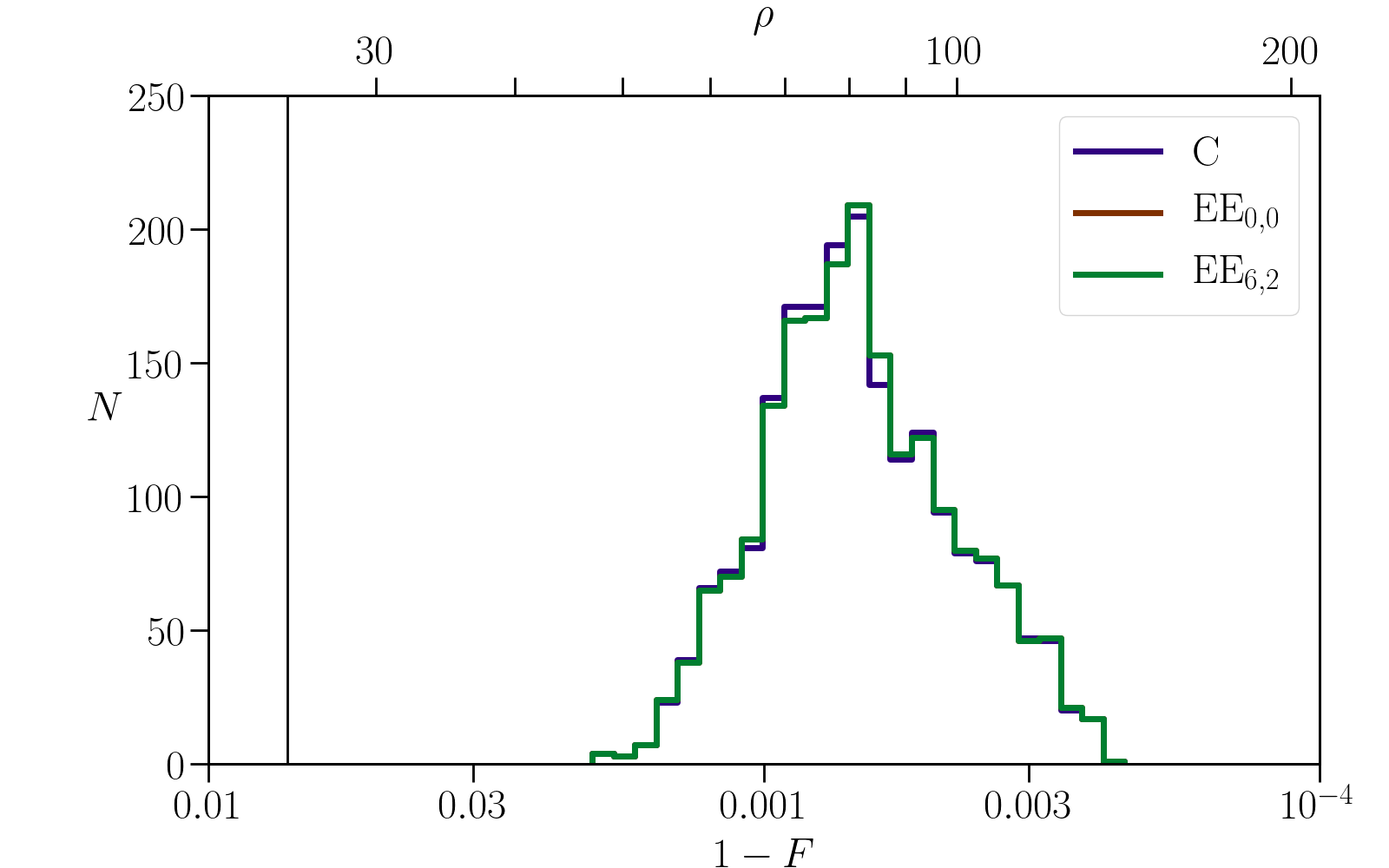}
	\caption{Same as Fig.~\ref{fig:faith-iva}, for early inspiral run (Vb) with starting 	
	eccentricity $e_0 = e\sub{min}$ and spin magnitudes $0 < \chi_i < 0.1$. The systems simulated 	
	here correspond to slowly spinning fully circularized binaries.}
	\label{fig:faith-vb}
\end{center}
\end{figure}

We present in Fig.~\ref{fig:faith-vb} the results from early inspiral run (Vb), with starting 
eccentricity $e_0 = e\sub{min}$ and spin magnitudes $0 < \chi_i < 0.1$. We can see that for these 
systems, circular waveforms have a faithfulness distribution almost identical to those of eccentric 
waveforms, indicating that when the spins are small and the binaries have fully circularized, the 
use of circular waveforms may be sufficient for unbiased parameter estimation.

\begin{figure}[th]
\begin{center}
	\includegraphics[width=0.45\textwidth]{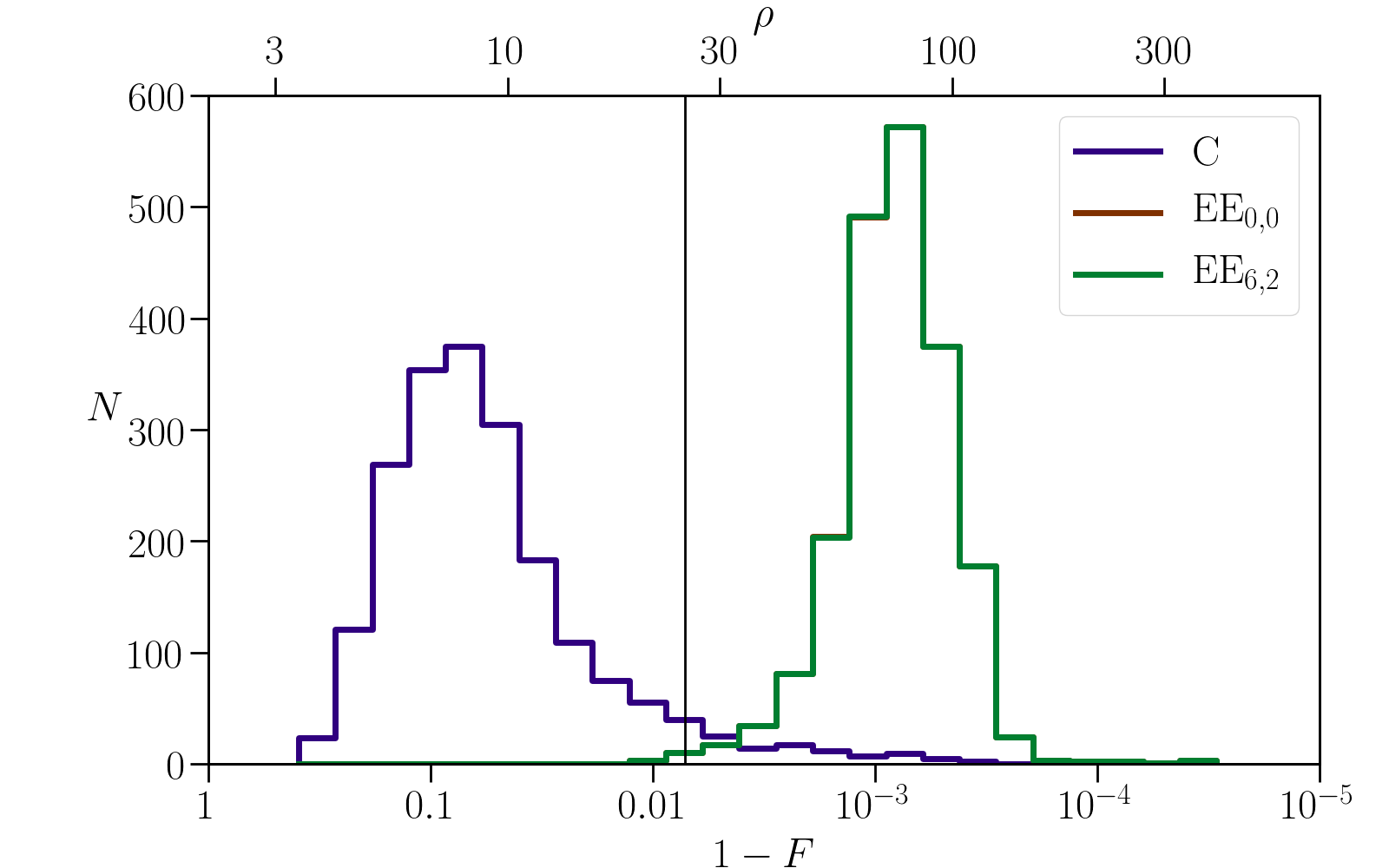}
	\caption{Same as Fig.~\ref{fig:faith-iva}, for early inspiral run (VIb) with starting 	
	eccentricity $e_0 = e\sub{min}$ and spin magnitudes $\chi_i = 1$. The systems simulated 	
	here correspond to maximally spinning fully circularized binaries.}
	\label{fig:faith-vib}
\end{center}
\end{figure}

We present in Fig.~\ref{fig:faith-vib} the results from early inspiral run (VIb), with starting 
eccentricity $e_0 = e\sub{min}$ and spin magnitudes $\chi_i = 1$. We observe in this figure that the faithfulness distribution for the circular waveforms has 94\% support below the threshold line, indicating that for highly spinning binaries, the use of eccentric waveforms will be crucial for unbiased parameter estimation. However, the distributions for the two eccentric waveforms 
EE$_{0,0}$ and EE$_{6,2}$ are indistinguishable also in this case, indicating that the precision of 
the waveform amplitude is of little importance. Thus, eccentricity and spins will be important 
to include in the analysis of stellar-origin black hole binaries with LISA to account for the 
possibility of high spins, even if the binaries have fully circularized. However, using accurate 
amplitudes might be unnecessary for those sources.

\begin{table}
\begin{center}
\begin{tabular}{c|ccc}
	\hline 
	\hline
	Waveform & $e_0\super{max,l}$ & & $e_0\super{max,e}$\\
	\hline
	C & $0.0078$ & & $< 7 \times 10^{-5}$ \\
	EE$_{0,0}$ & $0.056$ & & $0.086$ \\
	EE$_{6,0}$ & $0.036$ & & $0.04$ \\
	EE$_{0,2}$ & $0.056$ & & $0.086$ \\
	EE$_{2,2}$ & $0.23$ & & $0.32$ \\
	EE$_{4,2}$ & $0.29$ & & $> 0.4$ \\
	EE$_{6,2}$ & $0.32$ & & $> 0.4$ \\
	NE$_6$ & $0.33$ & & $> 0.4$ \\
	\hline
\end{tabular}
\end{center}
	\caption{For a few select waveforms, maximum initial eccentricity $e_0\super{max,l}$ for which 
	the median faithfulness in the late inspiral runs stays above the faithfulness threshold, and 
	the same for the early inspiral runs $e_0\super{max,e}$.\label{tab:medfaith}}
\end{table}

Comparing the different results, we find that the median faithfulness for each waveform is 
mainly influenced by the initial eccentricity, and the stage in the inspiral that they find 
themselves in. We summarize in Table~\ref{tab:medfaith} the initial eccentricity below which the 
median faithfulness falls above the threshold line for a few of the waveforms compared in our 
simulations. Interestingly, we find that waveform EE$_{0,0}$ performs slightly better than waveform 
EE$_{6,0}$. We find the same to be true by comparing EE$_{0,0}$ to any waveform EE$_{M,0}$ or 
EE$_{M,1}$ with $M>0$. We thus remark that in order for the inclusion of beyond-circular effects in 
the amplitudes to increase the accuracy of the waveform, one also needs to include periastron 
precession effects at least at second order.

\section{Conclusion}\label{sec:conc}

We have constructed two families of Fourier domain waveforms for spin-precessing binaries on 
eccentric orbits. These include phasing at the third nonspinning post-Newtonian order, including 
leading-order spin-orbit and spin-spin interactions, as well as instantaneous amplitudes at 
second post-Newtonian order as small eccentricity expansions. In this work, we have used amplitudes 
up to $\ord{e^6}$, but the extension to higher orders in the eccentricity would be trivial though 
lengthy. Through comparisons with a complete time domain waveform at consistent post-Newtonian 
order, we find that our new waveforms faithfully reproduce their Fourier transform for initial 
eccentricities up to $e_0 \sim 0.3$ for systems in the late inspiral, and at least up to $e_0 \sim 
0.4$ for systems in the early inspiral such as stellar-origin black hole binaries as observed by 
LISA. 

Comparing results, we find that using circular waveforms would likely lead to significant biases in 
parameter recovery, even for fully circularized binaries with a signal-to-noise ratio around 25, 
provided they are highly spinning. Indeed, a 2PN spin effect prevents the eccentricity of a binary 
system from vanishing completely unless the projections of the reduced spins in the orbital plane are 
exactly equal to each other.
We find that the use of circular waveforms can cause biases if fully circularized systems 
with large spin magnitudes and random orientations are observed in the late inspiral, but not if 
the spin magnitudes are small. This situation is made worse if binary systems are observed in the 
early inspiral, and we expect large biases with circular waveforms irrespective of the initial 
eccentricity for highly spinning systems, even if they are fully circularized. However, if the 
spins are sufficiently small and the binaries have circularized below an eccentricity of $10^{-4}$ 
when the observations start, we expect the use of circular waveforms to be appropriate for 
parameter estimation. 
Overall, we expect circular waveforms to be safe to use for parameter estimation in the late inspiral if the initial eccentricity falls below $10^{-2}$ and in the early inspiral when it falls below $10^{-4}$, but we would recommend the use of eccentric phasing in the waveform to describe highly spinning systems, even if they have fully circularized.

Those waveforms provide a step towards the inclusion of the eccentricity in gravitational wave data 
analysis such as that performed by the LIGO/Virgo community. We argue from the simulations described in 
this paper that the inclusion of spins and eccentricity might be of importance for reducing 
potential biases in the parameter recovery of binaries, even when they are fully circularized. 
While circular templates might be appropriate to describe slowly spinning systems, it can be 
important to include in the modeling of highly spinning systems. It is worth noting that the 
faithfulness measurements described in this work are not suitable to estimate the loss of events 
due to mismodeling, or the measurability of binary parameters, including the initial eccentricity. 
We leave those questions open for future work.

Some assumptions made in this work, particularly the neglect of orbital timescale effects in the 
spin-orbit precession dynamics, have to be more closely investigated. Furthermore, the inclusion of 
the merger and ringdown signals in our waveforms is also very important work for the future, and 
will have to be taken into account in the construction of waveform templates to use in current and 
future detectors. The waveform that we have presented in this work, while useful to describe 
inspiral-dominated signals such as stellar-origin black hole binaries in LISA or neutron star 
binaries in the LIGO/Virgo network, is inspiral-only and therefore cannot be used alone in the 
characterization of merger-dominated signals such as black hole binaries as observed by the 
LIGO/Virgo network.

\acknowledgments

We thank Katerina Chatziioannou and Eliu Huerta for useful comments. We thank our referees for useful comments.
A.~K. is supported by NSF CAREER Grant No. PHY-1055103, by FCT Contract No.~IF/00797/2014/CP1214/CT0012 
under the IF2014 Programme, and by H2020-MSCA-RISE-2015 Grant No. StronGrHEP-690904. This work was 
supported by the Centre National d'{\'E}tudes Spatiales. Y.~B. and L.~d.~V. are supported by the 
Swiss National Science Foundation. This work has made use of the Horizon Cluster, hosted by the 
Institut d’Astrophysique de Paris. We thank St{\'e}phane Rouberol for smoothly running this cluster 
for us.

\appendix

\begin{widetext}

\section{Quadrupole-monopole effects}\label{sec:quad-mono}

The 2PN part of the quasi-Keplerian parametrization found in~\cite{klein-2010} is based upon the 
reduced Lagrangian
\begin{align}
	\frac{\mathcal{L}}{\nu} &= \frac{\dvec{r}^2}{2} + \frac{1}{r} + \frac{1}{r^3} \bm{s}_1 \cdot 
		\bm{s}_2 - \frac{3}{r^3} \left(\uvec{r} \cdot \bm{s}_1 \right) \left( \uvec{r} \cdot 		
		\bm{s}_2 \right) \,,
\end{align}
where the reduced spins $\bm{s}_i = \bm{S}_i / m_i$. The quadrupole-monopole part of the reduced 
Lagrangian is~\cite{poisson-1998,keresztes-2005}
\begin{align}
	\frac{\mathcal{L}\sub{QM}}{\nu} &= \frac{1}{2r^3} \sum_{i=1}^2 q_i \left[\bm{s}_i^2 - 3 \left( 
		\uvec{r} \cdot \bm{s}_i \right)^2 \right] \,,
\end{align}
where the quadrupole parameter $q_i$ is defined in such a way that $q_i = 1$ for black holes. The 
total Lagrangian can then be written as
\begin{align}
	\frac{\mathcal{L}}{\nu} &= \frac{\dvec{r}^2}{2} + \frac{1}{r} + \frac{1}{2r^3} \bm{s}^2 - 
		\frac{3}{2r^3} \left(\uvec{r} \cdot \bm{s} \right)^2 + \frac{1}{2r^3} \sum_{i=1}^2 \left( 
		q_i - 1 \right) \left[\bm{s}_i^2 - 3 \left( \uvec{r} \cdot \bm{s}_i \right)^2 \right] \,,
\end{align}
where $\bm{s} = \bm{s}_1 + \bm{s}_2$.

Thus, a quasi-Keplerian description of the orbit including quadrupole-monopole terms can be found 
by adding the 2PN terms of~\cite{klein-2010}, using the substitutions 
$(\bm{s}_1 \to \bm{s}/\sqrt{2}, \bm{s}_2 \to \bm{s}/\sqrt{2})$, 
$(\bm{s}_1 \to \bm{s}_1 \sqrt{(q_1 - 1)/2}, \bm{s}_2 \to \bm{s}_1 \sqrt{(q_1 - 1)/2})$, 
and $(\bm{s}_1 \to \bm{s}_2 \sqrt{(q_2 - 1)/2}, \bm{s}_2 \to \bm{s}_2 \sqrt{(q_2 - 1)/2})$. It reads
\begin{subequations}
\begin{align}
	r &= a ( 1 - e_r \cos u) + f_r(v) \,, \\
	\phi &= (1 + k) v + f_{\phi,1}(v) + f_{\phi,2}(v) \,, \\
	\tan \frac{v}{2} &= \sqrt{\frac{1 + e_\phi}{1 - e_\phi}}\tan \frac{u}{2} \,,\\
	l &= u - e \sin u \,, \\
	\dot{l} &= n \,,
\end{align}
\end{subequations}
with
\begin{subequations}
\begin{align}
	a &= \frac{1}{\left(1 - e^2\right) y^2} \left[ 1 + \frac{1}{2} \left(1 + e^2 \right) \gamma_1 
		y^4 \right] \,, \\
	e_r^2 &= e^2 \left[ 1 + \left( 1 - e^2 \right) \gamma_1 y^4 \right] \,, \\
	k &= \frac{3}{2} \gamma_1 y^4 \,, \\
	e_\phi^2 &= e^2 \left[ 1 + 2\left( 1 - e^2 \right) \gamma_1 y^4 \right] \,, \\
	n &= \left(1 - e^2 \right)^{3/2} y ^3 \left(1 - \frac{3}{2} \gamma_1 y^4 \right) \,, \\
	\gamma_1 &= \frac{1}{2} \left\{ 3 \left(\uvec{L} \cdot \bm{s} \right)^2 - \bm{s}^2 + 
		\sum_{i=1}^2 \left(q_i - 1\right) \left[ 3 \left(\uvec{L} \cdot \bm{s}_i \right)^2 - 
		\bm{s}_i^2 \right] \right\} \,, \\
	f_r(v) &= - \frac{y^2}{4} \left[ \left| \uvec{L} \times \bm{s} \right|^2 \cos (2v - 2\psi) + 
		\sum_{i=1}^2 \left( q_i - 1 \right) \left| \uvec{L} \times \bm{s}_i \right|^2 \cos (2v - 
		2\psi_i) \right] \,, \\
	f_{\phi,1}(v) &= - \frac{y^4 e}{2} \left[ \left| \uvec{L} \times \bm{s} \right|^2 \sin (v - 
		2\psi) + \sum_{i=1}^2 \left( q_i - 1 \right) \left| \uvec{L} \times \bm{s}_i \right|^2 \sin 
		(v - 2\psi_i) \right] \,, \\
	f_{\phi,2}(v) &= - \frac{y^4}{8} \left[ \left| \uvec{L} \times \bm{s} \right|^2 \sin (2v 
		- 2\psi) + \sum_{i=1}^2 \left( q_i - 1 \right) \left| \uvec{L} \times \bm{s}_i \right|^2 
		\sin (2v - 2\psi_i) \right] \,,
\end{align}
\end{subequations}
where $\psi$ is the angle subtended by the total reduced spin $\bm{s}$ and the periastron line, 
$\psi_i$ is the angle subtended by the individual reduced spin $\bm{s}_i$ and the periastron line 
(see Fig. \ref{fig:angles}), and the periastron line is defined by the equation $v = u = l = 2 p 
\pi$, $p \in \mathbb{Z}$. We can then use this representation of the orbit together with the orbit 
averaged evolution equations for the energy and orbital angular momentum computed 
in~\cite{gergely-2002} to find
\begin{subequations}
\begin{align}
	M \frac{dy}{dt} =&\; \nu \left(1 - e^2 \right)^{3/2} \bigg\{\left(\frac{32}{5} + \frac{28}{5} 
		e^2 \right) y^9 + \sigma \bigg[ - \frac{84}{5} - \frac{228}{5} e^2 - \frac{33}{5} e^4, 
		\frac{242}{5} + \frac{654}{5} e^2 + \frac{381}{20} e^4, -\frac{447}{10} e^2 - \frac{93}{10} 
		e^4, \nonumber\\
		&\frac{88}{5} - 16 q + \left(48 - \frac{216}{5} q \right) e^2 + \left( \frac{69}{10} - 
		\frac{63}{10} q \right) e^4, - \frac{244}{5} + 48 q + \left(-132 + \frac{648}{5} q \right) 
		e^2 + \left(- \frac{96}{5} + \frac{189}{10} q \right) e^4, \nonumber\\
		&\left(1 - q\right) \left( \frac{447}{10} e^2 + \frac{93}{10} e^4 \right) \bigg] y^{13}
		\bigg\} \,, \\
	M \frac{de^2}{dt} =&\; - \nu \left( 1 - e^2 \right)^{3/2} \bigg\{\left(\frac{608}{15} e^2 + 
		\frac{242}{15} e^4 \right) y^8 + \sigma \bigg[ \frac{2}{3} - \frac{1961}{15} e^2 - 
		\frac{2527}{12} e^4 - \frac{157}{8} e^6, -\frac{2}{3} + \frac{5623}{15} e^2 + 
		\frac{2393}{4} e^4 \nonumber\\
		&+ \frac{447}{8} e^6, -\frac{5527}{30} e^2 - \frac{10117}{30} e^4 - \frac{5507}{160} e^6,
		- \frac{4}{3} + \left( \frac{682}{5} - \frac{1876}{15} q \right) e^2 + \left(\frac{1337}{6} 
		- \frac{595}{3} q \right) e^4 + \left(\frac{83}{4} - \frac{37}{2} q \right) e^6, \nonumber\\
		&\frac{4}{3} + \left( - \frac{5618}{15} + \frac{1876}{5} q \right) e^2 + \left( 
		-\frac{1203}{2} + 595 q \right) e^4 + \left( - \frac{225}{4} + \frac{111}{4} q\right) e^6 
		,\nonumber\\
		& \left( \frac{2764}{15} - \frac{921}{5} q \right) e^2 + \left(\frac{1687}{5} - 
		\frac{5056}{15} q \right) e^4 + \left( \frac{551}{16} - \frac{172}{5} q \right) e^6 \bigg] 
		y^{12} \bigg\} \,,
\end{align}
\end{subequations}
where
\begin{align}
	\sigma(a, b, c, a_1 + a_2 q, b_1 + b_2 q, c_1 + c_2 q) =&\; a \bm{s}^2 + b \left( \uvec{L} 
		\cdot \bm{s} \right)^2 + c \left| \uvec{L} \times \bm{s} \right|^2 \cos 2 \psi \nonumber\\
		&+ \sum_{i=1}^2 \left[ \left(a_1 + a_2 q_i \right) \bm{s}_i^2 + \left( b_1 + b_2 q_i 
		\right) \left( \uvec{L} \cdot \bm{s}_i \right)^2 + \left(c_1 + c_2 q_i \right) \left| 
		\uvec{L} \times \bm{s}_i \right|^2 \cos 2 \psi_i \right] \,.
\end{align}
We thus find the residual eccentricity found in~\cite{klein-2010} unchanged by quadrupole-monopole 
effects.

\section{Quasi-Keplerian parametrization}\label{sec:qK-param}

A full quasi-Keplerian parametrization of the orbit at 2PN order in harmonic coordinates is~\cite{memmesheimer-2004,klein-2010}
\begin{subequations}
\begin{align}
	r &= a ( 1 - e_r \cos u) + f_r(v) \,, \\
	\phi &= (1 + k) v + f_\phi(v) \,, \\
	\tan \frac{v}{2} &= \sqrt{\frac{1 + e_\phi}{1 - e_\phi}}\tan \frac{u}{2} \,,\\
	l &= u - e \sin u + f_t(u,v) \,,\\
	\dot{l} &= n \,,
\end{align}
\end{subequations}
with
\begin{subequations}
\begin{align}
	a =&\; \frac{1}{\left( 1 - e^2 \right) y^2} \bigg\{ 1 + \left[ -1 + \frac{\nu}{3} + \left( 3 - 
		\frac{\nu}{3} \right) e^2 \right] y^2 + \beta \left( \frac{2}{3} + 2 e^2, 1 + e^2\right) 
		y^3 + \bigg[5 + \frac{11}{4}\nu + \frac{\nu^2}{9} \nonumber \\
		&+ \bigg(\frac{21}{2} -\frac{73}{6} \nu - \frac{2}{9} \nu^2 \bigg) e^2 
		+ \left( 1 + \frac{5}{12}\nu + \frac{\nu^2}{9} \right) e^4 	+ \left(1 -	e^2\right)^{3/2} 
		\left(-5 + 2 \nu \right) + \frac{\gamma_1}{2} \left(1 + e^2\right)\bigg] y^4 \,, \\
	e_r^2 =&\; e^2 \bigg\{ 1 + \left(1 - e^2 \right) \bigg\{ \left(8 - 3\nu \right) y^2 + \beta 
		\left(4, 2\right) y^3 + \bigg[	32 - \frac{467}{12} \nu + 4 \nu^2 + \left(- 40 + 		
		\frac{371}{12} \nu - 4 \nu^2 \right) e^2 \nonumber\\
		& + \sqrt{1 - e^2} \left(15 - 6 \nu \right) + \gamma_1\bigg] y^4 \bigg\} \bigg\} \,,\\
	k =&\; 3 y^2 +\beta \left( 4, 3\right) y^3 + \left[ \frac{27}{2} - 7 \nu + \left( \frac{51}{4}- 
		\frac{13}{2} \nu \right) e^2 + \frac{3}{2} \gamma_1\right] y^4 \,,\\
	e_{\phi}^2 =&\; e^2 \bigg\{1 + \left(1 - e^2 \right) \bigg\{ \left(8 - 2\nu \right) y^2 + \beta 
		\left(4, 4 \right) y^3 + \bigg[ 42 - \frac{113}{12} \nu + \frac{11}{12} \nu^2 + \left(- 40 
		+ \frac{1043}{48} \nu - \frac{89}{48} \nu^2 \right) e^2 \nonumber\\
		&+ \sqrt{1 - e^2} \left(15 - 6 \nu \right) + 2\gamma_1 \bigg] y^4 \bigg\} \bigg\} \,, \\
	n =&\; \left(1 - e^2\right)^{3/2} y^3 \bigg\{ 1 - 3 y^2 - \beta\left( 4, 3 \right) y^3 + \left[ 
		-\frac{9}{2} + 7 \nu + \left( -\frac{51}{4} + \frac{13}{2} \nu \right) e^2 - \frac{3}{2} 
		\gamma_1\right] y^4 \bigg\} \,,
\end{align}
\end{subequations}
where
\begin{subequations}
\begin{align}
	\beta(a, b) &= -\left[ \left(a \mu_1 + b \mu_2 \right)\bm{s}_1 + \left(b\mu_1 + a\mu_2 \right) 
		\bm{s}_2 \right] \cdot \uvec{L} \,, \\
	\gamma_1 &= \frac{1}{2} \left\{ 3 \left(\uvec{L} \cdot \bm{s} \right)^2 - \bm{s}^2 + 
		\sum_{i=1}^2 \left(q_i - 1\right) \left[ 3 \left(\uvec{L} \cdot \bm{s}_i \right)^2 - 
		\bm{s}_i^2\right]\right\} \,.
\end{align}
\end{subequations}
The functions $f_r$, $f_{\phi}$, $f_t$, and $f_n$ are given by
\begin{subequations}
\begin{align}
	f_r(v) &= \sum_{i=0}^2 b_{r,i} \cos (2v - 2\psi_i) \,, \\
	f_\phi(v) &= \sum_{k=2}^3 a_{\phi,k} \sin(k v) + \sum_{k=1}^2 \sum_{i=0}^2 b_{\phi,k,i} 
		\sin(kv-2\psi_i) \,, \\
	f_t(u,v) &= g_t(u-v) + a_t \sin(v) \,,
\end{align}
\end{subequations}
with
\begin{subequations}
\begin{align}
	b_{r,i} &= - \frac{y^2}{4} F_i \left| \uvec{L} \times \bm{s}_i \right|^2 \,, \\ 
	a_{\phi,2} &= e^2 \left( \frac{\nu}{8} - \frac{3}{8} \nu^2 \right) y^4 \,, \\
	a_{\phi, 3} &= -e^3\frac{3}{32} \nu^2 y^4 \,,\\
	b_{\phi,1,i} &= - \frac{e}{2} F_i \left| \uvec{L} \times \bm{s}_i \right|^2 y^4 \,, \\
	b_{\phi,2,i} &= - \frac{1}{8} F_i \left| \uvec{L} \times \bm{s}_i \right|^2 y^4 \,,	\\
	g_t &= \left(1 - e^2 \right)^{3/2} \left( \frac{15}{2} - 3 \nu 	\right) y^4 \,, \\
	a_t &= e \left(1 - e^2 \right)^{3/2} \left( -\frac{\nu}{2} - \frac{\nu^2}{8} \right) y^4 \,,
\end{align}
\end{subequations}
where we defined, for convenience, $\bm{s}_0 = \bm{s}$, $\psi_0 = \psi$, $F_0 = 1$, $F_1 = q_1 - 1$, 
and $F_2 = q_2 - 1$.

\section{Evolution equations}\label{sec:evol-eq}

The evolution equations of $y$ and $e$ are given at 3PN order by 
~\cite{arun-2008,arun-2008-2,arun-2009,klein-2010}
\begin{subequations}
\begin{align}
	 M \frac{d y}{dt} &= \left( 1 - e^2	\right)^{3/2} \nu y^9 \left( a_0 + \sum_{n=2}^6 a_n y^n 
		 \right) \,, \\
	 M \frac{de^2}{dt} &= - \left( 1 - e^2 \right)^{3/2} \nu y^8 \left( b_0 + \sum_{n=2}^6 b_n y^n 
		\right) \,,
\end{align}
\end{subequations}
where
\begin{subequations}
\begin{align}
	a_0 =&\; \frac{32}{5} + \frac{28}{5} e^2 \,, \\
	a_2 =&\; -\frac{1486}{105} - \frac{88}{5} \nu + \left( \frac{12296}{105} - \frac{5258}{45}\nu 
		\right) e^2 + \left( \frac{3007}{84} - \frac{244}{9} \nu \right) e^4 \,, \\
	a_3 =&\; \frac{128 \pi}{5} \phi_y + \beta \left( \frac{904}{15} + \frac{2224}{15} e^2 + 
	\frac{99}{5} e^4, 40 + \frac{1916}{15} e^2 + \frac{314}{15}e^4 \right) \,, \\
	a_4 =&\; \frac{34103}{2835} + \frac{13661}{315} \nu + \frac{944}{45} \nu^2 + \left( 
		-\frac{256723}{945} - \frac{173587}{315} \nu + \frac{147443}{270} \nu^2 \right) e^2 + 
		\left( \frac{2095517}{7560} - \frac{589507}{504} \nu + \frac{34679}{45} \nu^2 \right) e^4 
		\nonumber\\ 
		&+ \left(\frac{53881}{2520} - \frac{7357}{90} \nu + \frac{9392}{135} \nu^2 \right) e^6 
		+ \frac{e^2}{1 - e^2} \left( \frac{85}{6} + \frac{1445}{6} \nu \right) + \frac{1 - \sqrt{1 
		- e^2}}{\sqrt{1 - e^2}} \bigg[ 16 - \frac{32}{5} \nu + \left(266 - \frac{532}{5}\nu \right) 
		e^2 \nonumber\\
		&+ \left( -\frac{859}{2} + \frac{859}{5} \nu \right) e^4 + \left(-65 + 26 \nu \right) e^6 
		\bigg] + \sigma \bigg[ - \frac{84}{5} - \frac{228}{5} e^2 - \frac{33}{5} e^4, \frac{242}{5} 
		+ \frac{654}{5} e^2 + \frac{381}{20} e^4, - \frac{447}{10} e^2 - \frac{93}{10} e^4, 
		\nonumber\\
		&\frac{88}{5} - 16 q + \left(48 - \frac{216}{5} q \right) e^2 + \left( \frac{69}{10} - 
		\frac{63}{10} q \right) e^4, - \frac{244}{5} + 48 q + \left(-132 + \frac{648}{5} q \right) 
		e^2 + \left(- \frac{96}{5} + \frac{189}{10} q \right) e^4, \nonumber\\
		&\left(1 - q\right) \left( \frac{447}{10} e^2 + \frac{93}{10} e^4 \right) \bigg] \,, \\
	a_5 =&\; \pi \left( -\frac{4159}{105} \psi_y - \frac{756}{5} \nu \zeta_y \right) \,, \\
	a_6 =&\; \frac{16447322263}{21829500} - \frac{54784}{525} \gamma_E + \frac{512}{15} \pi^2 + 
		\left(-\frac{56198689}{34020} + \frac{902}{15}\pi^2 \right) \nu + \frac{541}{140} \nu^2 - 
		\frac{1121}{81} \nu^3 + \bigg[ \frac{247611308999}{87318000} \nonumber\\
		&- \frac{392048}{525} \gamma_E + \frac{3664}{15} \pi^2 + \left(-\frac{2828420479}{680400} + 
		\frac{477}{4} \pi^2 \right) \nu + \frac{1070903}{315} \nu^2 - \frac{392945}{324} \nu^3 
		\bigg] e^2 + \bigg[-\frac{236637777001}{58212000} \nonumber\\
		&- \frac{93304}{175} \gamma_E + \frac{872}{5} \pi^2 + \left( \frac{2963572847}{453600} - 
		\frac{53131}{960} \pi^2 \right) \nu + \frac{44123941}{6048} \nu^2 - \frac{2198212}{405} 
		\nu^3 \bigg] e^4 + \bigg[ -\frac{28913792717}{6468000} \nonumber\\
		&- \frac{4922}{175} \gamma_E + \frac{46}{5} \pi^2 + \left( \frac{107275139}{30240} - 
		\frac{369}{80} \pi^2 \right) \nu + \frac{5155951}{1512} \nu^2 - \frac{44338}{15} \nu^3 
		\bigg] e^6 + \bigg(-\frac{243511057}{887040} + \frac{4179523}{15120} \nu \nonumber\\
		&+ \frac{83701}{3780} \nu^2 - \frac{1876}{15} \nu^3 \bigg) e^8 + \frac{e^2}{1 - e^2} \left[ 
		\frac{91284763}{378000} + \left( \frac{19505077}{5040} - \frac{595}{8} \pi^2 \right) \nu - 
		\frac{48569}{12} \nu^2 - \frac{730168}{23625 \left(1 +	\sqrt{1- e^2}\right)}\right] 
		\nonumber\\
		&+ \frac{1 - \sqrt{1 - e^2}}{\sqrt{1 - e^2}} \bigg\{-\frac{1425319}{3375} + \left( 
		\frac{9874}{315} - \frac{41}{30} \pi^2 \right) \nu + \frac{632}{15} \nu^2 + \bigg[ 
		\frac{2385427}{1050} + \left( -\frac{274234}{45} + \frac{4223}{240} \pi^2 \right) \nu + 
		\frac{70946}{45} \nu^2 \bigg] e^2 \nonumber\\
		&+ \bigg[\frac{8364697}{4200} + \bigg( \frac{1900517}{630} - \frac{32267}{960} \pi^2 
		\bigg) \nu - \frac{47443}{90} \nu^2 \bigg] e^4 + \bigg[-\frac{167385119}{25200} + \left( 
		\frac{4272491}{504} - \frac{123}{160} \pi^2 \right) \nu - \frac{43607}{18} \nu^2 \bigg] 
		e^6 \nonumber\\
		&+ \bigg( -\frac{65279}{168} + \frac{510361}{1260} \nu - \frac{5623}{45} \nu^2 \bigg) e^8
		\bigg\} + \frac{1284}{175} \kappa_y \nonumber\\
		&+ \left( \frac{54784}{525} + \frac{392048}{525} e^2 + \frac{93304}{175} e^4 + 
		\frac{4922}{175} e^6 \right) \log\left[ \frac{1 + \sqrt{1 - e^2}}{8 y \left(1 - e^2 
		\right)^{3/2}} \right] \,,\\
	b_0 =&\; \frac{608}{15} e^2 + \frac{242}{15} e^4 \,, \\
	b_2 =&\; \left( -\frac{1878}{35} - \frac{8168}{45} \nu \right) e^2 + \left( \frac{59834}{105} - 
		\frac{7753}{15} \nu \right) e^4 + \left( \frac{13929}{140} - \frac{3328}{45} \nu \right) 
		e^6 \,, \\
	b_3 =&\; \frac{788 \pi e^2}{3} \phi_e + \beta \left( \frac{19688}{45} e^2 +	\frac{28256}{45} 
	e^4 + \frac{263}{5} e^6, \frac{1448}{5} e^2 + \frac{1618}{3} e^4 + \frac{167}{3} e^6 \right) 
		\,, \\
	b_4 =&\; \left( -\frac{952397}{945} + \frac{5937}{7} \nu + \frac{1504}{5} \nu^2 \right) e^2 + 
		\left( -\frac{3113989}{1260} - \frac{388419}{140} \nu + \frac{64433}{20} \nu^2 \right) e^4 
		+ \bigg(\frac{4656611}{1512} - \frac{13057267}{2520} \nu \nonumber\\
		&+ \frac{127411}{45} \nu^2 \bigg) e^6 + \left( \frac{420727}{1680} - \frac{362071}{1260} 
		\nu + \frac{1642}{9} \nu^2 \right) e^8 + \sqrt{1 - e^2}\bigg[ \left(\frac{2672}{3} - 
		\frac{5344}{15} \nu \right) e^2 + \left( 2321 - \frac{4642}{5} \nu \right) e^4 \nonumber\\
		&+ \left( \frac{565}{3} - \frac{226}{3} \nu \right) e^6\bigg] + \sigma \bigg[ \frac{2}{3} 
		-\frac{1961}{15} e^2 - \frac{2527}{12}e^4 - \frac{157}{8} e^6, - \frac{2}{3} + 
		\frac{5623}{15} e^2 + \frac{2393}{4} e^4 + \frac{447}{8} e^6, -\frac{5527}{30} e^2 - 
		\frac{10117}{30} e^4 \nonumber\\
		&- \frac{5507}{160} e^6, -\frac{4}{3} + \left( \frac{682}{5} - \frac{1876}{15} q \right) 
		e^2 + \left(\frac{1337}{6} - \frac{595}{3} q \right) e^4 + \left( \frac{83}{4} - 
		\frac{37}{2} q \right) e^6,	\frac{4}{3} + \left( - \frac{5618}{15} + \frac{1876}{5} q 
		\right) e^2 \nonumber\\
		&+ \left(- \frac{1203}{2} + 595 q \right) e^4 + \left( - \frac{225}{4} + \frac{111}{4} q 
		\right) e^6 , \left( \frac{2764}{15} - \frac{921}{5} q \right) e^2 + \left(\frac{1687}{5} - 
		\frac{5056}{15} q \right) e^4 + \left( \frac{551}{16} - \frac{172}{5} q \right) e^6 \bigg] 
		\,, \\
	b_5 =&\; \pi\left( -\frac{55691}{105} \psi_e - \frac{610144}{315} \nu \zeta_e \right) e^2 \,, \\
	b_6 =&\; \left[ \frac{61655211971}{4365900} - \frac{2633056}{1575} \gamma_E + \frac{24608}{45} 
		\pi^2 + \left( \frac{43386337}{56700} + \frac{1017}{5} \pi^2 \right) \nu - 
		\frac{4148897}{1260} \nu^2 - \frac{61001}{243} \nu^3 \right] e^2 \nonumber\\
		&+ \left[\frac{64020009407}{21829500} - \frac{9525568}{1575} \gamma_E + \frac{89024}{45} 
		\pi^2 + \left(\frac{770214901}{12600} - \frac{15727}{96} \pi^2 \right) \nu - 
		\frac{80915371}{7560} \nu^2 - \frac{86910509}{9720} \nu^3 \right] e^4 \nonumber\\
		&+ \left[-\frac{1167012417073}{58212000} - \frac{4588588}{1575} \gamma_E + \frac{42884}{45} 
		\pi^2 + \left( \frac{8799500893}{453600} - \frac{295559}{960} \pi^2 \right) \nu + 
		\frac{351962207}{10080} \nu^2 - \frac{2223241}{90} \nu^3 \right] e^6 \nonumber\\
		&+ \left[\frac{120660628321}{12936000} - \frac{20437}{175} \gamma_E + \frac{191}{5} \pi^2 + 
		\left(- \frac{91818931}{5040} - \frac{6519}{320} \pi^2 \right) \nu + 		
		\frac{2495471}{126}\nu^2 - \frac{11792069}{1215} \nu^3 \right] e^8 \nonumber\\
		&+ \bigg(\frac{302322169}{887040} - \frac{1921387}{5040} \nu + \frac{41179}{108} \nu^2 - 
		\frac{386792}{1215} \nu^3 \bigg) e^{10} + \sqrt{1 - e^2} \bigg\{\bigg[ 
		-\frac{22713049}{7875} + \left( -\frac{11053982}{945} + \frac{8323}{90} \pi^2 \right) \nu 
		\nonumber\\
		&+\frac{108664}{45} \nu^2 \bigg] e^2 +\left[\frac{178791374}{7875} + \left( 
		-\frac{38295557}{630} + \frac{94177}{480} \pi^2 \right) \nu + \frac{681989}{45} \nu^2 
		\right] e^4 + \bigg[\frac{5321445613}{189000} \nonumber\\
		&+ \left( -\frac{26478311}{756} + \frac{2501}{1440} \pi^2 \right) \nu + \frac{450212}{45} 
		\nu^2 \bigg] e^6 + \left[\frac{186961}{168} - \frac{289691}{252}\nu + \frac{3197}{9} \nu^2 
		\right] e^8 \bigg\} + \frac{1460336}{23625} \left( 1 - \sqrt{1 - e^2} \right) \nonumber\\ 
		&+ \frac{428}{1575} e^2 \kappa_e + \left( \frac{2633056}{1575} e^2 + \frac{9525568}{1575} 
		e^4 + \frac{4588588}{1575} e^6 + \frac{20437}{175} e^8 \right) \log\left[ \frac{1 + \sqrt{1 
		- e^2}}{8 y \left(1 - e^2 \right)^{3/2}} \right] \,,
\end{align}
\end{subequations}
with the tail terms given, in terms of the functions found in~\cite{arun-2008,arun-2009}, by
\begin{subequations}
\begin{align}
	\phi_y &= \left(1 - e^2 \right)^{7/2} \tilde{\phi} \,,\\
	\phi_e &= \frac{192 \left(1 - e^2 \right)^{9/2}}{985 e^2} \left( \sqrt{1- e^2} \phi - 
		\tilde{\phi} \right) \,, \\
	\psi_y &= \left( 1 - e^2 \right)^{9/2} \left( -\frac{8064}{4159} \sqrt{1 - e^2} \phi + 
		\frac{4032}{4159} \tilde{\phi} + \frac{8191}{4159} \tilde{\psi} \right) \,, \\
	\zeta_y &= \left(1 - e^2 \right)^{7/2} \left[ \frac{160 \left(1 - e^2 \right)^{3/2}}{567} \phi 
		+ \left( -\frac{176}{567} + \frac{80}{567} e^2 \right) \tilde{\phi} + \frac{583 \left(1 - 
		e^2 \right)}{567} \tilde{\zeta} \right] \,, \\
	\psi_e &= \frac{16382 \left(1 - e^2\right)^{9/2}}{55691 e^2} \left[ \left( \frac{9408}{8191} - 
		\frac{14784}{8191} e^2 \right) \sqrt{1 - e^2} \phi+\left(-\frac{9408}{8191} + 
		\frac{4032}{8191} e^2 \right) \tilde{\phi} + \left(1 - e^2 \right) \left( \sqrt{1 - e^2} 
		\psi - \tilde{\psi} \right)\right] \,, \\
	\zeta_e &= \frac{12243 \left(1 - e^2\right)^{9/2}}{76268 e^2} \left[-\frac{16 \left(1 - e^2 
		\right)^{3/2}}{53} \phi	+ \left( \frac{16}{53} - \frac{80}{583} e^2 \right) \tilde{\phi} 
		+ \left(1 - e^2 \right) \left( \sqrt{1 - e^2} \zeta - \tilde{\zeta} \right)\right] \,, \\
	\kappa_y &= -\frac{ 934088 \left(1 - e^2 \right)^5}{33705} \left( \tilde{\kappa} - \tilde{F} 
		\right)	\,, \\
	\kappa_e &= -\frac{5604528 \left(1 - e^2 \right)^6}{3745 e^2} \left[ \sqrt{1 - e^2} \left( 
		\kappa - F \right) - \left(\tilde{\kappa} - \tilde{F} \right) \right] \,.
\end{align}
\end{subequations}
We chose to only include in the 3PN enhancement functions $\kappa_i$ the terms proportional to 
$\log n$, as the other ones are in finite number and can be combined with nontail terms. Using the 
formalism developed in~\cite{arun-2008,arun-2009}, we give them here at tenth order in the 
eccentricity:
\begin{subequations}
\begin{align}
	\phi_y =&\; 1 + \frac{97}{32} e^2 + \frac{49}{128} e^4 - \frac{49}{18432} e^6 - 
		\frac{109}{147456} e^8 - \frac{2567}{58982400} e^{10} + \ord{e^{12}} \,, \\
	\phi_e =&\; 1 + \frac{5969}{3940} e^2 + \frac{24217}{189120} e^4 + \frac{623}{4538880} e^6 - 
		\frac{96811}{363110400} e^8 - \frac{5971}{4357324800} e^{10} + \ord{e^{12}} \,, \\
	\psi_y =&\; 1 - \frac{207671}{8318} e^2 - \frac{8382869}{266176} e^4 - \frac{8437609}{4791168} 
		e^6 + \frac{10075915}{306634752} e^8 - \frac{38077159}{15331737600} e^{10} + \ord{e^{12}} 
		\,, \\
	\zeta_y =&\; 1 + \frac{113002}{11907} e^2 + \frac{6035543}{762048} e^4 + \frac{253177}{571536} 
		e^6 - \frac{850489}{877879296} e^8 - \frac{1888651}{10973491200} e^{10} + \ord{e^{12}} \,,\\
	\psi_e =&\; 1 - \frac{9904271}{891056} e^2 - \frac{101704075}{10692672} e^4 - 
		\frac{217413779}{513248256} e^6 + \frac{35703577}{6843310080} e^8 - 
		\frac{3311197679}{9854366515200} e^{10} + \ord{e^{12}} \,, \\
	\zeta_e =&\; 1 + \frac{11228233}{2440576} e^2 + \frac{37095275}{14643456} e^4 + 
		\frac{151238443}{1405771776} e^6 - \frac{118111}{611205120} e^8 - 		
		\frac{407523451}{26990818099300} e^{10} + \ord{e^{12}} \,, \\
	\kappa_y =&\; 244 \log 2 \left( e^2 - \frac{18881}{1098} e^4 + \frac{6159821}{39528} e^6 - 
		\frac{16811095}{19764} e^8 + \frac{446132351}{123525} e^{10} \right) - 243 \log 3 \bigg(e^2 
		- \frac{39}{4} e^4 + \frac{2735}{64}e^6 \nonumber\\
		&+ \frac{25959}{512} e^8 - \frac{638032239}{409600} e^{10}\bigg) - \frac{48828125 \log 
		5}{5184} \left( e^6 - \frac{83}{8} e^8 + \frac{12637}{256} e^{10} \right) - 
		\frac{4747561509943 \log 7}{33177600} e^{10} + \ord{e^{12}} \,, \\
	\kappa_e =&\; 6536 \log 2 \left(1 - \frac{22314}{817} e^2 + \frac{7170067}{19608}e^4 - 
		\frac{10943033}{4128} e^6 + \frac{230370959}{15480} e^8 - \frac{866124466133}{8823600} 
		e^{10} \right) \nonumber\\
		&- 6561 \log 3 \left( 1 - \frac{49}{4} e^2 + \frac{4369}{64} e^4 + \frac{214449}{512} e^6 - 
		\frac{623830739}{81920} e^8 + \frac{76513915569}{1638400} e^{10} \right) \nonumber\\
		&- \frac{48828125 \log 5}{64} \left( e^4 - \frac{293}{24} e^6 + \frac{159007}{2304} e^8 - 
		\frac{6631171}{27648} e^{10} \right) - \frac{4747561509943 \log 7}{245760} \left( e^8 - 
		\frac{259}{20} e^{10} \right) + \ord{e^{12}} \,.
\end{align}
\end{subequations}

It can be noted that those enhancement functions converge much more quickly than the ones presented 
in~\cite{arun-2008,arun-2009}. Indeed, because of the inclusion of factors of $\sqrt{1 - e^2}$ in 
them, the enhancement functions seem to converge in the parabolic limit $e \to 1$. We believe it to 
be related to the fact that the PN parameter $y$ we used here is related to the Newtonian orbital 
angular momentum and thus is finite and nonzero in this limit. In contrast, the PN parameter 
$(M\omega)^{1/3}$ is related to the energy and thus vanishes in this limit. In that case, in order 
for the tail effects to stay nonzero, the enhancement functions are forced to diverge.

\section{True and eccentric anomaly expansion}\label{sec:qK-inv}

The Fourier coefficients of the eccentric anomaly, true anomaly and orbital phase are given to 
order $\ord{y^4, e^5}$ by
\begin{subequations}
\begin{align}
	A_1 =&\; e-\frac{e^3}{8}+\frac{e^5}{192}+y^4 \left(e^3 \left(\frac{105}{8}-\frac{51\eta}{64} 
		-\frac{19 \eta ^2}{64}\right)+e \left(-\frac{15}{2}+\frac{9 \eta }{8} + 		
		\frac{\eta^2}{8}\right)+ e^5 \left(-\frac{735}{128}-\frac{489 \eta }{512}+\frac{111 
		\eta^2}{512} \right)\right) \,, \nonumber \\
	A_2 =&\; \frac{e^2}{2}-\frac{e^4}{6}+y^4 \left(e^4 \left(\frac{75}{4}-\frac{5 \eta}{32} 
		-\frac{47 \eta ^2}{96}\right)+e^2 \left(-\frac{75}{8}+\frac{15 \eta }{16}+\frac{3 
		\eta^2}{16} \right)\right) \,, \nonumber \\
	A_3 =&\; \frac{3 e^3}{8}-\frac{27 e^5}{128}+y^4 \left(e^5 \left(\frac{6825}{256}+\frac{705 \eta 
		}{1024} -\frac{775 \eta ^2}{1024}\right)+e^3 \left(-\frac{95}{8}+\frac{49 \eta}{64} 
		+\frac{17 \eta ^2}{64}\right)\right) \,, \nonumber \\
	A_4 =&\; \frac{e^4}{3}+e^4 y^4 \left(-\frac{975}{64}+\frac{35 \eta }{64}+\frac{71 \eta^2}{192} 
		\right) \,, \nonumber \\
	A_5 =&\; \frac{125 e^5}{384}+e^5 y^4 \left(-\frac{5049}{256}+\frac{1167 \eta }{5120}+\frac{523 
		\eta ^2}{1024} \right) \,, \\
	B_1 =&\; 2 e-\frac{e^3}{4}+\frac{5 e^5}{96}+y^2 \left(e (4-\eta )+e^5 \left( \frac{17}{48} - 
		\frac{17 \eta }{192}\right)+e^3 \left(-\frac{7}{2}+\frac{7 \eta}{8}\right)\right) 		
		\nonumber\\ 
		&+y^4 \left(e^3 \left(-\frac{39}{4}+\frac{197 \eta }{32}-\frac{7 \eta ^2}{32}\right)+e 
		\left(13-\frac{31 \eta }{12}+\frac{\eta ^2}{12}\right)+e^5 \left(-\frac{401}{192}-\frac{569 
		\eta }{144}+\frac{107 \eta ^2}{576}\right)\right) \,, \nonumber\\
	B_2 =&\; \frac{5 e^2}{4}-\frac{11 e^4}{24}+y^2 \left(e^2 (4-\eta )+e^4 	\left( -\frac{14}{3} + 
		\frac{7 \eta }{6}\right)\right) \nonumber\\ 
		&+ y^4 \left(e^4 \left(-\frac{217}{24}+\frac{3187 \eta }{288}-\frac{313 \eta^2}{288} 
		\right) + e^2 \left(\frac{91}{8}-\frac{101 \eta }{24}+\frac{11 \eta ^2}{24}\right)\right) 
		\,, \nonumber\\
	B_3 =&\; \frac{13 e^3}{12}-\frac{43 e^5}{64}+y^2 \left(e^3 \left(\frac{9}{2}-\frac{9 \eta}{8} 
		\right) + e^5 \left(-\frac{207}{32}+\frac{207 \eta }{128}\right)\right) \nonumber\\ 
		&+y^4 \left(e^5 \left(-\frac{1011}{128}+\frac{9471 \eta }{512}-\frac{1161 \eta^2}{512} 
		\right) + e^3 \left(\frac{43}{4}-\frac{51 \eta }{8}+\frac{7 \eta ^2}{8}\right)\right) \,, 
		\nonumber\\
	B_4 =&\; \frac{103 e^4}{96}+e^4 y^2 \left(\frac{16}{3}-\frac{4 \eta }{3}\right)+e^4 y^4 \left( 
		\frac{1969}{192}-\frac{2711 \eta }{288}+\frac{205 \eta ^2}{144}\right) \,, \nonumber\\
	B_5 =&\; \frac{1097 e^5}{960}+e^5 y^2 \left(\frac{625}{96}-\frac{625 \eta }{384}\right)+e^5 y^4 
		\left(\frac{3641}{384}-\frac{315821 \eta }{23040}+\frac{10039 \eta ^2}{4608}\right) \,,\\
	C_1 =&\; 2 e-\frac{e^3}{4}+\frac{5 e^5}{96}+y^2 \left(e (10-\eta)+e^5 \left( \frac{49}{96} - 
		\frac{17 \eta }{192}\right)+e^3 \left(-\frac{17}{4}+\frac{7 \eta}{8}\right)\right) 
		\nonumber\\
		&+ y^4 \left(e^5 \left(-\frac{317}{96}+\frac{2497 \eta}{2304}-\frac{1579 \eta^2}{2304} 
		\right) + e \left(52-\frac{235 \eta }{12}+\frac{\eta^2}{12}\right)+e^3 \left( \frac{13}{8} 
		- \frac{231 \eta }{32}+\frac{17 \eta^2}{32}\right)\right) \nonumber\\
	C_2 =&\; \frac{5 e^2}{4}-\frac{11 e^4}{24}+y^2 \left(e^2 \left(\frac{31}{4}-\eta \right)+e^4 
		\left(-\frac{145}{24}+\frac{7 \eta }{6}\right)\right) \nonumber\\
		&+y^4 \left(e^2 \left(\frac{323}{8}-\frac{163 \eta }{12}+\frac{\eta ^2}{12}\right)+e^4 
		\left(-\frac{331}{24}+\frac{\eta }{18}+\frac{25 \eta ^2}{36}\right)\right) \nonumber\\
	C_3 =&\; \frac{13 e^3}{12}-\frac{43 e^5}{64}+y^2 \left(e^3 \left(\frac{31}{4}-\frac{9 \eta}{8} 
		\right)+e^5 \left(-\frac{543}{64}+\frac{207 \eta }{128}\right)\right) \nonumber\\
		&+y^4 \left(e^3 \left(\frac{313}{8}-\frac{1205 \eta }{96}+\frac{\eta ^2}{32}\right)+e^5 
		\left(-\frac{749}{32}+\frac{7217 \eta }{1536}+\frac{567 \eta ^2}{512}\right)\right) 
		\nonumber\\
	C_4 =&\; \frac{103 e^4}{96}+e^4 y^2 \left(\frac{821}{96}-\frac{4 \eta }{3}\right)+e^4 y^4 
		\left(\frac{1975}{48}-\frac{118 \eta }{9}-\frac{11 \eta ^2}{144}\right) \nonumber\\
	C_5 =&\; \frac{1097 e^5}{960}+e^5 y^2 \left(\frac{9541}{960}-\frac{625 \eta }{384}\right)+e^5 
		y^4 \left(\frac{10813}{240}-\frac{338987 \eta }{23040}-\frac{1211 \eta ^2}{4608}\right) \,.
\end{align}
\end{subequations}

\section{Waveform amplitudes expansion}\label{sec:Gmn}

The amplitudes $G_{+,\times}^{(m,n)}$ in Eq.~\eqref{eq:waveformcoeff} are given to order 
$\ord{y^2, e}$ for $n < 0$, with $C_i = \cos i = \uvec{L} \cdot \uvec{N}$ and $S_i =\sin i$, by
\begin{subequations}
\begin{align}
	G_+^{(1,-1)} &= e \left(\frac{1}{2}-\frac{C_{\text{i}}^2}{2}\right)+e y^2 
		\left(-\frac{35}{16}+\frac{9 C_{\text{i}}^2}{4}-\frac{C_{\text{i}}^4}{16}+\eta 
		\left(-\frac{5}{48}-\frac{C_{\text{i}}^2}{12}+\frac{3 C_{\text{i}}^4}{16}\right)\right) \,,
		\nonumber\\ 
	G_+^{(1,-2)} &= e y \delta \left(-\frac{3 S_{\text{i}}}{2}+\frac{1}{2} C_{\text{i}}^2 
		S_{\text{i}}\right) \,,\nonumber\\ 
	G_+^{(1,-3)} &= e \left(-\frac{9}{4}-\frac{9 C_{\text{i}}^2}{4}\right)+e y^2 
		\left(-\frac{9}{8}-\frac{63 C_{\text{i}}^2}{8}+\eta \left(-\frac{75}{8}+\frac{87 
		C_{\text{i}}^2}{8}\right)\right) \,,\nonumber\\ 
	G_+^{(1,-4)} &= e y \delta \left(4 S_{\text{i}}+4 C_{\text{i}}^2 S_{\text{i}}\right) \,,
		\nonumber\\ 
	G_+^{(1,-5)} &= e y^2 \left(-\frac{625 S_{\text{i}}^2}{96}-\frac{625}{96} C_{\text{i}}^2 
		S_{\text{i}}^2+\eta \left(\frac{625 S_{\text{i}}^2}{32}+\frac{625}{32} C_{\text{i}}^2 
		S_{\text{i}}^2\right)\right) \,,\nonumber\\ 
	G_+^{(0,-1)} &= y \delta \left(-\frac{5 S_{\text{i}}}{8}-\frac{1}{8} C_{\text{i}}^2 
		S_{\text{i}}\right) \,,\nonumber\\ 
	G_+^{(0,-2)} &= -1-C_{\text{i}}^2+y^2 \left(\frac{19}{6}+\frac{3 
		C_{\text{i}}^2}{2}-\frac{C_{\text{i}}^4}{3}+\eta \left(-\frac{19}{6}+\frac{11 
		C_{\text{i}}^2}{6}+C_{\text{i}}^4\right)\right) \,,\nonumber\\ 
	G_+^{(0,-3)} &= y \delta \left(\frac{9 S_{\text{i}}}{8}+\frac{9}{8} C_{\text{i}}^2 
		S_{\text{i}}\right) \,,\nonumber\\ 
	G_+^{(0,-4)} &= y^2 \left(-\frac{4 S_{\text{i}}^2}{3}-\frac{4}{3} C_{\text{i}}^2 
		S_{\text{i}}^2+\eta \left(4 S_{\text{i}}^2+4 C_{\text{i}}^2 S_{\text{i}}^2\right)\right) 
		\,,\nonumber\\ 
	G_+^{(-1,-1)} &= e \left(\frac{3}{4}+\frac{3 C_{\text{i}}^2}{4}\right)+e y^2 
		\left(\frac{143}{24}+\frac{53 C_{\text{i}}^2}{8}+\frac{C_{\text{i}}^4}{12}+\eta 
		\left(\frac{7}{8}-\frac{9 C_{\text{i}}^2}{8}-\frac{C_{\text{i}}^4}{4}\right)\right) 
		\,,\nonumber\\ 
	G_+^{(-1,-2)} &= e y \delta \left(-\frac{3 S_{\text{i}}}{2}-\frac{3}{2} C_{\text{i}}^2 
		S_{\text{i}}\right) \,,\nonumber\\ 
	G_+^{(-1,-3)} &= e y^2 \left(\frac{81 S_{\text{i}}^2}{32}+\frac{81}{32} C_{\text{i}}^2 
		S_{\text{i}}^2+\eta \left(-\frac{243 S_{\text{i}}^2}{32}-\frac{243}{32} C_{\text{i}}^2 
		S_{\text{i}}^2\right)\right) \,,\\ 
	G_{\times}^{(1,-1)} &= i e y^2 \left(-\frac{1}{4} C_{\text{i}} S_{\text{i}}^2+\frac{3}{4} 
		\eta C_{\text{i}} S_{\text{i}}^2\right) \,,\nonumber\\ 
	G_{\times}^{(1,-2)} &= i e y \delta C_{\text{i}} S_{\text{i}} \,,\nonumber\\ 
	G_{\times}^{(1,-3)} &= i \left(\frac{9 e C_{\text{i}}}{2}+e y^2 \left(9 
		C_{\text{i}}-\frac{27}{8} C_{\text{i}} S_{\text{i}}^2+\eta \left(-\frac{3 
		C_{\text{i}}}{2}+\frac{81}{8} C_{\text{i}} S_{\text{i}}^2\right)\right)\right) 
		\,,\nonumber\\ 
	G_{\times}^{(1,-4)} &= -8 i e y \delta C_{\text{i}} S_{\text{i}} \,,\nonumber\\ 
	G_{\times}^{(1,-5)} &= i e y^2 \left(\frac{625}{48} C_{\text{i}} S_{\text{i}}^2-\frac{625}{16} 
		\eta C_{\text{i}} S_{\text{i}}^2\right) \,,\nonumber\\ 
	G_{\times}^{(0,-1)} &= \frac{3}{4} i y \delta C_{\text{i}} S_{\text{i}} \,,\nonumber\\ 
	G_{\times}^{(0,-2)} &= i \left(2 C_{\text{i}}+y^2 \left(-\frac{13 C_{\text{i}}}{3}-\frac{4}{3} 
		C_{\text{i}} S_{\text{i}}^2+\eta \left(\frac{C_{\text{i}}}{3}+4 C_{\text{i}} 
		S_{\text{i}}^2\right)\right)\right) \,,\nonumber\\ 
	G_{\times}^{(0,-3)} &= -\frac{9}{4} i y \delta C_{\text{i}} S_{\text{i}} \,,\nonumber\\ 
	G_{\times}^{(0,-4)} &= i y^2 \left(\frac{8}{3} C_{\text{i}} S_{\text{i}}^2-8 \eta C_{\text{i}} 
		S_{\text{i}}^2\right) \,,\nonumber\\ 
	G_{\times}^{(-1,-1)} &= i \left(-\frac{3 e C_{\text{i}}}{2}+e y^2 \left(-\frac{38 
		C_{\text{i}}}{3}+\frac{11}{24} C_{\text{i}} S_{\text{i}}^2+\eta 
		\left(\frac{C_{\text{i}}}{2}-\frac{11}{8} C_{\text{i}} S_{\text{i}}^2\right)\right)\right) 
		\,,\nonumber\\ 
	G_{\times}^{(-1,-2)} &= 3 i e y \delta C_{\text{i}} S_{\text{i}} \,,\nonumber\\ 
	G_{\times}^{(-1,-3)} &= i e y^2 \left(-\frac{81}{16} C_{\text{i}} S_{\text{i}}^2+\frac{243}{16} 
		\eta C_{\text{i}} S_{\text{i}}^2\right) \,.
\end{align}
\end{subequations}
Note that $G_{+,\times}^{(m,-n)} = \bar{G}_{+,\times}^{(-m,n)}$.

\end{widetext}

\bibliography{ewf-paper}

\begin{thebibliography}{66}%
\makeatletter
\providecommand \@ifxundefined [1]{%
 \@ifx{#1\undefined}
}%
\providecommand \@ifnum [1]{%
 \ifnum #1\expandafter \@firstoftwo
 \else \expandafter \@secondoftwo
 \fi
}%
\providecommand \@ifx [1]{%
 \ifx #1\expandafter \@firstoftwo
 \else \expandafter \@secondoftwo
 \fi
}%
\providecommand \natexlab [1]{#1}%
\providecommand \enquote  [1]{``#1''}%
\providecommand \bibnamefont  [1]{#1}%
\providecommand \bibfnamefont [1]{#1}%
\providecommand \citenamefont [1]{#1}%
\providecommand \href@noop [0]{\@secondoftwo}%
\providecommand \href [0]{\begingroup \@sanitize@url \@href}%
\providecommand \@href[1]{\@@startlink{#1}\@@href}%
\providecommand \@@href[1]{\endgroup#1\@@endlink}%
\providecommand \@sanitize@url [0]{\catcode `\\12\catcode `\$12\catcode
  `\&12\catcode `\#12\catcode `\^12\catcode `\_12\catcode `\%12\relax}%
\providecommand \@@startlink[1]{}%
\providecommand \@@endlink[0]{}%
\providecommand \url  [0]{\begingroup\@sanitize@url \@url }%
\providecommand \@url [1]{\endgroup\@href {#1}{\urlprefix }}%
\providecommand \urlprefix  [0]{URL }%
\providecommand \Eprint [0]{\href }%
\providecommand \doibase [0]{http://dx.doi.org/}%
\providecommand \selectlanguage [0]{\@gobble}%
\providecommand \bibinfo  [0]{\@secondoftwo}%
\providecommand \bibfield  [0]{\@secondoftwo}%
\providecommand \translation [1]{[#1]}%
\providecommand \BibitemOpen [0]{}%
\providecommand \bibitemStop [0]{}%
\providecommand \bibitemNoStop [0]{.\EOS\space}%
\providecommand \EOS [0]{\spacefactor3000\relax}%
\providecommand \BibitemShut  [1]{\csname bibitem#1\endcsname}%
\let\auto@bib@innerbib\@empty
\bibitem [{\citenamefont {{Aasi}}\ \emph {et~al.}(2015)\citenamefont {{Aasi}}
  \emph {et~al.}}]{LIGO}%
  \BibitemOpen
  \bibfield  {author} {\bibinfo {author} {\bibfnamefont {J.}~\bibnamefont
  {{Aasi}}} \emph {et~al.},\ }\href {\doibase 10.1088/0264-9381/32/11/115012}
  {\bibfield  {journal} {\bibinfo  {journal} {Classical and Quantum Gravity}\
  }\textbf {\bibinfo {volume} {32}},\ \bibinfo {eid} {115012} (\bibinfo {year}
  {2015})},\ \Eprint {http://arxiv.org/abs/1410.7764} {arXiv:1410.7764 [gr-qc]}
  \BibitemShut {NoStop}%
\bibitem [{\citenamefont {{Accadia}}\ \emph {et~al.}(2012)\citenamefont
  {{Accadia}} \emph {et~al.}}]{Virgo}%
  \BibitemOpen
  \bibfield  {author} {\bibinfo {author} {\bibfnamefont {T.}~\bibnamefont
  {{Accadia}}} \emph {et~al.},\ }\href {\doibase 10.1088/1748-0221/7/03/P03012}
  {\bibfield  {journal} {\bibinfo  {journal} {Journal of Instrumentation}\
  }\textbf {\bibinfo {volume} {7}},\ \bibinfo {pages} {3012} (\bibinfo {year}
  {2012})}\BibitemShut {NoStop}%
\bibitem [{\citenamefont {{Grote}}\ and\ \citenamefont {{LIGO Scientific
  Collaboration}}(2010)}]{GEO600}%
  \BibitemOpen
  \bibfield  {author} {\bibinfo {author} {\bibfnamefont {H.}~\bibnamefont
  {{Grote}}}\ and\ \bibinfo {author} {\bibnamefont {{LIGO Scientific
  Collaboration}}},\ }\href {\doibase 10.1088/0264-9381/27/8/084003} {\bibfield
   {journal} {\bibinfo  {journal} {Classical and Quantum Gravity}\ }\textbf
  {\bibinfo {volume} {27}},\ \bibinfo {eid} {084003} (\bibinfo {year}
  {2010})}\BibitemShut {NoStop}%
\bibitem [{\citenamefont {Abbott}\ \emph
  {et~al.}(2016{\natexlab{a}})\citenamefont {Abbott} \emph
  {et~al.}}]{GW150914}%
  \BibitemOpen
  \bibfield  {author} {\bibinfo {author} {\bibfnamefont {B.~P.}\ \bibnamefont
  {Abbott}} \emph {et~al.},\ }\href {\doibase 10.1103/PhysRevLett.116.061102}
  {\bibfield  {journal} {\bibinfo  {journal} {Phys. Rev. Lett.}\ }\textbf
  {\bibinfo {volume} {116}},\ \bibinfo {pages} {061102} (\bibinfo {year}
  {2016}{\natexlab{a}})},\ \Eprint {http://arxiv.org/abs/1602.03837}
  {arXiv:1602.03837 [gr-qc]} \BibitemShut {NoStop}%
\bibitem [{\citenamefont {Abbott}\ \emph
  {et~al.}(2016{\natexlab{b}})\citenamefont {Abbott} \emph
  {et~al.}}]{GW151226}%
  \BibitemOpen
  \bibfield  {author} {\bibinfo {author} {\bibfnamefont {B.~P.}\ \bibnamefont
  {Abbott}} \emph {et~al.},\ }\href {\doibase 10.1103/PhysRevLett.116.241103}
  {\bibfield  {journal} {\bibinfo  {journal} {Phys. Rev. Lett.}\ }\textbf
  {\bibinfo {volume} {116}},\ \bibinfo {pages} {241103} (\bibinfo {year}
  {2016}{\natexlab{b}})},\ \Eprint {http://arxiv.org/abs/1606.04855}
  {arXiv:1606.04855 [gr-qc]} \BibitemShut {NoStop}%
\bibitem [{\citenamefont {Abbott}\ \emph
  {et~al.}(2017{\natexlab{a}})\citenamefont {Abbott} \emph
  {et~al.}}]{GW170104}%
  \BibitemOpen
  \bibfield  {author} {\bibinfo {author} {\bibfnamefont {B.~P.}\ \bibnamefont
  {Abbott}} \emph {et~al.},\ }\href {\doibase 10.1103/PhysRevLett.118.221101}
  {\bibfield  {journal} {\bibinfo  {journal} {Phys. Rev. Lett.}\ }\textbf
  {\bibinfo {volume} {118}},\ \bibinfo {pages} {221101} (\bibinfo {year}
  {2017}{\natexlab{a}})},\ \Eprint {http://arxiv.org/abs/1706.01812}
  {arXiv:1706.01812 [gr-qc]} \BibitemShut {NoStop}%
\bibitem [{\citenamefont {Abbott}\ \emph
  {et~al.}(2017{\natexlab{b}})\citenamefont {Abbott} \emph
  {et~al.}}]{GW170814}%
  \BibitemOpen
  \bibfield  {author} {\bibinfo {author} {\bibfnamefont {B.~P.}\ \bibnamefont
  {Abbott}} \emph {et~al.},\ }\href {\doibase 10.1103/PhysRevLett.119.141101}
  {\bibfield  {journal} {\bibinfo  {journal} {Phys. Rev. Lett.}\ }\textbf
  {\bibinfo {volume} {119}},\ \bibinfo {pages} {141101} (\bibinfo {year}
  {2017}{\natexlab{b}})},\ \Eprint {http://arxiv.org/abs/1709.09660}
  {arXiv:1709.09660 [gr-qc]} \BibitemShut {NoStop}%
\bibitem [{\citenamefont {Abbott}\ \emph
  {et~al.}(2017{\natexlab{c}})\citenamefont {Abbott} \emph
  {et~al.}}]{GW170817}%
  \BibitemOpen
  \bibfield  {author} {\bibinfo {author} {\bibfnamefont {B.~P.}\ \bibnamefont
  {Abbott}} \emph {et~al.},\ }\href {\doibase 10.1103/PhysRevLett.119.161101}
  {\bibfield  {journal} {\bibinfo  {journal} {Phys. Rev. Lett.}\ }\textbf
  {\bibinfo {volume} {119}},\ \bibinfo {pages} {161101} (\bibinfo {year}
  {2017}{\natexlab{c}})},\ \Eprint {http://arxiv.org/abs/1710.05832}
  {arXiv:1710.05832 [gr-qc]} \BibitemShut {NoStop}%
\bibitem [{\citenamefont {Abbott}\ \emph
  {et~al.}(2017{\natexlab{d}})\citenamefont {Abbott} \emph
  {et~al.}}]{GW170608}%
  \BibitemOpen
  \bibfield  {author} {\bibinfo {author} {\bibfnamefont {B.~P.}\ \bibnamefont
  {Abbott}} \emph {et~al.},\ }\href {\doibase 10.3847/2041-8213/aa9f0c}
  {\bibfield  {journal} {\bibinfo  {journal} {The Astrophysical Journal
  Letters}\ }\textbf {\bibinfo {volume} {851}},\ \bibinfo {eid} {L35} (\bibinfo
  {year} {2017}{\natexlab{d}})},\ \Eprint {http://arxiv.org/abs/1711.05578}
  {arXiv:1711.05578 [astro-ph.HE]} \BibitemShut {NoStop}%
\bibitem [{\citenamefont {Peters}(1964)}]{peters-1964}%
  \BibitemOpen
  \bibfield  {author} {\bibinfo {author} {\bibfnamefont {P.~C.}\ \bibnamefont
  {Peters}},\ }\href {\doibase 10.1103/PhysRev.136.B1224} {\bibfield  {journal}
  {\bibinfo  {journal} {Phys. Rev.}\ }\textbf {\bibinfo {volume} {136}},\
  \bibinfo {pages} {B1224} (\bibinfo {year} {1964})}\BibitemShut {NoStop}%
\bibitem [{\citenamefont {Postnov}\ and\ \citenamefont
  {Yungelson}(2014)}]{postnov-lrr}%
  \BibitemOpen
  \bibfield  {author} {\bibinfo {author} {\bibfnamefont {K.~A.}\ \bibnamefont
  {Postnov}}\ and\ \bibinfo {author} {\bibfnamefont {L.~R.}\ \bibnamefont
  {Yungelson}},\ }\href {\doibase 10.12942/lrr-2014-3} {\bibfield  {journal}
  {\bibinfo  {journal} {Living Rev. Relativ.}\ }\textbf {\bibinfo {volume}
  {17}},\ \bibinfo {pages} {3} (\bibinfo {year} {2014})},\ \Eprint
  {http://arxiv.org/abs/1403.4754} {arXiv:1403.4754 [astro-ph]} \BibitemShut
  {NoStop}%
\bibitem [{\citenamefont {Shappee}\ and\ \citenamefont
  {Thompson}(2013)}]{shappee-2012}%
  \BibitemOpen
  \bibfield  {author} {\bibinfo {author} {\bibfnamefont {B.~J.}\ \bibnamefont
  {Shappee}}\ and\ \bibinfo {author} {\bibfnamefont {T.~A.}\ \bibnamefont
  {Thompson}},\ }\href {\doibase 10.1088/0004-637X/766/1/64} {\bibfield
  {journal} {\bibinfo  {journal} {Astrophys. J.}\ }\textbf {\bibinfo {volume}
  {766}},\ \bibinfo {pages} {64} (\bibinfo {year} {2013})},\ \Eprint
  {http://arxiv.org/abs/1204.1053} {arXiv:1204.1053 [astro-ph]} \BibitemShut
  {NoStop}%
\bibitem [{\citenamefont {Antonini}\ \emph {et~al.}(2016)\citenamefont
  {Antonini}, \citenamefont {Chatterjee}, \citenamefont {Rodriguez},
  \citenamefont {Morscher}, \citenamefont {Pattabiraman}, \citenamefont
  {Kalogera},\ and\ \citenamefont {Rasio}}]{antonini-2016}%
  \BibitemOpen
  \bibfield  {author} {\bibinfo {author} {\bibfnamefont {F.}~\bibnamefont
  {Antonini}}, \bibinfo {author} {\bibfnamefont {S.}~\bibnamefont
  {Chatterjee}}, \bibinfo {author} {\bibfnamefont {C.~L.}\ \bibnamefont
  {Rodriguez}}, \bibinfo {author} {\bibfnamefont {M.}~\bibnamefont {Morscher}},
  \bibinfo {author} {\bibfnamefont {B.}~\bibnamefont {Pattabiraman}}, \bibinfo
  {author} {\bibfnamefont {V.}~\bibnamefont {Kalogera}}, \ and\ \bibinfo
  {author} {\bibfnamefont {F.~A.}\ \bibnamefont {Rasio}},\ }\href {\doibase
  10.3847/0004-637X/816/2/65} {\bibfield  {journal} {\bibinfo  {journal}
  {Astrophys. J.}\ }\textbf {\bibinfo {volume} {816}},\ \bibinfo {pages} {65}
  (\bibinfo {year} {2016})},\ \Eprint {http://arxiv.org/abs/1509.05080}
  {arXiv:1509.05080 [astro-ph]} \BibitemShut {NoStop}%
\bibitem [{\citenamefont {Antonini}\ \emph {et~al.}(2017)\citenamefont
  {Antonini}, \citenamefont {Toonen},\ and\ \citenamefont
  {Hamers}}]{antonini-2017}%
  \BibitemOpen
  \bibfield  {author} {\bibinfo {author} {\bibfnamefont {F.}~\bibnamefont
  {Antonini}}, \bibinfo {author} {\bibfnamefont {S.}~\bibnamefont {Toonen}}, \
  and\ \bibinfo {author} {\bibfnamefont {A.~S.}\ \bibnamefont {Hamers}},\
  }\href {\doibase 10.3847/1538-4357/aa6f5e} {\bibfield  {journal} {\bibinfo
  {journal} {Astrophys. J.}\ }\textbf {\bibinfo {volume} {841}},\ \bibinfo
  {pages} {77} (\bibinfo {year} {2017})},\ \Eprint
  {http://arxiv.org/abs/1703.06614} {arXiv:1703.06614 [astro-ph.HE]}
  \BibitemShut {NoStop}%
\bibitem [{\citenamefont {Nishizawa}\ \emph {et~al.}(2016)\citenamefont
  {Nishizawa}, \citenamefont {Berti}, \citenamefont {Klein},\ and\
  \citenamefont {Sesana}}]{nishizawa-2016-1}%
  \BibitemOpen
  \bibfield  {author} {\bibinfo {author} {\bibfnamefont {A.}~\bibnamefont
  {Nishizawa}}, \bibinfo {author} {\bibfnamefont {E.}~\bibnamefont {Berti}},
  \bibinfo {author} {\bibfnamefont {A.}~\bibnamefont {Klein}}, \ and\ \bibinfo
  {author} {\bibfnamefont {A.}~\bibnamefont {Sesana}},\ }\href {\doibase
  10.1103/PhysRevD.94.064020} {\bibfield  {journal} {\bibinfo  {journal} {Phys.
  Rev. D}\ }\textbf {\bibinfo {volume} {94}},\ \bibinfo {pages} {064020}
  (\bibinfo {year} {2016})},\ \Eprint {http://arxiv.org/abs/1605.01341}
  {arXiv:1605.01341 [gr-qc]} \BibitemShut {NoStop}%
\bibitem [{\citenamefont {Nishizawa}\ \emph {et~al.}(2017)\citenamefont
  {Nishizawa}, \citenamefont {Sesana}, \citenamefont {Berti},\ and\
  \citenamefont {Klein}}]{nishizawa-2016-2}%
  \BibitemOpen
  \bibfield  {author} {\bibinfo {author} {\bibfnamefont {A.}~\bibnamefont
  {Nishizawa}}, \bibinfo {author} {\bibfnamefont {A.}~\bibnamefont {Sesana}},
  \bibinfo {author} {\bibfnamefont {E.}~\bibnamefont {Berti}}, \ and\ \bibinfo
  {author} {\bibfnamefont {A.}~\bibnamefont {Klein}},\ }\href {\doibase
  10.1093/mnras/stw2993} {\bibfield  {journal} {\bibinfo  {journal} {Mon. Not.
  R. Astron. Soc.}\ }\textbf {\bibinfo {volume} {465}},\ \bibinfo {pages}
  {4375} (\bibinfo {year} {2017})},\ \Eprint {http://arxiv.org/abs/1606.09295}
  {arXiv:1606.09295 [astro-ph]} \BibitemShut {NoStop}%
\bibitem [{\citenamefont {Breivik}\ \emph {et~al.}(2016)\citenamefont
  {Breivik}, \citenamefont {Rodriguez}, \citenamefont {Larson}, \citenamefont
  {Kalogera},\ and\ \citenamefont {Rasio}}]{breivik-2016}%
  \BibitemOpen
  \bibfield  {author} {\bibinfo {author} {\bibfnamefont {K.}~\bibnamefont
  {Breivik}}, \bibinfo {author} {\bibfnamefont {C.~L.}\ \bibnamefont
  {Rodriguez}}, \bibinfo {author} {\bibfnamefont {S.~L.}\ \bibnamefont
  {Larson}}, \bibinfo {author} {\bibfnamefont {V.}~\bibnamefont {Kalogera}}, \
  and\ \bibinfo {author} {\bibfnamefont {F.~A.}\ \bibnamefont {Rasio}},\ }\href
  {\doibase 10.3847/2041-8205/830/1/L18} {\bibfield  {journal} {\bibinfo
  {journal} {Astrophys. J.}\ }\textbf {\bibinfo {volume} {830}},\ \bibinfo
  {pages} {L18} (\bibinfo {year} {2016})},\ \Eprint
  {http://arxiv.org/abs/1606.09558} {arXiv:1606.09558 [astro-ph]} \BibitemShut
  {NoStop}%
\bibitem [{\citenamefont {Petrovich}\ and\ \citenamefont
  {Antonini}(2017)}]{petrovich-2017}%
  \BibitemOpen
  \bibfield  {author} {\bibinfo {author} {\bibfnamefont {C.}~\bibnamefont
  {Petrovich}}\ and\ \bibinfo {author} {\bibfnamefont {F.}~\bibnamefont
  {Antonini}},\ }\href {\doibase 10.3847/1538-4357/aa8628} {\bibfield
  {journal} {\bibinfo  {journal} {Astrophys. J.}\ }\textbf {\bibinfo {volume}
  {846}},\ \bibinfo {pages} {146} (\bibinfo {year} {2017})},\ \Eprint
  {http://arxiv.org/abs/1705.05848} {arXiv:1705.05848 [astro-ph.HE]}
  \BibitemShut {NoStop}%
\bibitem [{\citenamefont {Samsing}\ \emph {et~al.}(2018)\citenamefont
  {Samsing}, \citenamefont {MacLeod},\ and\ \citenamefont
  {Ramirez-Ruiz}}]{samsing-2017}%
  \BibitemOpen
  \bibfield  {author} {\bibinfo {author} {\bibfnamefont {J.}~\bibnamefont
  {Samsing}}, \bibinfo {author} {\bibfnamefont {M.}~\bibnamefont {MacLeod}}, \
  and\ \bibinfo {author} {\bibfnamefont {E.}~\bibnamefont {Ramirez-Ruiz}},\
  }\href {\doibase 10.3847/1538-4357/aaa715} {\bibfield  {journal} {\bibinfo
  {journal} {Astrophys. J.}\ }\textbf {\bibinfo {volume} {853}},\ \bibinfo
  {pages} {140} (\bibinfo {year} {2018})},\ \Eprint
  {http://arxiv.org/abs/1706.03776} {arXiv:1706.03776 [astro-ph.HE]}
  \BibitemShut {NoStop}%
\bibitem [{\citenamefont {{Rodriguez}}\ \emph {et~al.}(2018)\citenamefont
  {{Rodriguez}}, \citenamefont {{Amaro-Seoane}}, \citenamefont {{Chatterjee}},\
  and\ \citenamefont {{Rasio}}}]{rodriguez-2017}%
  \BibitemOpen
  \bibfield  {author} {\bibinfo {author} {\bibfnamefont {C.~L.}\ \bibnamefont
  {{Rodriguez}}}, \bibinfo {author} {\bibfnamefont {P.}~\bibnamefont
  {{Amaro-Seoane}}}, \bibinfo {author} {\bibfnamefont {S.}~\bibnamefont
  {{Chatterjee}}}, \ and\ \bibinfo {author} {\bibfnamefont {F.~A.}\
  \bibnamefont {{Rasio}}},\ }\href {\doibase 10.1103/PhysRevLett.120.151101}
  {\bibfield  {journal} {\bibinfo  {journal} {Phys. Rev. Lett.}\ }\textbf
  {\bibinfo {volume} {120}},\ \bibinfo {eid} {151101} (\bibinfo {year}
  {2018})},\ \Eprint {http://arxiv.org/abs/1712.04937} {arXiv:1712.04937
  [astro-ph.HE]} \BibitemShut {NoStop}%
\bibitem [{\citenamefont {Hoang}\ \emph {et~al.}(2018)\citenamefont {Hoang},
  \citenamefont {Naoz}, \citenamefont {Kocsis}, \citenamefont {Rasio},\ and\
  \citenamefont {Dosopoulou}}]{hoang-2018}%
  \BibitemOpen
  \bibfield  {author} {\bibinfo {author} {\bibfnamefont {B.-M.}\ \bibnamefont
  {Hoang}}, \bibinfo {author} {\bibfnamefont {S.}~\bibnamefont {Naoz}},
  \bibinfo {author} {\bibfnamefont {B.}~\bibnamefont {Kocsis}}, \bibinfo
  {author} {\bibfnamefont {F.~A.}\ \bibnamefont {Rasio}}, \ and\ \bibinfo
  {author} {\bibfnamefont {F.}~\bibnamefont {Dosopoulou}},\ }\href {\doibase
  10.3847/1538-4357/aaafce} {\bibfield  {journal} {\bibinfo  {journal}
  {Astrophys. J.}\ }\textbf {\bibinfo {volume} {856}},\ \bibinfo {pages} {140}
  (\bibinfo {year} {2018})},\ \Eprint {http://arxiv.org/abs/1706.09896}
  {arXiv:1706.09896 [astro-ph.HE]} \BibitemShut {NoStop}%
\bibitem [{\citenamefont {{Samsing}}\ \emph {et~al.}(2018)\citenamefont
  {{Samsing}}, \citenamefont {{D'Orazio}}, \citenamefont {{Askar}},\ and\
  \citenamefont {{Giersz}}}]{samsing-2018}%
  \BibitemOpen
  \bibfield  {author} {\bibinfo {author} {\bibfnamefont {J.}~\bibnamefont
  {{Samsing}}}, \bibinfo {author} {\bibfnamefont {D.~J.}\ \bibnamefont
  {{D'Orazio}}}, \bibinfo {author} {\bibfnamefont {A.}~\bibnamefont {{Askar}}},
  \ and\ \bibinfo {author} {\bibfnamefont {M.}~\bibnamefont {{Giersz}}},\
  }\href@noop {} {\bibfield  {journal} {\bibinfo  {journal} {ArXiv e-prints}\ }
  (\bibinfo {year} {2018})},\ \Eprint {http://arxiv.org/abs/1802.08654}
  {arXiv:1802.08654 [astro-ph.HE]} \BibitemShut {NoStop}%
\bibitem [{\citenamefont {Abbott}\ \emph
  {et~al.}(2017{\natexlab{e}})\citenamefont {Abbott} \emph
  {et~al.}}]{GW150914-systematics}%
  \BibitemOpen
  \bibfield  {author} {\bibinfo {author} {\bibfnamefont {B.~P.}\ \bibnamefont
  {Abbott}} \emph {et~al.},\ }\href {\doibase 10.1088/1361-6382/aa6854}
  {\bibfield  {journal} {\bibinfo  {journal} {Classical and Quantum Gravity}\
  }\textbf {\bibinfo {volume} {34}},\ \bibinfo {pages} {104002} (\bibinfo
  {year} {2017}{\natexlab{e}})},\ \Eprint {http://arxiv.org/abs/1611.07531}
  {arXiv:1611.07531 [gr-qc]} \BibitemShut {NoStop}%
\bibitem [{\citenamefont {Blaes}\ \emph {et~al.}(2002)\citenamefont {Blaes},
  \citenamefont {Lee},\ and\ \citenamefont {Socrates}}]{blaes-2002}%
  \BibitemOpen
  \bibfield  {author} {\bibinfo {author} {\bibfnamefont {O.}~\bibnamefont
  {Blaes}}, \bibinfo {author} {\bibfnamefont {M.~H.}\ \bibnamefont {Lee}}, \
  and\ \bibinfo {author} {\bibfnamefont {A.}~\bibnamefont {Socrates}},\ }\href
  {\doibase 10.1086/342655} {\bibfield  {journal} {\bibinfo  {journal}
  {Astrophys. J.}\ }\textbf {\bibinfo {volume} {578}},\ \bibinfo {pages} {775}
  (\bibinfo {year} {2002})},\ \Eprint {http://arxiv.org/abs/astro-ph/0203370}
  {arXiv:astro-ph/0203370 [astro-ph]} \BibitemShut {NoStop}%
\bibitem [{\citenamefont {Hoffman}\ and\ \citenamefont
  {Loeb}(2007)}]{hoffman-2007}%
  \BibitemOpen
  \bibfield  {author} {\bibinfo {author} {\bibfnamefont {L.}~\bibnamefont
  {Hoffman}}\ and\ \bibinfo {author} {\bibfnamefont {A.}~\bibnamefont {Loeb}},\
  }\href {\doibase 10.1111/j.1365-2966.2007.11694.x} {\bibfield  {journal}
  {\bibinfo  {journal} {Mon. Not. R. Astron. Soc.}\ }\textbf {\bibinfo {volume}
  {377}},\ \bibinfo {pages} {957} (\bibinfo {year} {2007})},\ \Eprint
  {http://arxiv.org/abs/astro-ph/0612517} {arXiv:astro-ph/0612517 [astro-ph]}
  \BibitemShut {NoStop}%
\bibitem [{\citenamefont {{Amaro-Seoane}}\ \emph {et~al.}(2010)\citenamefont
  {{Amaro-Seoane}}, \citenamefont {{Sesana}}, \citenamefont {{Hoffman}},
  \citenamefont {{Benacquista}}, \citenamefont {{Eichhorn}}, \citenamefont
  {{Makino}},\ and\ \citenamefont {{Spurzem}}}]{amaro-seoane-2010}%
  \BibitemOpen
  \bibfield  {author} {\bibinfo {author} {\bibfnamefont {P.}~\bibnamefont
  {{Amaro-Seoane}}}, \bibinfo {author} {\bibfnamefont {A.}~\bibnamefont
  {{Sesana}}}, \bibinfo {author} {\bibfnamefont {L.}~\bibnamefont {{Hoffman}}},
  \bibinfo {author} {\bibfnamefont {M.}~\bibnamefont {{Benacquista}}}, \bibinfo
  {author} {\bibfnamefont {C.}~\bibnamefont {{Eichhorn}}}, \bibinfo {author}
  {\bibfnamefont {J.}~\bibnamefont {{Makino}}}, \ and\ \bibinfo {author}
  {\bibfnamefont {R.}~\bibnamefont {{Spurzem}}},\ }\href {\doibase
  10.1111/j.1365-2966.2009.16104.x} {\bibfield  {journal} {\bibinfo  {journal}
  {Mon. Not. R. Astron. Soc.}\ }\textbf {\bibinfo {volume} {402}},\ \bibinfo
  {pages} {2308} (\bibinfo {year} {2010})},\ \Eprint
  {http://arxiv.org/abs/0910.1587} {arXiv:0910.1587 [astro-ph.CO]} \BibitemShut
  {NoStop}%
\bibitem [{\citenamefont {{Bonetti}}\ \emph {et~al.}(2017)\citenamefont
  {{Bonetti}}, \citenamefont {{Barausse}}, \citenamefont {{Faye}},
  \citenamefont {{Haardt}},\ and\ \citenamefont {{Sesana}}}]{bonetti-2017-1}%
  \BibitemOpen
  \bibfield  {author} {\bibinfo {author} {\bibfnamefont {M.}~\bibnamefont
  {{Bonetti}}}, \bibinfo {author} {\bibfnamefont {E.}~\bibnamefont
  {{Barausse}}}, \bibinfo {author} {\bibfnamefont {G.}~\bibnamefont {{Faye}}},
  \bibinfo {author} {\bibfnamefont {F.}~\bibnamefont {{Haardt}}}, \ and\
  \bibinfo {author} {\bibfnamefont {A.}~\bibnamefont {{Sesana}}},\ }\href
  {\doibase 10.1088/1361-6382/aa8da5} {\bibfield  {journal} {\bibinfo
  {journal} {Classical and Quantum Gravity}\ }\textbf {\bibinfo {volume}
  {34}},\ \bibinfo {eid} {215004} (\bibinfo {year} {2017})},\ \Eprint
  {http://arxiv.org/abs/1707.04902} {arXiv:1707.04902 [gr-qc]} \BibitemShut
  {NoStop}%
\bibitem [{\citenamefont {{Bonetti}}\ \emph
  {et~al.}(2018{\natexlab{a}})\citenamefont {{Bonetti}}, \citenamefont
  {{Haardt}}, \citenamefont {{Sesana}},\ and\ \citenamefont
  {{Barausse}}}]{bonetti-2017-2}%
  \BibitemOpen
  \bibfield  {author} {\bibinfo {author} {\bibfnamefont {M.}~\bibnamefont
  {{Bonetti}}}, \bibinfo {author} {\bibfnamefont {F.}~\bibnamefont {{Haardt}}},
  \bibinfo {author} {\bibfnamefont {A.}~\bibnamefont {{Sesana}}}, \ and\
  \bibinfo {author} {\bibnamefont {{Barausse}}},\ }\href {\doibase
  10.1093/mnras/sty896} {\bibfield  {journal} {\bibinfo  {journal} {Mon. Not.
  Roy. Astron. Soc.}\ }\textbf {\bibinfo {volume} {477}},\ \bibinfo {pages}
  {3910} (\bibinfo {year} {2018}{\natexlab{a}})},\ \Eprint
  {http://arxiv.org/abs/1709.06088} {arXiv:1709.06088 [astro-ph]} \BibitemShut
  {NoStop}%
\bibitem [{\citenamefont {{Bonetti}}\ \emph
  {et~al.}(2018{\natexlab{b}})\citenamefont {{Bonetti}}, \citenamefont
  {{Sesana}}, \citenamefont {{Barausse}},\ and\ \citenamefont
  {{Haardt}}}]{bonetti-2017-3}%
  \BibitemOpen
  \bibfield  {author} {\bibinfo {author} {\bibfnamefont {M.}~\bibnamefont
  {{Bonetti}}}, \bibinfo {author} {\bibfnamefont {A.}~\bibnamefont {{Sesana}}},
  \bibinfo {author} {\bibfnamefont {E.}~\bibnamefont {{Barausse}}}, \ and\
  \bibinfo {author} {\bibfnamefont {F.}~\bibnamefont {{Haardt}}},\ }\href
  {\doibase 10.1093/mnras/sty874} {\bibfield  {journal} {\bibinfo  {journal}
  {Mon. Not. Roy. Astron. Soc.}\ }\textbf {\bibinfo {volume} {477}},\ \bibinfo
  {pages} {2599} (\bibinfo {year} {2018}{\natexlab{b}})},\ \Eprint
  {http://arxiv.org/abs/1709.06095} {arXiv:1709.06095 [astro-ph]} \BibitemShut
  {NoStop}%
\bibitem [{\citenamefont {Klein}\ and\ \citenamefont
  {Jetzer}(2010)}]{klein-2010}%
  \BibitemOpen
  \bibfield  {author} {\bibinfo {author} {\bibfnamefont {A.}~\bibnamefont
  {Klein}}\ and\ \bibinfo {author} {\bibfnamefont {P.}~\bibnamefont {Jetzer}},\
  }\href {\doibase 10.1103/PhysRevD.81.124001} {\bibfield  {journal} {\bibinfo
  {journal} {Phys. Rev. D}\ }\textbf {\bibinfo {volume} {81}},\ \bibinfo
  {pages} {124001} (\bibinfo {year} {2010})},\ \Eprint
  {http://arxiv.org/abs/1005.2046} {arXiv:1005.2046 [gr-qc]} \BibitemShut
  {NoStop}%
\bibitem [{\citenamefont {Memmesheimer}\ \emph {et~al.}(2004)\citenamefont
  {Memmesheimer}, \citenamefont {Gopakumar},\ and\ \citenamefont
  {Sch{\"a}fer}}]{memmesheimer-2004}%
  \BibitemOpen
  \bibfield  {author} {\bibinfo {author} {\bibfnamefont {R.-M.}\ \bibnamefont
  {Memmesheimer}}, \bibinfo {author} {\bibfnamefont {A.}~\bibnamefont
  {Gopakumar}}, \ and\ \bibinfo {author} {\bibfnamefont {G.}~\bibnamefont
  {Sch{\"a}fer}},\ }\href {\doibase 10.1103/PhysRevD.70.104011} {\bibfield
  {journal} {\bibinfo  {journal} {Phys. Rev. D}\ }\textbf {\bibinfo {volume}
  {70}},\ \bibinfo {pages} {104011} (\bibinfo {year} {2004})},\ \Eprint
  {http://arxiv.org/abs/gr-qc/0407049} {arXiv:gr-qc/0407049 [gr-qc]}
  \BibitemShut {NoStop}%
\bibitem [{\citenamefont {Damour}\ \emph {et~al.}(2004)\citenamefont {Damour},
  \citenamefont {Gopakumar},\ and\ \citenamefont {Iyer}}]{damour-2004}%
  \BibitemOpen
  \bibfield  {author} {\bibinfo {author} {\bibfnamefont {T.}~\bibnamefont
  {Damour}}, \bibinfo {author} {\bibfnamefont {A.}~\bibnamefont {Gopakumar}}, \
  and\ \bibinfo {author} {\bibfnamefont {B.~R.}\ \bibnamefont {Iyer}},\ }\href
  {\doibase 10.1103/PhysRevD.70.064028} {\bibfield  {journal} {\bibinfo
  {journal} {Phys. Rev. D}\ }\textbf {\bibinfo {volume} {70}},\ \bibinfo
  {pages} {064028} (\bibinfo {year} {2004})},\ \Eprint
  {http://arxiv.org/abs/gr-qc/0404128} {arXiv:gr-qc/0404128 [gr-qc]}
  \BibitemShut {NoStop}%
\bibitem [{\citenamefont {K{\"o}nigsd{\"o}rffer}\ and\ \citenamefont
  {Gopakumar}(2006)}]{koenigsdoerffer-2006}%
  \BibitemOpen
  \bibfield  {author} {\bibinfo {author} {\bibfnamefont {C.}~\bibnamefont
  {K{\"o}nigsd{\"o}rffer}}\ and\ \bibinfo {author} {\bibfnamefont
  {A.}~\bibnamefont {Gopakumar}},\ }\href {\doibase 10.1103/PhysRevD.73.124012}
  {\bibfield  {journal} {\bibinfo  {journal} {Phys. Rev. D}\ }\textbf {\bibinfo
  {volume} {73}},\ \bibinfo {pages} {124012} (\bibinfo {year} {2006})},\
  \Eprint {http://arxiv.org/abs/gr-qc/0603056} {arXiv:gr-qc/0603056 [gr-qc]}
  \BibitemShut {NoStop}%
\bibitem [{\citenamefont {Arun}\ \emph
  {et~al.}(2008{\natexlab{a}})\citenamefont {Arun}, \citenamefont {Blanchet},
  \citenamefont {Iyer},\ and\ \citenamefont {Qusailah}}]{arun-2008}%
  \BibitemOpen
  \bibfield  {author} {\bibinfo {author} {\bibfnamefont {K.~G.}\ \bibnamefont
  {Arun}}, \bibinfo {author} {\bibfnamefont {L.}~\bibnamefont {Blanchet}},
  \bibinfo {author} {\bibfnamefont {B.~R.}\ \bibnamefont {Iyer}}, \ and\
  \bibinfo {author} {\bibfnamefont {M.~S.~S.}\ \bibnamefont {Qusailah}},\
  }\href {\doibase 10.1103/PhysRevD.77.064034} {\bibfield  {journal} {\bibinfo
  {journal} {Phys. Rev. D}\ }\textbf {\bibinfo {volume} {77}},\ \bibinfo
  {pages} {064034} (\bibinfo {year} {2008}{\natexlab{a}})},\ \Eprint
  {http://arxiv.org/abs/0711.0250} {arXiv:0711.0250 [gr-qc]} \BibitemShut
  {NoStop}%
\bibitem [{\citenamefont {Arun}\ \emph
  {et~al.}(2008{\natexlab{b}})\citenamefont {Arun}, \citenamefont {Blanchet},
  \citenamefont {Iyer},\ and\ \citenamefont {Qusailah}}]{arun-2008-2}%
  \BibitemOpen
  \bibfield  {author} {\bibinfo {author} {\bibfnamefont {K.~G.}\ \bibnamefont
  {Arun}}, \bibinfo {author} {\bibfnamefont {L.}~\bibnamefont {Blanchet}},
  \bibinfo {author} {\bibfnamefont {B.~R.}\ \bibnamefont {Iyer}}, \ and\
  \bibinfo {author} {\bibfnamefont {M.~S.~S.}\ \bibnamefont {Qusailah}},\
  }\href {\doibase 10.1103/PhysRevD.77.064035} {\bibfield  {journal} {\bibinfo
  {journal} {Phys. Rev. D}\ }\textbf {\bibinfo {volume} {77}},\ \bibinfo
  {pages} {064035} (\bibinfo {year} {2008}{\natexlab{b}})},\ \Eprint
  {http://arxiv.org/abs/0711.0302} {arXiv:0711.0302 [gr-qc]} \BibitemShut
  {NoStop}%
\bibitem [{\citenamefont {Arun}\ \emph {et~al.}(2009)\citenamefont {Arun},
  \citenamefont {Blanchet}, \citenamefont {Iyer},\ and\ \citenamefont
  {Sinha}}]{arun-2009}%
  \BibitemOpen
  \bibfield  {author} {\bibinfo {author} {\bibfnamefont {K.~G.}\ \bibnamefont
  {Arun}}, \bibinfo {author} {\bibfnamefont {L.}~\bibnamefont {Blanchet}},
  \bibinfo {author} {\bibfnamefont {B.~R.}\ \bibnamefont {Iyer}}, \ and\
  \bibinfo {author} {\bibfnamefont {S.}~\bibnamefont {Sinha}},\ }\href
  {\doibase 10.1103/PhysRevD.80.124018} {\bibfield  {journal} {\bibinfo
  {journal} {Phys. Rev. D}\ }\textbf {\bibinfo {volume} {80}},\ \bibinfo
  {pages} {124018} (\bibinfo {year} {2009})},\ \Eprint
  {http://arxiv.org/abs/0908.3854} {arXiv:0908.3854 [gr-qc]} \BibitemShut
  {NoStop}%
\bibitem [{\citenamefont {Mishra}\ \emph {et~al.}(2015)\citenamefont {Mishra},
  \citenamefont {Arun},\ and\ \citenamefont {Iyer}}]{mishra-2015}%
  \BibitemOpen
  \bibfield  {author} {\bibinfo {author} {\bibfnamefont {C.~K.}\ \bibnamefont
  {Mishra}}, \bibinfo {author} {\bibfnamefont {K.~G.}\ \bibnamefont {Arun}}, \
  and\ \bibinfo {author} {\bibfnamefont {B.~R.}\ \bibnamefont {Iyer}},\ }\href
  {\doibase 10.1103/PhysRevD.91.084040} {\bibfield  {journal} {\bibinfo
  {journal} {Phys. Rev. D}\ }\textbf {\bibinfo {volume} {91}},\ \bibinfo
  {pages} {084040} (\bibinfo {year} {2015})},\ \Eprint
  {http://arxiv.org/abs/1501.07096} {arXiv:1501.07096 [gr-qc]} \BibitemShut
  {NoStop}%
\bibitem [{\citenamefont {Gergely}\ \emph {et~al.}(1998)\citenamefont
  {Gergely}, \citenamefont {Perj{\'e}s},\ and\ \citenamefont
  {Vas{\'u}th}}]{gergely-1998}%
  \BibitemOpen
  \bibfield  {author} {\bibinfo {author} {\bibfnamefont {L.~{\'A}.}\
  \bibnamefont {Gergely}}, \bibinfo {author} {\bibfnamefont {Z.~I.}\
  \bibnamefont {Perj{\'e}s}}, \ and\ \bibinfo {author} {\bibfnamefont
  {M.}~\bibnamefont {Vas{\'u}th}},\ }\href {\doibase
  10.1103/PhysRevD.58.124001} {\bibfield  {journal} {\bibinfo  {journal} {Phys.
  Rev. D}\ }\textbf {\bibinfo {volume} {58}},\ \bibinfo {pages} {124001}
  (\bibinfo {year} {1998})}\BibitemShut {NoStop}%
\bibitem [{\citenamefont {Gergely}(1999)}]{gergely-1999}%
  \BibitemOpen
  \bibfield  {author} {\bibinfo {author} {\bibfnamefont {L.~{\'A}.}\
  \bibnamefont {Gergely}},\ }\href {\doibase 10.1103/PhysRevD.61.024035}
  {\bibfield  {journal} {\bibinfo  {journal} {Phys. Rev. D}\ }\textbf {\bibinfo
  {volume} {61}},\ \bibinfo {pages} {024035} (\bibinfo {year} {1999})},\
  \Eprint {http://arxiv.org/abs/gr-qc/991182} {arXiv:gr-qc/991182 [gr-qc]}
  \BibitemShut {NoStop}%
\bibitem [{\citenamefont {Gergely}\ and\ \citenamefont
  {Keresztes}(2003)}]{gergely-2002}%
  \BibitemOpen
  \bibfield  {author} {\bibinfo {author} {\bibfnamefont {L.~{\'A}.}\
  \bibnamefont {Gergely}}\ and\ \bibinfo {author} {\bibfnamefont
  {Z.}~\bibnamefont {Keresztes}},\ }\href {\doibase 10.1103/PhysRevD.67.024020}
  {\bibfield  {journal} {\bibinfo  {journal} {Phys. Rev. D}\ }\textbf {\bibinfo
  {volume} {67}},\ \bibinfo {pages} {024020} (\bibinfo {year} {2003})},\
  \Eprint {http://arxiv.org/abs/gr-qc/0211027} {arXiv:gr-qc/0211027 [gr-qc]}
  \BibitemShut {NoStop}%
\bibitem [{\citenamefont {Mik{\'o}czi}\ \emph {et~al.}(2005)\citenamefont
  {Mik{\'o}czi}, \citenamefont {Vas{\'u}th},\ and\ \citenamefont
  {Gergely}}]{mikoczi-2005}%
  \BibitemOpen
  \bibfield  {author} {\bibinfo {author} {\bibfnamefont {B.}~\bibnamefont
  {Mik{\'o}czi}}, \bibinfo {author} {\bibfnamefont {M.}~\bibnamefont
  {Vas{\'u}th}}, \ and\ \bibinfo {author} {\bibfnamefont {L.~{\'A}.}\
  \bibnamefont {Gergely}},\ }\href {\doibase 10.1103/PhysRevD.71.124043}
  {\bibfield  {journal} {\bibinfo  {journal} {Phys. Rev. D}\ }\textbf {\bibinfo
  {volume} {71}},\ \bibinfo {pages} {124043} (\bibinfo {year} {2005})},\
  \Eprint {http://arxiv.org/abs/gr-qc/0504538} {arXiv:gr-qc/0504538 [gr-qc]}
  \BibitemShut {NoStop}%
\bibitem [{\citenamefont {Keresztes}\ \emph {et~al.}(2005)\citenamefont
  {Keresztes}, \citenamefont {Mik{\'o}czi},\ and\ \citenamefont
  {Gergely}}]{keresztes-2005}%
  \BibitemOpen
  \bibfield  {author} {\bibinfo {author} {\bibfnamefont {Z.}~\bibnamefont
  {Keresztes}}, \bibinfo {author} {\bibfnamefont {B.}~\bibnamefont
  {Mik{\'o}czi}}, \ and\ \bibinfo {author} {\bibfnamefont {L.~{\'A}.}\
  \bibnamefont {Gergely}},\ }\href {\doibase 10.1103/PhysRevD.72.104022}
  {\bibfield  {journal} {\bibinfo  {journal} {Phys. Rev. D}\ }\textbf {\bibinfo
  {volume} {72}},\ \bibinfo {pages} {104022} (\bibinfo {year} {2005})},\
  \Eprint {http://arxiv.org/abs/astro-ph/0510602} {arXiv:astro-ph/0510602
  [astro-ph]} \BibitemShut {NoStop}%
\bibitem [{\citenamefont {Yunes}\ \emph {et~al.}(2009)\citenamefont {Yunes},
  \citenamefont {Arun}, \citenamefont {Berti},\ and\ \citenamefont
  {Will}}]{yunes-2009}%
  \BibitemOpen
  \bibfield  {author} {\bibinfo {author} {\bibfnamefont {N.}~\bibnamefont
  {Yunes}}, \bibinfo {author} {\bibfnamefont {K.~G.}\ \bibnamefont {Arun}},
  \bibinfo {author} {\bibfnamefont {E.}~\bibnamefont {Berti}}, \ and\ \bibinfo
  {author} {\bibfnamefont {C.~M.}\ \bibnamefont {Will}},\ }\href {\doibase
  10.1103/PhysRevD.80.084001} {\bibfield  {journal} {\bibinfo  {journal} {Phys.
  Rev. D}\ }\textbf {\bibinfo {volume} {80}},\ \bibinfo {pages} {084001}
  (\bibinfo {year} {2009})},\ \Eprint {http://arxiv.org/abs/0906.0313}
  {arXiv:0906.0313 [gr-qc]} \BibitemShut {NoStop}%
\bibitem [{\citenamefont {Cornish}\ and\ \citenamefont
  {Key}(2010)}]{cornish-2011}%
  \BibitemOpen
  \bibfield  {author} {\bibinfo {author} {\bibfnamefont {N.~J.}\ \bibnamefont
  {Cornish}}\ and\ \bibinfo {author} {\bibfnamefont {J.~S.}\ \bibnamefont
  {Key}},\ }\href {\doibase 10.1103/PhysRevD.82.044028} {\bibfield  {journal}
  {\bibinfo  {journal} {Phys. Rev. D}\ }\textbf {\bibinfo {volume} {82}},\
  \bibinfo {pages} {044028} (\bibinfo {year} {2010})},\ \Eprint
  {http://arxiv.org/abs/1004.5322} {arXiv:1004.5322 [gr-qc]} \BibitemShut
  {NoStop}%
\bibitem [{\citenamefont {Cornish}\ and\ \citenamefont
  {Key}(2011)}]{cornish-err}%
  \BibitemOpen
  \bibfield  {author} {\bibinfo {author} {\bibfnamefont {N.~J.}\ \bibnamefont
  {Cornish}}\ and\ \bibinfo {author} {\bibfnamefont {J.~S.}\ \bibnamefont
  {Key}},\ }\href {\doibase 10.1103/PhysRevD.84.029901} {\bibfield  {journal}
  {\bibinfo  {journal} {Phys. Rev. D}\ }\textbf {\bibinfo {volume} {84}},\
  \bibinfo {pages} {029901(E)} (\bibinfo {year} {2011})}\BibitemShut {NoStop}%
\bibitem [{\citenamefont {Key}\ and\ \citenamefont {Cornish}(2011)}]{key-2011}%
  \BibitemOpen
  \bibfield  {author} {\bibinfo {author} {\bibfnamefont {J.~S.}\ \bibnamefont
  {Key}}\ and\ \bibinfo {author} {\bibfnamefont {N.~J.}\ \bibnamefont
  {Cornish}},\ }\href {\doibase 10.1103/PhysRevD.83.083001} {\bibfield
  {journal} {\bibinfo  {journal} {Phys. Rev. D}\ }\textbf {\bibinfo {volume}
  {83}},\ \bibinfo {pages} {083001} (\bibinfo {year} {2011})},\ \Eprint
  {http://arxiv.org/abs/1006.3759} {arXiv:1006.3759 [gr-qc]} \BibitemShut
  {NoStop}%
\bibitem [{\citenamefont {Gopakumar}\ and\ \citenamefont
  {Sch{\"a}fer}(2011)}]{gopakumar-2011}%
  \BibitemOpen
  \bibfield  {author} {\bibinfo {author} {\bibfnamefont {A.}~\bibnamefont
  {Gopakumar}}\ and\ \bibinfo {author} {\bibfnamefont {G.}~\bibnamefont
  {Sch{\"a}fer}},\ }\href {\doibase 10.1103/PhysRevD.84.124007} {\bibfield
  {journal} {\bibinfo  {journal} {Phys. Rev. D}\ }\textbf {\bibinfo {volume}
  {84}},\ \bibinfo {pages} {124007} (\bibinfo {year} {2011})}\BibitemShut
  {NoStop}%
\bibitem [{\citenamefont {Huerta}\ \emph {et~al.}(2014)\citenamefont {Huerta},
  \citenamefont {Kumar}, \citenamefont {McWilliams}, \citenamefont
  {O'Shaughnessy},\ and\ \citenamefont {Yunes}}]{huerta-2014}%
  \BibitemOpen
  \bibfield  {author} {\bibinfo {author} {\bibfnamefont {E.~A.}\ \bibnamefont
  {Huerta}}, \bibinfo {author} {\bibfnamefont {P.}~\bibnamefont {Kumar}},
  \bibinfo {author} {\bibfnamefont {S.~T.}\ \bibnamefont {McWilliams}},
  \bibinfo {author} {\bibfnamefont {R.}~\bibnamefont {O'Shaughnessy}}, \ and\
  \bibinfo {author} {\bibfnamefont {N.}~\bibnamefont {Yunes}},\ }\href
  {\doibase 10.1103/PhysRevD.90.084016} {\bibfield  {journal} {\bibinfo
  {journal} {Phys. Rev. D}\ }\textbf {\bibinfo {volume} {90}},\ \bibinfo
  {pages} {084016} (\bibinfo {year} {2014})},\ \Eprint
  {http://arxiv.org/abs/1408.3406} {arXiv:1408.3406 [gr-qc]} \BibitemShut
  {NoStop}%
\bibitem [{\citenamefont {Tanay}\ \emph {et~al.}(2016)\citenamefont {Tanay},
  \citenamefont {Haney},\ and\ \citenamefont {Gopakumar}}]{tanay-2016}%
  \BibitemOpen
  \bibfield  {author} {\bibinfo {author} {\bibfnamefont {S.}~\bibnamefont
  {Tanay}}, \bibinfo {author} {\bibfnamefont {M.}~\bibnamefont {Haney}}, \ and\
  \bibinfo {author} {\bibfnamefont {A.}~\bibnamefont {Gopakumar}},\ }\href
  {\doibase 10.1103/PhysRevD.93.064031} {\bibfield  {journal} {\bibinfo
  {journal} {Phys. Rev. D}\ }\textbf {\bibinfo {volume} {93}},\ \bibinfo
  {pages} {064031} (\bibinfo {year} {2016})},\ \Eprint
  {http://arxiv.org/abs/1602.03081} {arXiv:1602.03081 [gr-qc]} \BibitemShut
  {NoStop}%
\bibitem [{\citenamefont {{Moore}}\ \emph {et~al.}(2016)\citenamefont
  {{Moore}}, \citenamefont {{Favata}}, \citenamefont {{Arun}},\ and\
  \citenamefont {{Mishra}}}]{moore-2016}%
  \BibitemOpen
  \bibfield  {author} {\bibinfo {author} {\bibfnamefont {B.}~\bibnamefont
  {{Moore}}}, \bibinfo {author} {\bibfnamefont {M.}~\bibnamefont {{Favata}}},
  \bibinfo {author} {\bibfnamefont {K.~G.}\ \bibnamefont {{Arun}}}, \ and\
  \bibinfo {author} {\bibfnamefont {C.~K.}\ \bibnamefont {{Mishra}}},\ }\href
  {\doibase 10.1103/PhysRevD.93.124061} {\bibfield  {journal} {\bibinfo
  {journal} {\prd}\ }\textbf {\bibinfo {volume} {93}},\ \bibinfo {eid} {124061}
  (\bibinfo {year} {2016})},\ \Eprint {http://arxiv.org/abs/1605.00304}
  {arXiv:1605.00304 [gr-qc]} \BibitemShut {NoStop}%
\bibitem [{\citenamefont {Huerta}\ \emph {et~al.}(2017)\citenamefont {Huerta},
  \citenamefont {Kumar}, \citenamefont {Agarwal}, \citenamefont {George},
  \citenamefont {Schive}, \citenamefont {Pfeiffer}, \citenamefont {Haas},
  \citenamefont {Ren}, \citenamefont {Chu}, \citenamefont {Boyle},
  \citenamefont {Hemberger}, \citenamefont {Kidder}, \citenamefont {Scheel},\
  and\ \citenamefont {Szilagyi}}]{huerta-2017-1}%
  \BibitemOpen
  \bibfield  {author} {\bibinfo {author} {\bibfnamefont {E.~A.}\ \bibnamefont
  {Huerta}}, \bibinfo {author} {\bibfnamefont {P.}~\bibnamefont {Kumar}},
  \bibinfo {author} {\bibfnamefont {B.}~\bibnamefont {Agarwal}}, \bibinfo
  {author} {\bibfnamefont {D.}~\bibnamefont {George}}, \bibinfo {author}
  {\bibfnamefont {H.-Y.}\ \bibnamefont {Schive}}, \bibinfo {author}
  {\bibfnamefont {H.~P.}\ \bibnamefont {Pfeiffer}}, \bibinfo {author}
  {\bibfnamefont {R.}~\bibnamefont {Haas}}, \bibinfo {author} {\bibfnamefont
  {W.}~\bibnamefont {Ren}}, \bibinfo {author} {\bibfnamefont {T.}~\bibnamefont
  {Chu}}, \bibinfo {author} {\bibfnamefont {M.}~\bibnamefont {Boyle}}, \bibinfo
  {author} {\bibfnamefont {D.~A.}\ \bibnamefont {Hemberger}}, \bibinfo {author}
  {\bibfnamefont {L.~E.}\ \bibnamefont {Kidder}}, \bibinfo {author}
  {\bibfnamefont {M.~A.}\ \bibnamefont {Scheel}}, \ and\ \bibinfo {author}
  {\bibfnamefont {B.}~\bibnamefont {Szilagyi}},\ }\href {\doibase
  10.1103/PhysRevD.95.024038} {\bibfield  {journal} {\bibinfo  {journal} {Phys.
  Rev. D}\ }\textbf {\bibinfo {volume} {95}},\ \bibinfo {pages} {024038}
  (\bibinfo {year} {2017})},\ \Eprint {http://arxiv.org/abs/1609.05933}
  {arXiv:1609.05933 [gr-qc]} \BibitemShut {NoStop}%
\bibitem [{\citenamefont {{Huerta}}\ \emph {et~al.}(2018)\citenamefont
  {{Huerta}}, \citenamefont {{Moore}}, \citenamefont {{Kumar}}, \citenamefont
  {{George}}, \citenamefont {{Chua}}, \citenamefont {{Haas}}, \citenamefont
  {{Wessel}}, \citenamefont {{Johnson}}, \citenamefont {{Glennon}},
  \citenamefont {{Rebei}}, \citenamefont {{Holgado}}, \citenamefont {{Gair}},\
  and\ \citenamefont {{Pfeiffer}}}]{huerta-2017-2}%
  \BibitemOpen
  \bibfield  {author} {\bibinfo {author} {\bibfnamefont {E.~A.}\ \bibnamefont
  {{Huerta}}}, \bibinfo {author} {\bibfnamefont {C.~J.}\ \bibnamefont
  {{Moore}}}, \bibinfo {author} {\bibfnamefont {P.}~\bibnamefont {{Kumar}}},
  \bibinfo {author} {\bibfnamefont {D.}~\bibnamefont {{George}}}, \bibinfo
  {author} {\bibfnamefont {A.~J.~K.}\ \bibnamefont {{Chua}}}, \bibinfo {author}
  {\bibfnamefont {R.}~\bibnamefont {{Haas}}}, \bibinfo {author} {\bibfnamefont
  {E.}~\bibnamefont {{Wessel}}}, \bibinfo {author} {\bibfnamefont
  {D.}~\bibnamefont {{Johnson}}}, \bibinfo {author} {\bibfnamefont
  {D.}~\bibnamefont {{Glennon}}}, \bibinfo {author} {\bibfnamefont
  {A.}~\bibnamefont {{Rebei}}}, \bibinfo {author} {\bibfnamefont {A.~M.}\
  \bibnamefont {{Holgado}}}, \bibinfo {author} {\bibfnamefont {J.~R.}\
  \bibnamefont {{Gair}}}, \ and\ \bibinfo {author} {\bibfnamefont {H.~P.}\
  \bibnamefont {{Pfeiffer}}},\ }\href {\doibase 10.1103/PhysRevD.97.024031}
  {\bibfield  {journal} {\bibinfo  {journal} {Phys. Rev. D}\ }\textbf {\bibinfo
  {volume} {97}},\ \bibinfo {eid} {024031} (\bibinfo {year} {2018})},\ \Eprint
  {http://arxiv.org/abs/1711.06276} {arXiv:1711.06276 [gr-qc]} \BibitemShut
  {NoStop}%
\bibitem [{\citenamefont {{Hinder}}\ \emph {et~al.}(2018)\citenamefont
  {{Hinder}}, \citenamefont {{Kidder}},\ and\ \citenamefont
  {{Pfeiffer}}}]{hinder-2017}%
  \BibitemOpen
  \bibfield  {author} {\bibinfo {author} {\bibfnamefont {I.}~\bibnamefont
  {{Hinder}}}, \bibinfo {author} {\bibfnamefont {L.~E.}\ \bibnamefont
  {{Kidder}}}, \ and\ \bibinfo {author} {\bibfnamefont {H.~P.}\ \bibnamefont
  {{Pfeiffer}}},\ }\href {\doibase 10.1103/PhysRevD.98.044015} {\bibfield
  {journal} {\bibinfo  {journal} {Phys. Rev. D}\ }\textbf {\bibinfo {volume}
  {98}},\ \bibinfo {eid} {044015} (\bibinfo {year} {2018})},\ \Eprint
  {http://arxiv.org/abs/1709.02007} {arXiv:1709.02007 [gr-qc]} \BibitemShut
  {NoStop}%
\bibitem [{\citenamefont {Hinderer}\ and\ \citenamefont
  {Babak}(2017)}]{hinderer-2017}%
  \BibitemOpen
  \bibfield  {author} {\bibinfo {author} {\bibfnamefont {T.}~\bibnamefont
  {Hinderer}}\ and\ \bibinfo {author} {\bibfnamefont {S.}~\bibnamefont
  {Babak}},\ }\href {\doibase 10.1103/PhysRevD.96.104048} {\bibfield  {journal}
  {\bibinfo  {journal} {Phys. Rev. D}\ }\textbf {\bibinfo {volume} {96}},\
  \bibinfo {pages} {104048} (\bibinfo {year} {2017})},\ \Eprint
  {http://arxiv.org/abs/1707.08426} {arXiv:1707.08426 [gr-qc]} \BibitemShut
  {NoStop}%
\bibitem [{\citenamefont {Cao}\ and\ \citenamefont {Han}(2017)}]{cao-2017}%
  \BibitemOpen
  \bibfield  {author} {\bibinfo {author} {\bibfnamefont {Z.}~\bibnamefont
  {Cao}}\ and\ \bibinfo {author} {\bibfnamefont {W.-B.}\ \bibnamefont {Han}},\
  }\href {\doibase 10.1103/PhysRevD.96.044028} {\bibfield  {journal} {\bibinfo
  {journal} {Phys. Rev. D}\ }\textbf {\bibinfo {volume} {96}},\ \bibinfo
  {pages} {044028} (\bibinfo {year} {2017})},\ \Eprint
  {http://arxiv.org/abs/1708.00166} {arXiv:1708.00166 [gr-qc]} \BibitemShut
  {NoStop}%
\bibitem [{\citenamefont {Klein}\ \emph {et~al.}(2014)\citenamefont {Klein},
  \citenamefont {Cornish},\ and\ \citenamefont {Yunes}}]{klein-2014}%
  \BibitemOpen
  \bibfield  {author} {\bibinfo {author} {\bibfnamefont {A.}~\bibnamefont
  {Klein}}, \bibinfo {author} {\bibfnamefont {N.}~\bibnamefont {Cornish}}, \
  and\ \bibinfo {author} {\bibfnamefont {N.}~\bibnamefont {Yunes}},\ }\href
  {\doibase 10.1103/PhysRevD.90.124029} {\bibfield  {journal} {\bibinfo
  {journal} {Phys. Rev. D}\ }\textbf {\bibinfo {volume} {90}},\ \bibinfo
  {pages} {124029} (\bibinfo {year} {2014})},\ \Eprint
  {http://arxiv.org/abs/1408.5158} {arXiv:1408.5158 [gr-qc]} \BibitemShut
  {NoStop}%
\bibitem [{\citenamefont {Barker}\ and\ \citenamefont
  {O'Connell}(1975)}]{barker-1975}%
  \BibitemOpen
  \bibfield  {author} {\bibinfo {author} {\bibfnamefont {B.~M.}\ \bibnamefont
  {Barker}}\ and\ \bibinfo {author} {\bibfnamefont {R.~F.}\ \bibnamefont
  {O'Connell}},\ }\href {\doibase 10.1103/PhysRevD.12.329} {\bibfield
  {journal} {\bibinfo  {journal} {Phys. Rev. D}\ }\textbf {\bibinfo {volume}
  {12}},\ \bibinfo {pages} {329} (\bibinfo {year} {1975})}\BibitemShut
  {NoStop}%
\bibitem [{\citenamefont {Cutler}\ \emph {et~al.}(1994)\citenamefont {Cutler},
  \citenamefont {Kennefick},\ and\ \citenamefont {Poisson}}]{cutler-1994}%
  \BibitemOpen
  \bibfield  {author} {\bibinfo {author} {\bibfnamefont {C.}~\bibnamefont
  {Cutler}}, \bibinfo {author} {\bibfnamefont {D.}~\bibnamefont {Kennefick}}, \
  and\ \bibinfo {author} {\bibfnamefont {E.}~\bibnamefont {Poisson}},\ }\href
  {\doibase 10.1103/PhysRevD.50.3816} {\bibfield  {journal} {\bibinfo
  {journal} {Phys. Rev. D}\ }\textbf {\bibinfo {volume} {50}},\ \bibinfo
  {pages} {3816} (\bibinfo {year} {1994})}\BibitemShut {NoStop}%
\bibitem [{\citenamefont {Loutrel}\ \emph {et~al.}(2018)\citenamefont
  {Loutrel}, \citenamefont {Liebersbach}, \citenamefont {Yunes},\ and\
  \citenamefont {Cornish}}]{loutrel-2018}%
  \BibitemOpen
  \bibfield  {author} {\bibinfo {author} {\bibfnamefont {N.}~\bibnamefont
  {Loutrel}}, \bibinfo {author} {\bibfnamefont {S.}~\bibnamefont
  {Liebersbach}}, \bibinfo {author} {\bibfnamefont {N.}~\bibnamefont {Yunes}},
  \ and\ \bibinfo {author} {\bibfnamefont {N.}~\bibnamefont {Cornish}},\
  }\href@noop {} {\bibfield  {journal} {\bibinfo  {journal} {ArXiv e-prints}\ }
  (\bibinfo {year} {2018})},\ \Eprint {http://arxiv.org/abs/1801.09009}
  {arXiv:1801.09009 [gr-qc]} \BibitemShut {NoStop}%
\bibitem [{\citenamefont {Racine}(2008)}]{racine-2008}%
  \BibitemOpen
  \bibfield  {author} {\bibinfo {author} {\bibfnamefont {E.}~\bibnamefont
  {Racine}},\ }\href {\doibase 10.1103/PhysRevD.78.044021} {\bibfield
  {journal} {\bibinfo  {journal} {Phys. Rev. D}\ }\textbf {\bibinfo {volume}
  {78}},\ \bibinfo {pages} {044021} (\bibinfo {year} {2008})},\ \Eprint
  {http://arxiv.org/abs/0803.1820} {arXiv:0803.1820 [gr-qc]} \BibitemShut
  {NoStop}%
\bibitem [{\citenamefont {Cutler}(1998)}]{cutler-1998}%
  \BibitemOpen
  \bibfield  {author} {\bibinfo {author} {\bibfnamefont {C.}~\bibnamefont
  {Cutler}},\ }\href {\doibase 10.1103/PhysRevD.57.7089} {\bibfield  {journal}
  {\bibinfo  {journal} {Phys. Rev. D}\ }\textbf {\bibinfo {volume} {57}},\
  \bibinfo {pages} {7089} (\bibinfo {year} {1998})}\BibitemShut {NoStop}%
\bibitem [{\citenamefont {Apostolatos}\ \emph {et~al.}(1994)\citenamefont
  {Apostolatos}, \citenamefont {Cutler}, \citenamefont {Sussman},\ and\
  \citenamefont {Thorne}}]{apostolatos-1994}%
  \BibitemOpen
  \bibfield  {author} {\bibinfo {author} {\bibfnamefont {T.~A.}\ \bibnamefont
  {Apostolatos}}, \bibinfo {author} {\bibfnamefont {C.}~\bibnamefont {Cutler}},
  \bibinfo {author} {\bibfnamefont {G.~J.}\ \bibnamefont {Sussman}}, \ and\
  \bibinfo {author} {\bibfnamefont {K.~S.}\ \bibnamefont {Thorne}},\ }\href
  {\doibase 10.1103/PhysRevD.49.6274} {\bibfield  {journal} {\bibinfo
  {journal} {Phys. Rev. D}\ }\textbf {\bibinfo {volume} {49}},\ \bibinfo
  {pages} {6274} (\bibinfo {year} {1994})}\BibitemShut {NoStop}%
\bibitem [{\citenamefont {{Boetzel}}\ \emph {et~al.}(2017)\citenamefont
  {{Boetzel}}, \citenamefont {{Susobhanan}}, \citenamefont {{Gopakumar}},
  \citenamefont {{Klein}},\ and\ \citenamefont {{Jetzer}}}]{boetzel-2017}%
  \BibitemOpen
  \bibfield  {author} {\bibinfo {author} {\bibfnamefont {Y.}~\bibnamefont
  {{Boetzel}}}, \bibinfo {author} {\bibfnamefont {A.}~\bibnamefont
  {{Susobhanan}}}, \bibinfo {author} {\bibfnamefont {A.}~\bibnamefont
  {{Gopakumar}}}, \bibinfo {author} {\bibfnamefont {A.}~\bibnamefont
  {{Klein}}}, \ and\ \bibinfo {author} {\bibfnamefont {P.}~\bibnamefont
  {{Jetzer}}},\ }\href {\doibase 10.1103/PhysRevD.96.044011} {\bibfield
  {journal} {\bibinfo  {journal} {\prd}\ }\textbf {\bibinfo {volume} {96}},\
  \bibinfo {eid} {044011} (\bibinfo {year} {2017})},\ \Eprint
  {http://arxiv.org/abs/1707.02088} {arXiv:1707.02088 [gr-qc]} \BibitemShut
  {NoStop}%
\bibitem [{\citenamefont {Bender}\ and\ \citenamefont {Orszag}(1999)}]{Bender}%
  \BibitemOpen
  \bibfield  {author} {\bibinfo {author} {\bibfnamefont {C.~M.}\ \bibnamefont
  {Bender}}\ and\ \bibinfo {author} {\bibfnamefont {S.~A.}\ \bibnamefont
  {Orszag}},\ }\href@noop {} {\emph {\bibinfo {title} {Advanced mathematical
  methods for scientists and engineers I: Asymptotic methods and perturbation
  theory}}}\ (\bibinfo  {publisher} {Springer},\ \bibinfo {address} {New
  York},\ \bibinfo {year} {1999})\BibitemShut {NoStop}%
\bibitem [{\citenamefont {Chatziioannou}\ \emph {et~al.}(2017)\citenamefont
  {Chatziioannou}, \citenamefont {Klein}, \citenamefont {Yunes},\ and\
  \citenamefont {Cornish}}]{chatziioannou-2017}%
  \BibitemOpen
  \bibfield  {author} {\bibinfo {author} {\bibfnamefont {K.}~\bibnamefont
  {Chatziioannou}}, \bibinfo {author} {\bibfnamefont {A.}~\bibnamefont
  {Klein}}, \bibinfo {author} {\bibfnamefont {N.}~\bibnamefont {Yunes}}, \ and\
  \bibinfo {author} {\bibfnamefont {N.~J.}\ \bibnamefont {Cornish}},\ }\href
  {\doibase 10.1103/PhysRevD.95.104004} {\bibfield  {journal} {\bibinfo
  {journal} {Phys. Rev. D}\ }\textbf {\bibinfo {volume} {95}},\ \bibinfo
  {pages} {104004} (\bibinfo {year} {2017})},\ \Eprint
  {http://arxiv.org/abs/1703.03967} {arXiv:1703.03967 [gr-qc]} \BibitemShut
  {NoStop}%
\bibitem [{\citenamefont {Poisson}(1998)}]{poisson-1998}%
  \BibitemOpen
  \bibfield  {author} {\bibinfo {author} {\bibfnamefont {E.}~\bibnamefont
  {Poisson}},\ }\href {\doibase 10.1103/PhysRevD.57.5287} {\bibfield  {journal}
  {\bibinfo  {journal} {Phys. Rev. D}\ }\textbf {\bibinfo {volume} {57}},\
  \bibinfo {pages} {5287} (\bibinfo {year} {1998})},\ \Eprint
  {http://arxiv.org/abs/gr-qc/9709032} {arXiv:gr-qc/9709032 [gr-qc]}
  \BibitemShut {NoStop}%
\end{thebibliography}%

\end{document}